\documentclass[preprint,3p,times,10pt]{elsarticle}

\usepackage{titlesec}

\titleformat{\section}
  {\normalfont\fontsize{13}{18}\selectfont\bfseries}
  {\thesection}{1em}{}
\titlespacing*{\section}
  {0pt}{2.5ex plus 1ex minus .2ex}{1.5ex plus .2ex}

\titleformat{\subsection}
  {\normalfont\fontsize{12}{16}\selectfont\bfseries}
  {\thesubsection}{1em}{}
\titlespacing*{\subsection}
  {0pt}{2ex plus 0.5ex minus .2ex}{1.2ex plus .2ex}

\titleformat{\subsubsection}
  {\normalfont\fontsize{11}{14}\selectfont\bfseries}
  {\thesubsubsection}{1em}{}
\titlespacing*{\subsubsection}
  {0pt}{1.5ex plus 0.5ex minus .2ex}{1ex plus .2ex}

\titleformat{\paragraph}[runin]
  {\normalfont\normalsize\bfseries}
  {\theparagraph}{1em}{}
\titlespacing*{\paragraph}
  {0pt}{1ex plus 0.5ex}{1em}

\usepackage{indentfirst}
\usepackage{amsmath}
\usepackage{graphicx}
\usepackage{hyperref}
\usepackage{amssymb}
\usepackage{xcolor}
\usepackage{algorithm} 
\usepackage{algpseudocode} 
\usepackage{adjustbox}
\usepackage{multirow}
\usepackage{array}

\usepackage{float} 
\usepackage{enumitem}

\usepackage{bm}
\usepackage{caption}
\usepackage{subcaption}  
\usepackage{cleveref}
\crefname{figure}{Figure.}{Figure.}

\journal{Comput. Methods Appl. Mech. Engrg.}

\begin{document}

\begin{frontmatter}

\title{Point-wise Diffusion Models for Physical Systems with Shape Variations: Application to Spatio-temporal and Large-scale system}

\author[inst1]{Jiyong Kim} \ead{sbo05050@kaist.ac.kr}

\affiliation[inst1]{organization={Cho Chun Shik Graduate School of Mobility},
            addressline={Korea Advanced Institute of Science and Technology}, 
            city={Daejeon},
            postcode={34051}, 
            country={Republic of korea}}

\author[inst1]{Sunwoong Yang\corref{cor1}} \ead{sunwoongy@kaist.ac.kr}
\author[inst1,inst2]{Namwoo Kang\corref{cor1}} \ead{nwkang@kaist.ac.kr}

\cortext[cor1]{Co-corresponding authors.}

\affiliation[inst2]{organization={Narnia Labs},
            addressline={193, Munji-ro}, 
            city={Daejeon},
            postcode={34051}, 
            country={Republic of korea}}

\begin{abstract}
This study introduces a novel point-wise diffusion model that processes spatio-temporal points independently to efficiently predict complex physical systems with shape variations. This methodological contribution lies in applying forward and backward diffusion processes at individual spatio-temporal points, coupled with a point-wise diffusion transformer architecture for denoising. Unlike conventional image-based diffusion models that operate on structured data representations, this framework enables direct processing of any data formats including meshes and point clouds while preserving geometric fidelity. We validate our approach across three distinct physical domains with complex geometric configurations: 2D spatio-temporal systems including cylinder fluid flow and OLED drop impact test, and 3D large-scale system for road-car external aerodynamics. To justify the necessity of our point-wise approach for real-time prediction applications, we employ denoising diffusion implicit models (DDIM) for efficient deterministic sampling, requiring only 5-10 steps compared to traditional 1000-step diffusion procedures and providing computational speedup of 100 to 200 times during inference without compromising accuracy. In addition, our proposed model achieves superior performance compared to image-based diffusion model: reducing training time by 94.4\% and requiring 89.0\% fewer parameters while achieving over 28\% improvement in prediction accuracy. Comprehensive comparisons against established data-flexible surrogate models including DeepONet and Meshgraphnet demonstrate consistent superiority of our approach across all three physical systems explored in this study, with performance improvements ranging from 30-90\% error reduction. To further refine the proposed model, we investigate two key aspects: 1) comparison of final physical states prediction or incremental change prediction, and 2) computational efficiency evaluation across varying subsampling ratios (10\%-100\%). Our refined model shows that incremental change prediction outperforms final physical states prediction especially for position prediction in the drop impact system, and maintains superior performance even when using only 30\% of the original point samples while requiring significantly less computational resources during training.
\end{abstract}


\end{frontmatter}

\section{Introduction}
\label{sec:introduction}

Scientific machine learning (SciML) has emerged as a powerful alternative to traditional numerical methods for simulating physical systems. Therefore, it can offer two principal advantages during shape design processes: a substantial reduction in computational cost during design iterations involving shape variations, and the ability to make reliable real-time predictions for unseen geometries \citep{kang2025generative}. These advances have proven particularly valuable in disciplines that require rapid simulation of complex physical systems, such as fluid dynamics \citep{yang2024long, yang2025physics, du2024conditional, fan2025neural, li2025latent, zhou2024text2pde}, structural mechanics \citep{jadhav2023stressd, xie2025spatiotemporal, shin2023wheel, cheng2025attention}, and climate modeling \citep{lam2023learning, kochkov2024neural, alet2025skillful}.

A core challenge in SciML techniques handling physical systems with varying geometries lies in the effective representation of diverse geometries. This challenge has recently prompted extensive research into a variety of data representations, including images \citep{jadhav2023stressd, ogoke2024deep, ruhling2023dyffusion, kohl2023benchmarking, song2023surrogate, jiang2023transcfd}, meshes \citep{yang2024long, pfaff2020learning, fortunato2022multiscale, nabian2024x, cao2023efficient, kim2025physics, han2022predicting}, and point clouds \citep{alkin2024universal, zhdanov2025erwin, serrano2024aroma, lu2021learning, he2023novel, he2024geom, kim2025decoupled}. However, regular grid-based image representations often struggle to accurately capture irregular geometries and fail to preserve important topological information. In contrast, mesh- and point-based representations have been particularly effective for handling irregular grids, which are common in real-world applications. While these approaches also face challenges including increased computational overhead and memory requirements, they demonstrate strong flexibility in explicitly capturing critical geometric features (e.g. sharp edges, corners, and singularities) compared to regular grids, without requiring burdensome pre- or post-processing.

Building upon these data representations, mesh-based graph neural networks, particularly Meshgraphnet (MGNs),  have emerged as a SciML framework for effectively predicting large-scale physical systems with different mesh geometries, demonstrating successful results in fluid dynamics, structural mechanics, and weather forecasting \citep{lam2023learning, alet2025skillful, pfaff2020learning, fortunato2022multiscale, nabian2024x, kim2025physics}. However, MGNs suffer from significant computational overhead due to their inherent message-passing mechanism, making them highly sensitive to mesh density. As mesh resolution increases, each node must iteratively exchange messages with a larger number of neighbors to aggregate sufficient information about the global system state \citep{nabian2024x}. Furthermore, MGNs face additional challenge in temporal modeling: when predicting spatio-temporal physical systems, MGNs adopt autoregressive schemes that predict the next time step based on the current state, with these predictions serving as inputs for subsequent time steps. This sequential dependency causes errors to accumulate over time steps, hindering the model's ability to capture long-term temporal dynamics.

In parallel to mesh-based approaches, point-based methodologies have gained significant attention for their capability to handle irregular geometries and complex boundary conditions without mesh connectivity requirements \citep{alkin2024universal, zhdanov2025erwin, serrano2024aroma, lu2021learning}. Most notably, DeepONet has become the most widely-adopted coordinate-based framework for learning solution mappings in infinite-dimensional function spaces by operating in a point-based manner \citep{lu2021learning}. It can address parametric PDEs with varying boundary conditions and geometries through a dual-network architecture: a branch network for encoding input functions (e.g., initial, boundary conditions) and a trunk network for spatio-temporal coordinates. Moreover, by predicting entire solution trajectories using coordinate-based querying in trunk network, DeepONet avoids autoregressive inference and associated temporal error accumulation. However, DeepONet suffers from two fundamental architectural limitations. First, its fully connected architecture exhibits spectral bias, prioritizing low-frequency patterns while failing to capture high-frequency components \citep{rahaman2019spectral}. Second, compounding this limitation, the simple dot product between branch and trunk outputs provides simple linear combinations of features, fundamentally limiting the modeling of nonlinear geometry-dependent interactions and necessitating extensive task-specific tuning that severely restricts generalization to unseen geometries \citep{oommen2025integrating, yang2025physics}.

Recently, generative model-based approaches, such as generative adversarial networks (GAN) \citep{goodfellow2014generative}, variational autoencoders (VAE) \citep{kingma2013auto} and diffusion models \citep{sohl2015deep, ho2020denoising, song2020denoising}, have been explored for physical field prediction \citep{liu2024cfd, jiang2021stressgan, kang2022physics, shu2023physics, ogoke2024inexpensive}.  In particular, diffusion models have shown strong capabilities in learning complex data distributions through iterative denoising, enabling accurate reconstruction of high-frequency physical features \citep{oommen2025integrating, shu2023physics}, without the mode collapse and blurred outputs that limit GANs and VAEs \citep{drygala2024comparison}. However, most diffusion-based models for physical field prediction are built upon frameworks originally designed for image generation. This leads to physical fields being typically represented as regular grid-based images, treating the field as a fixed-size structured array \citep{jadhav2023stressd, shu2023physics, ogoke2024inexpensive}. For example, Jadhav et al. \citep{jadhav2023stressd} proposed StressD, a diffusion-based framework designed to predict von Mises stress distributions on regular grid-based representations for 2D static analysis, to address the high computational cost incurred by repeated finite element analysis (FEA) in design optimization involving geometric variations. The framework demonstrates superior performance within this regular grid-based approaches, achieving a mean absolute error (MAE) approximately 79.1\% lower than StressNet \citep{nie2020stress} and 78.0\% lower than StressGAN \citep{jiang2021stressgan}, while also showing improved computational efficiency compared to conventional FEA. However, StressD has several limitations: (1) It assumes a regular grid-based image representation of stress fields, which makes it difficult to directly apply to real-world engineering problems involving complex irregular geometric representations. (2) The framework is focused on 2D static stress analysis, limiting its applicability to 3D structures or time-dependent stress analysis.

Therefore, recent studies have explored alternative diffusion-based frameworks that support flexible data representations that include unstructured geometries in spatio-temporal domains \citep{du2024conditional, zhou2024text2pde, gao2025generative}. Gao et al. \citep{gao2025generative} proposed a diffusion model framework that incorporates gradient guidance and virtual observations to simulate flow fields governed by parametric PDEs. The framework was applied to two case studies: 2D laminar cylinder flow on an unstructured mesh and 3D incompressible turbulent channel flow on structured grids, demonstrating high-fidelity spatio-temporal predictions across a range of Reynolds numbers with strong physical consistency. The approach achieved more than 350 times speed-up compared to conventional numerical simulations. Moreover, the use of virtual observations enabled improved model accuracy even in the presence of sparse or incomplete data. However, since the diffusion backbone relies on convolutional neural networks designed for image-based representations, it requires compression into fixed-size latent spaces through specialized encoder-decoder architectures. This encoder-decoder framework necessitates different architectural designs for each mesh type (graph neural networks for unstructured meshes and convolutional neural networks for structured grids) and introduces geometric information loss during the dimensionality reduction process, where complex spatial features may be inadequately represented in the compressed latent space.

Additionally, Zhou et al. \citep{zhou2024text2pde} proposed the Text2PDE framework to enhance the accessibility and usability of deep learning-based PDE solvers. This framework generates complete spatio-temporal physical simulations at once to mitigate autoregressive error accumulation by employing a latent diffusion model and a mesh autoencoder, where the mesh autoencoder is designed to handle irregular grids and diverse geometries. Experimental results demonstrate that Text2PDE achieves higher predictive accuracy than traditional deterministic surrogate models (e.g., Fourier Neural Operator (FNO) \citep{li2020fourier}, Geometry-Informed Neural Operator (GINO) \citep{li2023geometry}, etc.) and supports flexible conditioning via either text or initial physical fields. However, the framework has several limitations. First, the inherent ambiguity of natural language can lead to inaccuracies in the generated results, including potential hallucinations caused by imprecise or underspecified physical descriptions. Second, while the mesh autoencoder enables the handling of irregular grid data, forcing uniform latent representations may result in information loss, potentially limiting the model's ability to reconstruct fine-scale physical details.

To address the limitations of existing diffusion models, such as their reliance on regular grids and limited flexibility in handling domains with varying shapes or resolutions, this study proposes a point-wise diffusion model that operates directly on geometries of arbitrary structure and resolution. The proposed model performs diffusion process by perturbing and denoising physical quantities at individual spatio-temporal point, allowing it to process any data formats---including pixel-based images, irregular meshes, and point cloud---without the need for data preprocessing. Architecturally, the model adapts the point-wise diffusion transformer architecture to operate in a point-wise manner. Unlike conventional diffusion models that apply noise and denoising operations to each snapshot image, the proposed approach enables it to learn the denoising process directly at the level of individual points. Furthermore, to condition the model on physical and geometric context, such as boundary conditions or shape parameters, adaptive layer normalization with zero initialization (adaLN-Zero) is employed to inject conditional information effectively. Furthermore, the denoising diffusion implicit model (DDIM) is employed to provide a deterministic alternative to the stochastic sampling process of traditional denoising diffusion probabilistic model (DDPM), ensuring reproducible results while significantly reducing inference time and preserving high fidelity to the target numerical solution.

Our proposed framework is validated within three scenarios according to different physical systems: (1) \textbf{[Eulerian] Cylinder fluid flow}: a spatio-temporal flow field around 2D cylinders of various sizes and locations. For its modeling, the Eulerian method that models temporal changes of physical quantities in a fixed coordinate system is adopted. (2) \textbf{[Lagrangian] Drop impact}: a spatio-temporal system that tracks stress and displacement over time as a ball falls on multi-layered OLED display panels with varying geometric configurations. The system applies the Lagrangian method to dynamically model time-varying node positions and states. (3) \textbf{[Large-scale] Road-car external aerodynamics}: a large-scale physical system consisted of the surface pressure and wall shear stress fields on complex 3D vehicle geometries. The simulation datasets consist of high-fidelity, large-scale data encompassing a wide range of vehicle geometries, enabling comprehensive evaluation across diverse aerodynamic configurations. Throughout the above datasets, our model adapts to various physical scenarios by simply modifying problem-specific parameters (coordinate systems, geometric configurations, initial conditions and boundary conditions) without architectural changes. 

\vspace{1em}
\noindent The main contributions of this paper can be summarized as follows: 

\begin{enumerate}

    \item \textbf{A novel point-wise diffusion model agnostic to spatial data types:} We propose a point-wise diffusion model that processes each point independently, without relying on any structured spatial or temporal sequences. This eliminates the need for data preprocessing steps such as grid conversion or transformation into predetermined representations, thereby preserving geometric fidelity and enabling the direct handling of complex real-world geometries without geometric information loss.

    \item \textbf{Validated through diverse physical systems:} We present a unified framework capable of addressing different physical systems, including Eulerian spatio-temporal systems, Lagrangian spatio-temporal systems and large-scale 3D complex geometry systems, demonstrating superior adaptability and performance across problem domains compared to state-of-the-art methods such as DeepONet and Meshgraphnets, known for their data flexibility.

    \item \textbf{Improved temporal modeling capabilities via non-autoregressive approach:} The proposed point-wise diffusion model achieves non-autoregressive prediction by directly querying spatio-temporal coordinates with flexible conditioning via adaLN-Zero, eliminating temporal error accumulation and enabling stable long-term predictions for complex spatio-temporal physical systems.
    
    \item \textbf{Comprehensive experimental validation of model efficiency and superiority:} Based on DDIM sampling, we establish the model's computational efficiency and confirm significantly consistent prediction results across different random noise initializations, achieving deterministic reproducibility comparable to traditional numerical solvers. We also establish the superiority of our point-wise approach over conventional image-based diffusion methods through systematic comparative analysis.
    
    \item \textbf{Validation of geometric generalization for shape design applications:} We demonstrate robust performance across both diverse geometric configurations and different physical systems, confirming the model's generalization capabilities. The inherent shape flexibility of our point-based approach, validated through these diverse applications, enables straightforward and successful extension to shape design applications.

    \item \textbf{Model optimization strategies for performance refinement:} We suggest additional refinement strategies to optimize proposed model performance, through comparative analysis between direct and residual prediction approaches in spatio-temporal dynamics and computational efficiency evaluation across varying point sampling ratios, ensuring scalability for large-scale 3D systems.

\end{enumerate}

The remainder of this paper is organized as follows. \Cref{sec:methodology} presents the methodology of our point-wise diffusion model, detailing the forward-backward diffusion process applied to individual points (\Cref{sec: point-wise forward-backward diffusion process for physical system modeling}) and the point-wise diffusion model architecture (\Cref{sec: point-wise diffusion model architecture}). \Cref{sec:implementation_details} describes the implementation details across three diverse physical systems. \Cref{sec:Preliminary analysis for verifying efficiency and superiority over conventional diffusion approaches} provides preliminary analysis, including validation of the DDIM sampling for deterministic physics simulation (\Cref{subsection:Validation of DDIM sampling for deterministic physics simulation}), and comparative analysis between the image-based and point-wise approaches (\Cref{subsection:Comparative analysis between image-based and point-wise approaches}). \Cref{sec:Performance investigation: extensive comparison with existing data-flexible surrogate models} presents comparative analysis with existing surrogate models across the three physical systems. \Cref{sec:Further refinement towards optimization of proposed point-wise diffusion model} explores further optimization strategies to enhance model performance, particularly examining direct versus residual prediction strategies (\Cref{subsec:Direct versus residual prediction schemes in spatio-temporal physical systems: a comparison}) and performance across varying point sampling ratios for computational scalability (\Cref{subsec:Efficiency analysis across different sampling ratios for computational scalability}). Finally, \Cref{sec:conclusion} concludes the paper with a discussion of contributions and future research directions.

\section{Methodology}
\label{sec:methodology}
\begin{figure}[H]
    \captionsetup{font=normalsize}
    \centering
    \includegraphics[width=1\linewidth]{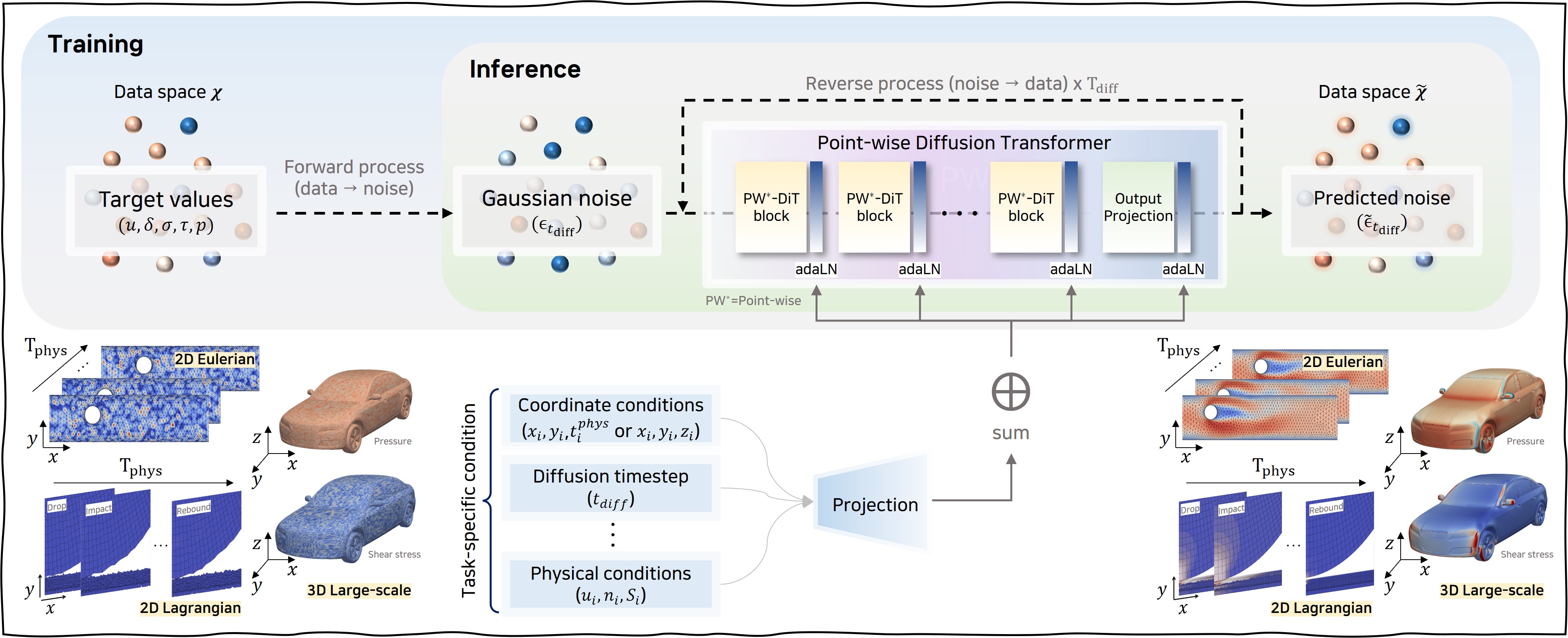}
    \caption{Point-wise diffusion model framework for simulating spatio-temporal and large-scale systems with shape variations.}
    \label{fig:Fig.1}
\end{figure}

We introduce a novel point-wise diffusion model (\Cref{fig:Fig.1}) capable of predicting complex physical systems with shape variations. In contrast to conventional diffusion models that add noise to entire images at each diffusion timestep, our method performs the diffusion process at the point level, injecting and denoising each point individually with different diffusion timesteps. This point-wise formulation ensures compatibility with any unstructured data format, including meshes and point clouds.

Building upon the standard diffusion framework, our approach adapts the conventional two-stage process: (1) a forward process that progressively adds noise to data, gradually transforming it into Gaussian noise; and (2) a reverse process that learns to systematically remove this noise to recover the original data. However, rather than applying this process globally at the snapshot level, we perform diffusion operations on individual points. This point-wise formulation enables flexible control over complex geometric structures and allows the model to condition geometric features and physical information into the denoising process at each point.

This section proceeds as follows: \Cref{sec: point-wise forward-backward diffusion process for physical system modeling} introduces the diffusion process applied to an individual point, which is a novel approach of this work. Then, \Cref{sec: point-wise diffusion model architecture} describes model architectural details of our proposed point-wise diffusion.

\subsection{Point-wise forward-backward diffusion process}
\label{sec: point-wise forward-backward diffusion process for physical system modeling}
\begin{figure}[H]   
    \captionsetup{font=normalsize}
    \centering
    \includegraphics[width=1\linewidth]{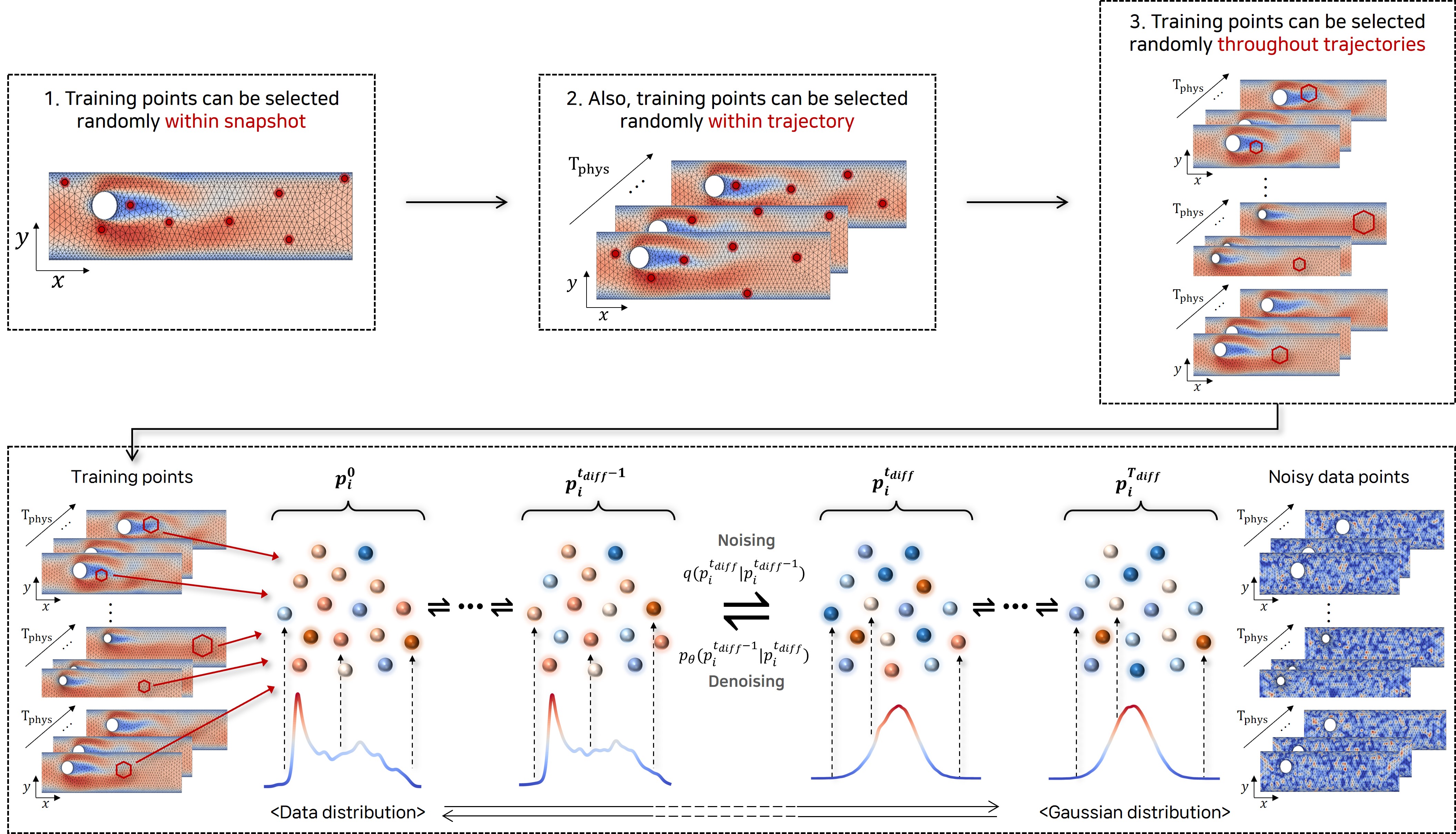}
    \caption{Point-wise forward-backward diffusion process for physical system modeling.}
    \label{fig:Fig.2}
\end{figure}
In this section, we introduce a diffusion process that progressively adds noise in the forward direction and denoises in the backward direction in a point-wise manner, enabling the model to learn the underlying data distribution through noise-based generative modeling. As shown in \Cref{fig:Fig.2}, our point-wise diffusion framework demonstrates flexible training strategies where points can be selected randomly within individual snapshots, within specific trajectories, or throughout complete trajectories, providing adaptable sampling approaches across different spatio-temporal scales. The point-wise diffusion process is applied to the physical quantity $p^{t_{\text{diff}}}_i$ at each individual point $i$ from these selected training points, where the framework processes them through a forward noising process and backward denoising process to learn accurate physical system predictions. Here, the superscript $t_{\text{diff}}$ denotes the current diffusion timestep, and $T_{\text{diff}}$ is the maximum diffusion timestep corresponding to pure Gaussian noise.

\paragraph{Forward process: Noising.} Each point gradually transitions from its original state $p_i^0$ to a noisy state $p_i^{T_{\text{diff}}}$ by progressively adding Gaussian noise. We denote this forward diffusion process as $q(p_i^{t_{\text{diff}}}|p_i^{t_{\text{diff}}-1)}$, which defines the conditional probability distribution for adding noise to each point $i$. And its process across all diffusion timesteps is defined as follows:

\begin{equation}
q(p_i^{t_{\text{diff}}}|p_i^0) = \mathcal{N}\left(p_i^{t_{\text{diff}}}; \sqrt{\bar{\alpha}_{t_{\text{diff}}}}p_i^0,\ (1-\bar{\alpha}_{t_{\text{diff}}})\mathbf{I} \right), \quad \forall i \in \{1,...,N\}
\label{eq:forward_process_dist}
\end{equation}
where $N$ represents the total number of points across all trajectories, $\bar{\alpha}_{t_{\text{diff}}} = \prod_{s=1}^{t_{\text{diff}}} \alpha_s$ is the cumulative product of the noise scheduling coefficients ${\alpha_s}$ with $\alpha_s = 1 - \beta_s$, and $\beta_s$ is the noise variance schedule that controls the amount of noise added at diffusion timestep $s$.  The sequence ${\beta_1, \beta_2, ..., \beta_{T_{\text{diff}}}}$ typically follows a predefined schedule (e.g., linear or cosine), ensuring that the original signal contribution diminishes while noise contribution increases as $t_{\text{diff}}$ progresses.

\vspace{1em}
This distribution implies that the noisy sample \( p_i^{t_{\text{diff}}} \) can be obtained by:
\begin{equation}
p_i^{t_{\text{diff}}} = \sqrt{\bar{\alpha}_{t_{\text{diff}}}} p_i^0 + \sqrt{1 - \bar{\alpha}_{t_{\text{diff}}}}\, \epsilon_i
\label{eq:forward_process_sample}
\end{equation}
where $\boldsymbol{\epsilon} \sim \mathcal{N}(0, \mathbf{I})$ represents the Gaussian noise generated for randomly selected training points throughout the trajectories, and $\epsilon_i$ is the noise component at point $i$ extracted from this coherent noise field $\boldsymbol{\epsilon}$. Therefore, the forward process provides an explicit construction of the noisy target values $p_i^{t_{\text{diff}}}$ by linearly blending each original physical data point $p_i^0$ with its corresponding noise component $\epsilon_i$. This formulation allows for controllable noise injection while maintaining spatial and temporal coherence across the selected training points, which eventually leads each $p_i^{T_{\text{diff}}}$ to approach a standard Gaussian distribution.

\paragraph{Backward process: Denoising.} Our approach adopts DDIM \citep{song2020denoising} as a deterministic sampling strategy. It enables efficient recovery of high-dimensional probability distributions with significantly fewer computational steps, making it well-suited for fast prediction of physical systems compared to DDPM's stochastic sampling procedure.

DDIM ensures that the same initial noise field for randomly selected training points consistently produces identical outputs, preserving deterministic behavior essential for physical system modeling. The deterministic feature of DDIM is particularly important for physical systems where reproducibility and consistency of predictions are paramount for validation and deployment. In \Cref{subsubsection:Model consistency evaluation across different noise initializations}, we experimentally validate that our point-wise diffusion model maintains consistent physical quantities when different random seeds generate varying initial noise fields.

\vspace{1em}

Technically, during the backward process, each noisy point $p_i^{t_{\text{diff}}}$ from the randomly selected training set is denoised to $p_i^{t_{\text{diff}}-1}$ as:
\begin{equation}
p_i^{t_{\text{diff}}-1} = \sqrt{\bar{\alpha}_{t_{\text{diff}}-1}} \cdot \hat{p}_i^{0} + \sqrt{1-\bar{\alpha}_{t_{\text{diff}}-1}} \cdot \epsilon_{\theta,i}(p_i^{t_{\text{diff}}}, \mathbf{c}_i)
\label{eq:backward_process}
\end{equation}
where $\epsilon_{\theta,i}(p_i^{t_{\text{diff}}}, \mathbf{c}_i)$ is the noise component at point $i$ predicted by the point-wise diffusion model with conditioning $\mathbf{c}_i$ (coordinate, diffusion timestep, and physical conditions), and $\hat{p}_i^{0} = \frac{p_i^{t_{\text{diff}}} - \sqrt{1-\bar{\alpha}_{t_{\text{diff}}}} \epsilon_{\theta,i}(p_i^{t_{\text{diff}}}, \mathbf{c}_i)}{\sqrt{\bar{\alpha}_{t_{\text{diff}}}}}$ represents the estimated clean data point derived from the predicted noise.

\vspace{1em}

A key advantage of DDIM lies in its ability to accelerate sampling by leveraging a deterministic non-Markovian process, interpreted as a discretization of a continuous-time ordinary differential equation (ODE). Unlike DDPM, which requires fine-grained sequential denoising over hundreds to thousands of steps, DDIM enables direct transitions from $t_{\text{diff}}$ to $t_{\text{diff}} - s$ for any step size $s > 1$. Specifically, rather than the single-step denoising ($s=1$) described in Equation \ref{eq:backward_process}, DDIM can skip multiple timesteps ($s>1$) during inference. This is formulated as:
\begin{equation}
p_i^{t_{\text{diff}}-s} = \sqrt{\bar{\alpha}_{t_{\text{diff}}-s}} \cdot \hat{p}_i^{0} + \sqrt{1-\bar{\alpha}_{t_{\text{diff}}-s}} \cdot \epsilon_{\theta,i}(p_i^{t_{\text{diff}}}, \mathbf{c}_i)
\label{eq:ddim_backward_process}
\end{equation}
This formulation allows sampling with as few as $5$-$10$ steps instead of the typical thousand steps required by standard diffusion models. For example, when using $10$ sampling steps from a model trained with $1000$ diffusion steps, the step size becomes $s=100$, proceeding through the sequence $t_{\text{diff}} = 1000, 900, 800, \ldots, 100$. Therefore, DDIM offers significant computational speedup, making it particularly well-suited for real-time physical system prediction where fast inference is essential.

\subsection{Point-wise diffusion model architecture}
\label{sec: point-wise diffusion model architecture}

\begin{figure}[H]
    \captionsetup{font=normalsize}
    \centering
    \includegraphics[width=1\linewidth]{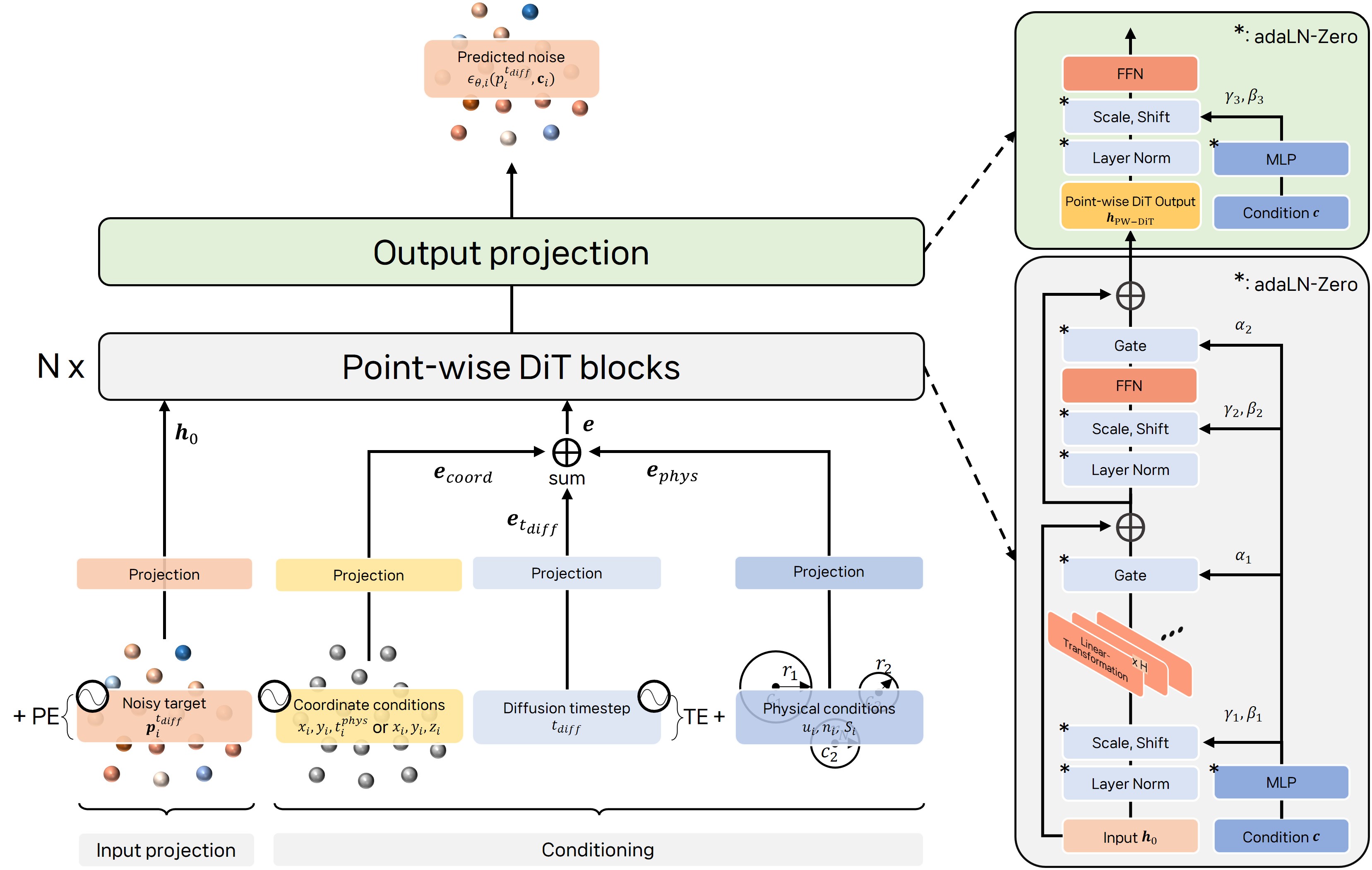}
    \caption{Point-wise diffusion model architecture for predicting noise within noisy target values via the denoising process.}
    \label{fig:Fig.3}
\end{figure}

To implement the denoising process described in \Cref{sec: point-wise forward-backward diffusion process for physical system modeling}, we introduce a point-wise diffusion model based on the diffusion transformer (DiT)-inspired architecture \citep{peebles2023scalable}. As shown in \Cref{fig:Fig.3}, the architecture follows a sequential processing pipeline: input and conditioning information are first projected into appropriate feature spaces, then processed through $N$ stacked point-wise DiT blocks for feature transformation, and finally passed through output projection to predict the noise ${\epsilon_{\theta,t_{\text{diff}}}}$ that is added to the target physical quantities at diffusion timestep $t_{\text{diff}}$. By learning to accurately estimate this added noise, the network can effectively remove it during inference, reconstructing the original physical quantities $\textbf{p}_i$ at each individual point, even for unseen geometries not included in the training dataset. Before we introduce the model architecture in detail, we first present two fundamental components that are essential to describe both input projection and conditioning: the positional encoding that richly injects spatial information for the noisy target and coordinates variables, and the time embedding that encodes diffusion timesteps.

\paragraph{Positional encoding.} 
In our point-wise diffusion architecture, positional encoding plays a crucial role in both input projection and conditioning stages. Since our model processes unstructured point clouds with varying geometries, we need to provide rich spatial information that enables the network to understand complex geometric relationships between points. Without proper spatial encoding, neural networks suffer from spectral bias, favoring low-frequency functions and failing to capture fine-grained spatial details essential for accurate physical system modeling.

To address this challenge, we adopt the positional encoding technique from neural radiance fields \citep{mildenhall2021nerf}, which transforms low-dimensional coordinates into high-dimensional feature representations. As described in Algorithm~\ref{alg:PE}, the encoding process follows four procedures: First, we compute logarithmically spaced frequency bands $\omega_i = 2^i$ for $i \in [0, n_{\text{freqs}}-1]$, which enable capturing spatial variations at multiple scales from coarse geometric structures to fine-grained details. Second, the input coordinates $\mathbf{x}$ are expanded with each frequency band through tensor product $\mathbf{x} \otimes \omega$, creating a multi-scale representation. Third, we apply sinusoidal transformations to generate $E_{\sin} = \sin(\pi \cdot \mathbf{x}_{\text{expand}})$ and $E_{\cos} = \cos(\pi \cdot \mathbf{x}_{\text{expand}})$, creating smooth, periodic features that provide high-frequency components necessary to overcome spectral bias. Finally, we concatenate all components: $\text{PE}(\mathbf{x}) = [\mathbf{x}, E_{\sin}, E_{\cos}]$, preserving original spatial information while enriching it with multi-scale frequency representations.

\begin{algorithm}[H]
\caption{Positional encoding (PE)}
\label{alg:PE}
\textbf{Input:} Inputs $\mathbf{x} \in \mathbb{R}^{B \times d}$, number of frequencies $n_{\text{freqs}}$ \\
\hspace{1.8em} 
\textbf{Output:} Positional Encoding $\text{PE}(\mathbf{x}) \in \mathbb{R}^{B \times (d \cdot (2 \cdot n_{\text{freqs}} + 1))}$
    \begin{algorithmic}[1]
    \State Compute frequency bands $\omega_i$:
    \[
    \omega_i = 2^i, \quad i \in [0, n_{\text{freqs}}-1]
    \]
    \State Expand inputs with frequency bands:
    \[
    \mathbf{x}_{\text{expand}} = \mathbf{x} \otimes \omega \in \mathbb{R}^{B \times d \times n_{\text{freqs}}}
    \]
    \State Apply sine and cosine transformations:
    \[
    E_{\sin} = \sin(\pi \cdot \mathbf{x}_{\text{expand}}), \quad
    E_{\cos} = \cos(\pi \cdot \mathbf{x}_{\text{expand}})
    \]
    where $E_{\sin}, E_{\cos} \in \mathbb{R}^{B \times d \times n_{\text{freqs}}}$.
    \State Concatenate original input, $E_{\sin}$, and $E_{\cos}$:
    \[
    \text{PE}(\mathbf{x}) = \Big[\mathbf{x}, E_{\sin}, E_{\cos}\Big]
    \]
    \end{algorithmic}
\end{algorithm}

\paragraph{Time embedding.}
Time embedding serves as a critical component that injects temporal information into the diffusion model, enabling it to distinguish between different noise levels during the progressive denoising process. Since the diffusion process operates across multiple diffusion timesteps with varying noise scales, the network must understand which denoising step it is currently performing to apply appropriate noise removal strategies. As described in Algorithm~\ref{alg:TE}, the time embedding process involves five procedures: First, we initialize the half-frequency dimension. Second, we generate exponentially decaying frequency bands $\omega_i$ that capture temporal patterns on multiple scales. Third, we compute phase arguments by expanding diffusion timesteps with these frequencies. Fourth, we apply sinusoidal encoding to create unique, continuous embeddings $\mathbf{f}_{\text{sin}} = [\cos(\phi), \sin(\phi)]$ for each diffusion timestep. Finally, these sinusoidal features are processed through a multi-layer perceptron to generate learnable temporal representations that can be seamlessly integrated with the model's hidden dimensions, allowing the model to adjust its behavior according to the current diffusion timestep.

\begin{algorithm}[H]
\caption{Time embedding (TE)}
\label{alg:TE}
\raggedright
\textbf{Input:} Diffusion timesteps $t_{\text{diff}} \in \mathbb{R}^{B}$, Frequency embedding dimension $d_{\text{freq}}$, Hidden dimension $d_{\text{hidden}}$, Maximum period $T_{\text{max}} = 10000$ \newline
\textbf{Output:} Time embedding $\text{TE}(t) \in \mathbb{R}^{B \times d_{\text{hidden}}}$
\begin{algorithmic}[1]
   \State Initialize half dimension: $h = \lfloor d_{\text{freq}} / 2 \rfloor$
   \State Generate exponentially decaying frequency bands: 
   \begin{equation*}
   \omega_i = \exp\left(-\frac{\ln(T_{\text{max}}) \cdot i}{h}\right), \quad i = 0, 1, \ldots, h-1
   \end{equation*}
   \State Compute phase arguments for all diffusion timesteps: 
   \begin{equation*}
   \phi = t_{\text{diff}} \otimes \omega \in \mathbb{R}^{B \times h}
   \end{equation*}
   \State Generate sinusoidal temporal features:
   \begin{equation*}
   \mathbf{f}_{\text{sin}} = \big[\cos(\phi), \sin(\phi)\big] \in \mathbb{R}^{B \times 2h}
   \end{equation*}
   
   \State Transform to learnable temporal representation:
   \begin{align*}
   \text{TE}(t) &= W_2 \cdot \sigma(W_1 \cdot \mathbf{f}_{\text{sin}} + b_1) + b_2
   \end{align*}
   where $W_1 \in \mathbb{R}^{d_{\text{hidden}} \times d_{\text{freq}}}$, $W_2 \in \mathbb{R}^{d_{\text{hidden}} \times d_{\text{hidden}}}$
\end{algorithmic}
\end{algorithm}

\vspace{1em}

\paragraph{Input projection.} The noisy target value $p_i^{t_{\text{diff}}}$ at a given diffusion timestep is first transformed into the model's feature space through a learnable linear projection:
\begin{equation}
\mathbf{h}_0  = \mathbf{W}_{\text{proj}} \cdot \text{PE}(p_i^{t_{\text{diff}}}) + \mathbf{b}_{\text{proj}}
\end{equation}
where $\mathbf{W}_{\text{proj}} \in \mathbb{R}^{d_{\text{model}} \times d_{\text{input}}}$ and $\mathbf{b}_{\text{proj}} \in \mathbb{R}^{d_{\text{model}}}$ are learnable weight matrix and bias vector, respectively, and $\mathbf{h}_0 \in \mathbb{R}^{d_{\text{model}}}$ represents the projected feature representation that serves as input to point-wise DiT blocks.

\paragraph{Conditioning.} To incorporate essential physical information into the model, we project multiple conditioning inputs at each individual point \textit{i}, including coordinate conditions, diffusion timestep $t_{\text{diff}}$, and physical conditions. Each conditioning component is embedded through a two-layer multilayer perceptron (MLP) with nonlinear activation:
\begin{align}
\mathbf{e}_{\text{coord}} &= f_{\text{coord}}(\text{PE}(c_{coord})) \\
\mathbf{e}_{\text{time}} &= f_{\text{time}}(\text{TE}(c_{t_{\text{diff}}})) \\
\mathbf{e}_{\text{phys}} &= f_{\text{phys}}(c_\text{phys})
\end{align}
where $[\cdot]$ denotes concatenation and $f_{\text{coord}}$, $f_{\text{time}}$, $f_{\text{phys}}$ represent the respective two-layer MLPs. The resulting condition embeddings are then summed to form a unified conditioning vector:
\begin{equation}
\mathbf{e} = \mathbf{e}_{\text{coord}} + \mathbf{e}_{\text{time}} + \mathbf{e}_{\text{phys}}
\end{equation}
This condition embedding vector $\mathbf{e}$ is subsequently injected into each point-wise DiT block through adaptive layer normalization \textit{(adaLN)}, enabling the model to incorporate domain-specific physical knowledge during the denoising process.

\paragraph{Point-wise DiT blocks.} The projected input features and condition embeddings are processed through $N$ number of point-wise DiT blocks (\Cref{fig:Fig.3}), where each block learns point-wise physical representations by incorporating the conditioning information. Each point-wise DiT block consists of self-attention and feed-forward networks (FFN) \citep{peebles2023scalable}. Notably, since our approach processes each spatio-temporal point independently, we set the sequence length to 1 for the self-attention mechanism. With this configuration, the attention score between query (Q) and key (K) becomes a constant value of 1, effectively simplifying the multi-head attention into a collection of value (V) transformations operating in parallel. This strategic adaptation reduces the computational complexity of the attention mechanism to a set of parallel linear transformations, significantly decreasing computational cost while focusing the model's capacity on individual point characteristics rather than interactions with neighboring points.

Furthermore, we employ adaptive layer normalization with zero initialization (\textit{adaLN-Zero}) \citep{peebles2023scalable} twice in each point-wise DiT block to ensure stable and effective conditioning injection (see \Cref{fig:Fig.3}).  This scheme provides more effective condition integration by dynamically modulating the feature representations through learnable scaling and shifting parameters, rather than simply concatenating or adding conditioning information. The conditioning process follows Eq. \ref{eq:adaLN-zero}:
\begin{equation}
\begin{aligned}
\hat{\mathbf{h}} &= \text{LayerNorm}(\mathbf{h}_{0}) \\
(\boldsymbol{\gamma},\boldsymbol{\beta} , \boldsymbol{\alpha}) &= f_{\text{cond}}(\mathbf{e}) \\
\mathbf{h}_{\text{mod}} &= \boldsymbol{\gamma} \odot \hat{\mathbf{h}} + \boldsymbol{\beta} \\
\mathbf{h}_{\text{out}} &= \boldsymbol{\alpha} \odot \mathbf{h}_{\text{mod}}
\end{aligned}
\label{eq:adaLN-zero}
\end{equation}
First, we apply layer normalization to standardize input features $\mathbf{h}_{0}$ obtained from input projection, with the inherent learnable affine transformation parameters are zero-initialized to eliminate unconditional scaling and shifting effects. The condition embedding vector $\mathbf{e}$ is then processed through the conditioning network $f_{\text{cond}}$, which consists of a nonlinear activation followed by a linear projection, generating three types of modulation parameters: scaling parameters $\boldsymbol{\gamma}$, shifting parameters $\boldsymbol{\beta}$, and gating parameters $\boldsymbol{\alpha}$. The normalized features $\hat{\mathbf{h}}$ are then conditionally modulated through element-wise scaling and shifting operations, resulting in $\mathbf{h}_{\text{mod}}$. Finally, the gating parameters $\boldsymbol{\alpha}$ control the intensity of conditional modulation by determining how much of the conditioning effect is applied to the network output $\mathbf{h}_{\text{out}}$. Here, the zero-initialization of $f_{\text{cond}}$ ensures that the model begins with identity modulation (no conditioning effect) and gradually learns condition-specific behaviors, maintaining training stability while enhancing adaptability to each physical scenario.

\paragraph{Output projection.} 
Following the series of point-wise DiT blocks, the processed features undergo a final conditional modulation step using the same \textit{adaLN-Zero} mechanism described in Eq. \ref{eq:adaLN-zero}. As shown in \Cref{fig:Fig.3}, the output from the final point-wise DiT block passes through layer normalization, receives conditional modulation (scale and shift), and is then transformed through a feed-forward network to generate the predicted noise $\boldsymbol{\epsilon}_{\theta,t_{\text{diff}}}$:

\begin{equation}
\begin{aligned}
\hat{\mathbf{h}}_{\text{final}} &= \text{LayerNorm}(\mathbf{h}_{\text{PW-DiT}}) \\
(\boldsymbol{\gamma}_{\text{out}}, \boldsymbol{\beta}_{\text{out}}) &= f_{\text{cond}}(\mathbf{c}) \\
\mathbf{h}_{\text{out}} &= \boldsymbol{\gamma}_{\text{out}} \odot \hat{\mathbf{h}}_{\text{final}} + \boldsymbol{\beta}_{\text{out}} \\
\boldsymbol{\epsilon}_{\theta,i}(p_i^{t_{\text{diff}}}, \mathbf{c}_i) &= \mathbf{W}_{\text{out}} \mathbf{h}_{\text{out}} + \mathbf{b}_{\text{out}}
\end{aligned}
\end{equation}
where $\mathbf{h}_{\text{PW-DiT}}$ represents the output from the final point-wise DiT block, $\mathbf{W}_{\text{out}} \in \mathbb{R}^{d_{\text{output}} \times d_{\text{model}}}$ and $\mathbf{b}_{\text{out}} \in \mathbb{R}^{d_{\text{output}}}$ are learnable parameters of the feed-forward network that performs the final linear transformation to produce noise predictions with appropriate output dimensions.

\section{Implementation details}
\label{sec:implementation_details}

\subsection{Experimental setup}

\paragraph*{Model training.} The models were trained using the Adam optimizer with a learning rate of 1e-4. For batch configuration, we used batch sizes of 8192 for the spatio-temporal system and 100,000 for the large-scale system, where each data point in a batch represents a randomly selected training point. Specifically, for the spatio-temporal systems, data points are randomly sampled from the entire dataset spanning all geometries and all physical timesteps, while for the large-scale system, data points are randomly sampled from all geometries. 

The diffusion process was discretized into 1000 time steps ($t_{\text{diff}} = 1, 2, \ldots, 1000$) with a linear noise schedule for $\beta_t$. For the loss function, we employed mean squared error (MSE) loss between the predicted noise $\epsilon_{\theta,i}(p_i^{t_{\text{diff}}}, \mathbf{c}_i)$ and the target noise $\epsilon_i$ to train the model parameters. Training was conducted on an Nvidia RTX 3090 (24GB) for the spatio-temporal systems and an Nvidia A100 (80GB) for the large-scale automotive system.

\paragraph*{Model evaluation.} We evaluated our point-wise diffusion model for different physical systems using multiple error metrics across the entire test dataset. For spatio-temporal systems (cylinder fluid flow and drop impact), we employed mean absolute error (MAE) and root mean square error (RMSE) calculated as:
\begin{align}
\text{MAE} &= \frac{1}{N} \sum_{i=1}^{N} |p_i - \hat{p}_i| \\
\text{RMSE} &= \sqrt{\frac{1}{N} \sum_{i=1}^{N} (p_i - \hat{p}_i)^2}
\end{align}
where $N$ represents the total number of points across all test trajectories, $\hat{p}_i$ is the predicted physical quantity (velocity, position, or stress) at point $i$, and $p_i$ is the corresponding ground truth value.

For the large-scale automotive aerodynamics system, we used relative error metrics to account for the wide range of physical quantities:
\begin{align}
\text{Relative L1} &= \frac{\sum_{i=1}^{N} | p_i - \hat{p}_i|}{\sum_{i=1}^{N} |p_i|} \\
\text{Relative L2} &= \frac{\sqrt{\sum_{i=1}^{N} ( p_i - \hat{p}_i)^2}}{\sqrt{\sum_{i=1}^{N} (p_i)^2}}
\end{align}
These relative metrics provide normalized comparisons across different vehicle geometries and physical quantities (surface pressure and wall shear stress components), enabling fair evaluation despite varying magnitude scales.

\subsection{Details on used datasets and model inference}

\subsubsection{Incompressible cylinder flow}  

\paragraph*{Datasets.} The cylinder flow dataset used in this study was obtained from Meshgraphnet \citep{pfaff2020learning}. We used 50 trajectories for training and 10 trajectories for inference. Each trajectory has a different geometry, with variations in both cylinder diameter and position to capture diverse flow conditions. These trajectories contain 600 temporal snapshots with a physical time interval of $\Delta{t^{\text{phys}}} = 0.01s$; however, the first 100 snapshots per configuration were selected to reduce computational burden during training.

\paragraph*{Conditions.} To capture incompressible flow phenomena around cylinders with different geometric parameters, we incorporate three distinct conditions: coordinate conditions, diffusion timestep, and physical conditions. 

\begin{itemize}
    \item \textit{Coordinate conditions.} The coordinate conditions consist of spatio-temporal coordinates $(x_i, y_i, t^{\text{phys}}_i)$ for time-dependent 2D flow field.
    \item \textit{Diffusion timestep.} The condition includes diffusion timestep $t_{\text{diff}}$ for the denoising process.
    \item \textit{Physical conditions.} The physical conditions comprise three components: (1) the initial shape condition $\textit{S}_i$, which specifies cylinder geometry through center coordinates $c$ and radius $r$, enabling the model to handle cylinders with different sizes and positions; (2) the boundary conditions $n_i$ encoded using one-hot representation to distinguish different boundary types (e.g., fluid nodes, wall nodes and inflow/outflow boundary nodes) ; and (3) the initial velocity field $u_i$, which provides the starting flow state.
\end{itemize} 

\paragraph*{Model inference.}  In this system, the output target is defined as the velocity difference (residual) between each target physical timestep ($t^{\text{phys}}=1, \ldots, T$) and the initial state ($t^{\text{phys}}=0$). During inference, this residual prediction is denoised through our diffusion model. The velocity field at each physical timestep is then reconstructed by adding the predicted residual to the initial velocity field ($t^{\text{phys}}=0$). The impact of this residual-based prediction approach on model performance will be analyzed in detail in \Cref{subsec:Direct versus residual prediction schemes in spatio-temporal physical systems: a comparison}.

\subsubsection{Drop impact test on OLED display panel} 

\paragraph*{Datasets.} The objective of this study is to predict the deformation behavior and stress distributions when a ball impacts multi-layered OLED display panels with different geometric configurations, specifically varying optically clear adhesive (OCA) thicknesses. Therefore, we utilized the displacement and stress field dataset from a drop impact simulation study \citep{kim2025physics}. \ref{sec:Data generation in drop impact system} presents both the material properties (\Cref{table:material_properties}) and the geometric configuration (\Cref{[Appendix] Fig.17}) of the ball and multi-layered OLED display panels used in this drop impact simulation study. From the given complete dataset of 150 trajectories, 100 were used for training and 50 for inference. Each trajectory consists of 100 physical timesteps, with a time interval of $\Delta{t^{\text{phys}}} = 4 \times 10^{-3}s$.

\paragraph*{Conditions.} For modeling drop impact dynamics on multi-layered OLED display panels, we also define three types of condition parameters: coordinate conditions, diffusion timestep, and physical conditions. 

\begin{itemize}
    \item \textit{Coordinate conditions.} The coordinate conditions consist of spatio-temporal coordinates $(x_i, y_i, t^{\text{phys}}_i)$ for time-dependent 2D impact simulation.
    \item \textit{Diffusion timestep.} The condition includes diffusion timestep $t_{\text{diff}}$ for the denoising process.
    \item \textit{Physical conditions.} The physical conditions also comprise three components: (1) the initial shape condition $S_i$, which specifies the geometric configuration through varying optically clear adhesive (OCA) thicknesses, enabling the model to handle display panels with different OCA thickness configurations; (2) the boundary conditions $n_i$ encoded using one-hot representation to distinguish different boundary types (e.g., ball, display panel components, fixed condition, symmetric condition); and (3) the initial position state $u_i$, which provides the starting position state of each trajectory.
\end{itemize} 

\paragraph*{Model inference.}  Within this Lagrangian framework, the output targets consist of two components: the absolute displacement $\delta$, defined as the difference between the target position ($t^{\text{phys}} =1, \ldots, T$) and the initial position ($t^{\text{phys}}=0$), and the stress values $\sigma$. In the inference phase, the predicted absolute displacement is added to the initial position to reconstruct the target position, whereas stress values are directly predicted by denoising. We also provide a detailed analysis of how residual prediction affects our model performance in \Cref{subsec:Direct versus residual prediction schemes in spatio-temporal physical systems: a comparison}.

\subsubsection{Road-car external aerodynamics} 

\paragraph*{Datasets.} In this system, we used the DrivAerML dataset \citep{ashton2024drivaerml} to predict high-fidelity CFD results (surface pressure and wall shear stress fields) for various 3D car shape configurations. This dataset was generated using 16 design parameters, which are described in \Cref{table:morphing-parameters}. Furthermore, the numerical analysis of the dataset was conducted using a hybrid Reynolds-averaged Navier-Stokes–Large Eddy Simulation (RANS-LES), which can be considered a high-fidelity CFD solver for industrial applications. The corresponding dataset consists of four physical quantities on the surface of cars: surface pressure and XYZ-wall shear stresses. From a total of 479 vehicle configurations, we used 380 for training and 99 for testing.

\paragraph*{Conditions.} To describe aerodynamic behavior across diverse vehicle geometries, we define multiple conditions that can capture the geometric features of the automotive system:

\begin{itemize}

   \item \textit{Coordinate conditions.} The coordinate conditions consist of positions $(x_i, y_i, z_i)$ for the 3D vehicle surface mesh and normal vectors $(n_x, n_y, n_z)$ corresponding to each spatial axis to capture surface orientation information.
   \item \textit{Diffusion timestep.} The diffusion condition incorporates the timestep $t_{\text{diff}}$ for the denoising process.
   \item \textit{Physical conditions.} The physical conditions include the shape parameter $S_i$, which encompasses the 16 morphing parameters outlined in \Cref{table:morphing-parameters}. These parameters define comprehensive vehicle geometry variations including overall dimensions (length, width, height, and various angular configurations) affecting aerodynamic performance.

\end{itemize}

\begin{table}[H]
\captionsetup{font=normalsize}
\centering
\caption{List of 16 vehicle morphing parameters and their value ranges}
\label{table:morphing-parameters}
\begin{adjustbox}{width=0.55\textwidth,center}
\begin{tabular}{ll|ll}
\hline
\textbf{Parameter} & \textbf{Range (mm)} & \textbf{Parameter} & \textbf{Range (mm)} \\
\hline
Vehicle Length         & -150 to +200    & Vehicle Width           & -100 to +100 \\
Vehicle Height         & -100 to +100    & Front Overhang          & -150 to +100 \\
Rear Overhang          & -150 to +100    & Hood Angle              & -50 to +50 \\
Approach Angle         & -40 to +30      & Windscreen Angle        & -150 to +150 \\
Backlight Angle        & -100 to +200    & Decklid Height          & -50 to +50 \\
Greenhouse Tapering    & -100 to +100    & Rear-end Tapering       & -90 to +70 \\
Front Planview         & -75 to +75      & Rear Diffuser Angle & -50 to +50 \\
Vehicle Ride Height    & -50 to +50      & Vehicle Pitch           & -1° to +1° \\
\hline
\end{tabular}
\end{adjustbox}
\end{table}

\paragraph*{Model inference.} For the road-car external aerodynamics system, we perform direct prediction of steady-state flow fields on vehicle surfaces with varying geometric configurations. This approach generates time-averaged aerodynamic solutions (surface pressure and wall shear stress fields) for each vehicle shape without requiring temporal evolution. However, accurately predicting such complex 3D high-fidelity aerodynamic systems across diverse vehicle geometries presents significant computational challenges.

\section{Preliminary analysis for verifying efficiency and superiority over conventional diffusion approaches }
\label{sec:Preliminary analysis for verifying efficiency and superiority over conventional diffusion approaches}

Traditional diffusion models suffer from prohibitively slow inference times due to iterative denoising procedures, making real-time physics prediction computationally infeasible. Moreover, image-based diffusion approaches necessitate grid interpolation that destroys geometric information when processing irregular meshes and point clouds common in engineering simulations. Therefore, we demonstrate the necessity of our point-wise diffusion framework through two preliminary analyses. \Cref{subsection:Validation of DDIM sampling for deterministic physics simulation} evaluates the computational efficiency of DDIM sampling across different sampling steps to determine optimal settings for real-time physics inference. Additionally, this subsection examines model consistency across different noise initializations to assess the deterministic behavior crucial for physics simulations. \Cref{subsection:Comparative analysis between image-based and point-wise approaches} investigates the advantages of point-wise processing over conventional image-based approaches in terms of prediction accuracy and computational efficiency. These analyses validate our proposed framework and provide the foundation for performance comparisons with existing surrogate models in the following sections.

\subsection{Validation of DDIM sampling for deterministic physics simulation}
\label{subsection:Validation of DDIM sampling for deterministic physics simulation}

\subsubsection{Analyzing computational efficiency across different sampling steps}
\label{subsubsection:Analyzing computational efficiency across different sampling steps}

For deterministic numerical simulations that require consistent and efficient predictions, we employ DDIM instead of DDPM. While DDPM relies on stochastic sampling that introduces randomness and requires hundreds to thousands of denoising steps, DDIM offers a deterministic alternative that achieves high-quality outputs with significantly fewer sampling steps through its non-Markovian deterministic process. This deterministic nature ensures that the same initial noise input always produces identical outputs, while the reduced sampling steps enable faster inference without computationally expensive iterative process. However, the effectiveness of DDIM for physical predictions still needs to be validated.

\begin{table}[H]
\captionsetup{font=normalsize}
\centering
\caption{Model performance across varying sampling steps in cylinder fluid flow}
\label{table:Model performance across varying sampling steps in cylinder fluid flow}
\begin{adjustbox}{width=0.68\textwidth, center}
\begin{tabular}{c|c|c|c|c}
\hline
\textbf{Physical system}& \textbf{Sampling step}& \textbf{Inference time[$s$]}& \textbf{Velocity MAE}& \textbf{Velocity RMSE} \\ \hline
\multirow{5}{*}{Cylinder} & 1 & 0.19 & 1.338 & 1.765 \\
 & \textbf{5}& 0.66 & \textbf{0.035} & \textbf{0.065} \\ 
 & 10 & 1.25 & 0.035 & 0.065 \\ 
 & 100 & 12.16 & 0.035 & 0.065 \\ 
 & 1000 & 122.03 & 0.035 & 0.065 \\ 
\hline
\end{tabular}
\end{adjustbox}
\end{table}
\vspace{-5mm}
\begin{table}[H]
\captionsetup{font=normalsize}
\centering
\caption{Model performance across varying sampling steps in drop impact}
\label{table:Model performance across varying sampling steps in drop impact}
\begin{adjustbox}{width=0.68\textwidth,center}
\begin{tabular}{c|c|c|c c|c c}
\hline
\multirow{2}{*}{\textbf{Physical system}} & \multirow{2}{*}{\textbf{Sampling step}} & \multirow{2}{*}{\textbf{Inference time [s]}} & \multicolumn{2}{c|}{\textbf{Position [mm]}} & \multicolumn{2}{c}{\textbf{Stress [MPa]}} \\
& & & \textbf{MAE} & \textbf{RMSE} & \textbf{MAE} & \textbf{RMSE} \\ \hline
\multirow{5}{*}{Drop impact} & 1 & 0.23 & 18.5 & 17.2 & 27.1 & 56.6 \\
 & 5 & 1.07 & 0.018 & 0.019 & 0.076 & 0.454 \\ 
 & \textbf{10}& 2.11 & \textbf{0.017} & \textbf{0.018} & \textbf{0.073} & \textbf{0.445} \\ 
 & 100 & 21.47 & 0.017 & 0.018 & 0.075 & 0.481 \\ 
 & 1000 & 215.17 & 0.017 & 0.018 & 0.077 & 0.499 \\ 
\hline
\end{tabular}
\end{adjustbox}
\end{table}
\vspace{-5mm}
\begin{table}[H]
\captionsetup{font=normalsize}
\centering
\caption{Model performance across varying sampling steps in road-car external aerodynamics (Rel: Relative)}
\label{table:Model performance across varying sampling steps in road-car aerodynamics}
\begin{adjustbox}{width=\textwidth,center}
\begin{tabular}{c|c|c|c c|c c|c c|c c}
\hline
\multirow{3}{*}{\textbf{Physical system}} & \multirow{3}{*}{\textbf{Sampling step}} & \multirow{3}{*}{\textbf{Inference time [s]}} & \multicolumn{2}{c|}{\multirow{2}{*}{\textbf{Surface Pressure}}} & \multicolumn{6}{c}{\textbf{Shear Stresses}} \\
\cline{6-11}
& & & \multicolumn{2}{c|}{} & \multicolumn{2}{c|}{\textbf{X-Wall}} & \multicolumn{2}{c|}{\textbf{Y-Wall}} & \multicolumn{2}{c}{\textbf{Z-Wall}} \\
\cline{4-11}
& & & \textbf{Rel-L2} & \textbf{Rel-L1} & \textbf{Rel-L2} & \textbf{Rel-L1} & \textbf{Rel-L2} & \textbf{Rel-L1} & \textbf{Rel-L2} & \textbf{Rel-L1} \\
\hline
\multirow{5}{*}{\begin{tabular}{c}Road-car external\\aerodynamics\end{tabular}} & 1 & 0.78 & 1.024 & 0.842 & 0.983 & 0.896 & 1.404 & 2.075 & 1.386 & 1.816 \\
 & \textbf{5}& 3.94 & \textbf{0.078} & \textbf{0.035} & \textbf{0.084} & \textbf{0.043} & \textbf{0.187} & \textbf{0.137} & \textbf{0.176} & \textbf{0.115} \\ 
 & 10 & 7.91 & 0.080 & 0.035 & 0.086 & 0.042 & 0.192 & 0.137 & 0.182 & 0.116 \\ 
 & 100 & 79.38 & 0.084 & 0.036 & 0.092 & 0.045 & 0.203 & 0.148 & 0.192 & 0.125 \\ 
 & 1000 & 794.86 & 0.085 & 0.038 & 0.093 & 0.047 & 0.205 & 0.153 & 0.195 & 0.129 \\ 
\hline
\end{tabular}
\end{adjustbox}
\end{table}
We evaluate the prediction accuracy across different sampling steps to determine the optimal number of steps that DDIM requires for each physical system. The fundamental advantage of DDIM lies in its non-Markovian sampling process, which enables direct transitions between non-consecutive timesteps. Rather than requiring sequential denoising through every timestep, DDIM can directly sample from strategically selected timesteps. For example, using only 5 sampling steps, DDIM transitions directly between $t_{\text{diff}} = 1000 \rightarrow 800 \rightarrow 600 \rightarrow 400 \rightarrow 200 \rightarrow 0$, bypassing hundreds of intermediate timesteps while maintaining prediction accuracy. The results in \Cref{table:Model performance across varying sampling steps in cylinder fluid flow,table:Model performance across varying sampling steps in drop impact,table:Model performance across varying sampling steps in road-car aerodynamics} demonstrate that across three different physical systems, each system shows comparable prediction accuracy with significantly fewer sampling steps (5-10) compared to the full 1000-step procedure through deterministic characteristics. For the cylinder fluid flow (\Cref{table:Model performance across varying sampling steps in cylinder fluid flow}), only 5 sampling step yielded identical velocity predictions (MAE: 0.035) to the full 1000 sampling step, while the drop impact simulation required only 10 steps to optimize both position (MAE: 0.017) and stress (MAE: 0.073) predictions (\Cref{table:Model performance across varying sampling steps in drop impact}). Similarly, the complex road-car aerodynamics system achieved its best performance with merely 5 sampling steps across surface pressure and wall shear stresses metrics (\Cref{table:Model performance across varying sampling steps in road-car aerodynamics}). These findings translate into dramatic computational accelerations, ranging from 100- to 200-fold reductions in inference time while maintaining prediction quality. 

\begin{figure}[H]
    \captionsetup{font=normalsize}
    \centering
    \includegraphics[width=1\linewidth]{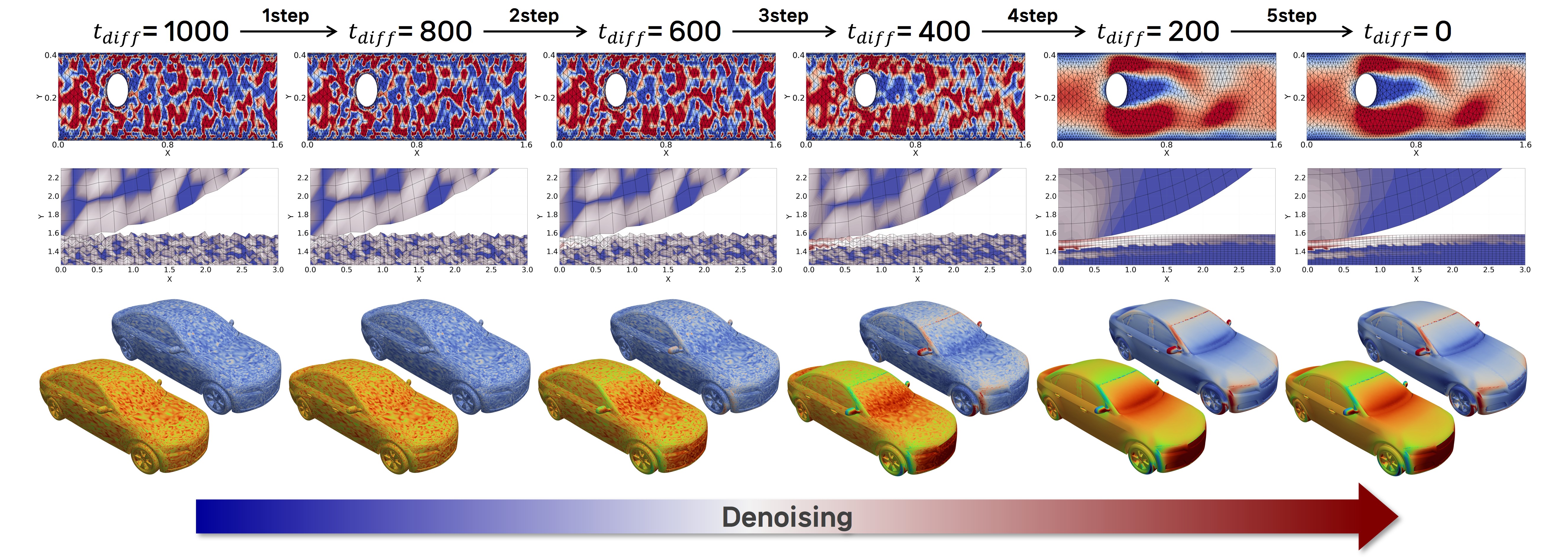}
\caption{Visualization of DDIM process with 5 sampling steps across different physical systems: (top row) cylinder fluid flow, (middle row) drop impact, and (bottom row) road-car external aerodynamics.}
    \label{fig:Fig.6}
\end{figure}

For further investigation, we visualize the progressive denoising sequence when a 5 sampling steps is adopted. \Cref{fig:Fig.6} captures key transition points in this process, revealing how physical features gradually emerge and sharpen as denoising progresses. At $t_{\text{diff}} = 1000$, the field represents pure Gaussian noise. From $t_{\text{diff}} = 800$ to $600$, the field still exhibits noisy patterns. As denoising progresses from $t_{\text{diff}} = 600$ to $400$, the main physical features begin to emerge while still retaining some noise. By $t_{\text{diff}} = 200$, the physical patterns start to be well-defined, and at $t_{\text{diff}} = 0$ (final prediction), we observe clean, physically accurate representations of the physical fields. These results demonstrate the effectiveness of DDIM's deterministic sampling approach, showing that our model achieves essentially comparable prediction accuracy with just 5 sampling steps compared to 1000 sampling steps. This deterministic characteristic reduces computational time by 100-200 times, enabling real-time physics simulations without compromising accuracy. Therefore, based on these computational efficiency gains and deterministic properties, we adopt DDIM as our primary sampling method for physics simulation tasks. 

\subsubsection{Model consistency evaluation across different noise initializations}
\label{subsubsection:Model consistency evaluation across different noise initializations}

Based on the demonstrated computational efficiency and deterministic sampling properties of DDIM, we adopt DDIM as our primary sampling procedure for physics simulation predictions. However, while DDIM guarantees identical outputs under identical initial noise, it can produce different outputs when initialized with different random noise. This contrasts with traditional deterministic numerical solvers that produce identical results with identical physics conditions (boundary conditions, initial conditions, and geometry). Therefore, we examine whether our model maintains robust performance by generating consistent results across different random noise initializations under identical physics conditions.

We evaluated the denoising results from three different initial noise vectors $x_T$ generated by setting random seed configurations in the PyTorch library (referred to as seed 1, seed 2, and seed 3).
\begin{figure}[H]
    \captionsetup{font=normalsize}
    \centering
    \includegraphics[width=1\linewidth]{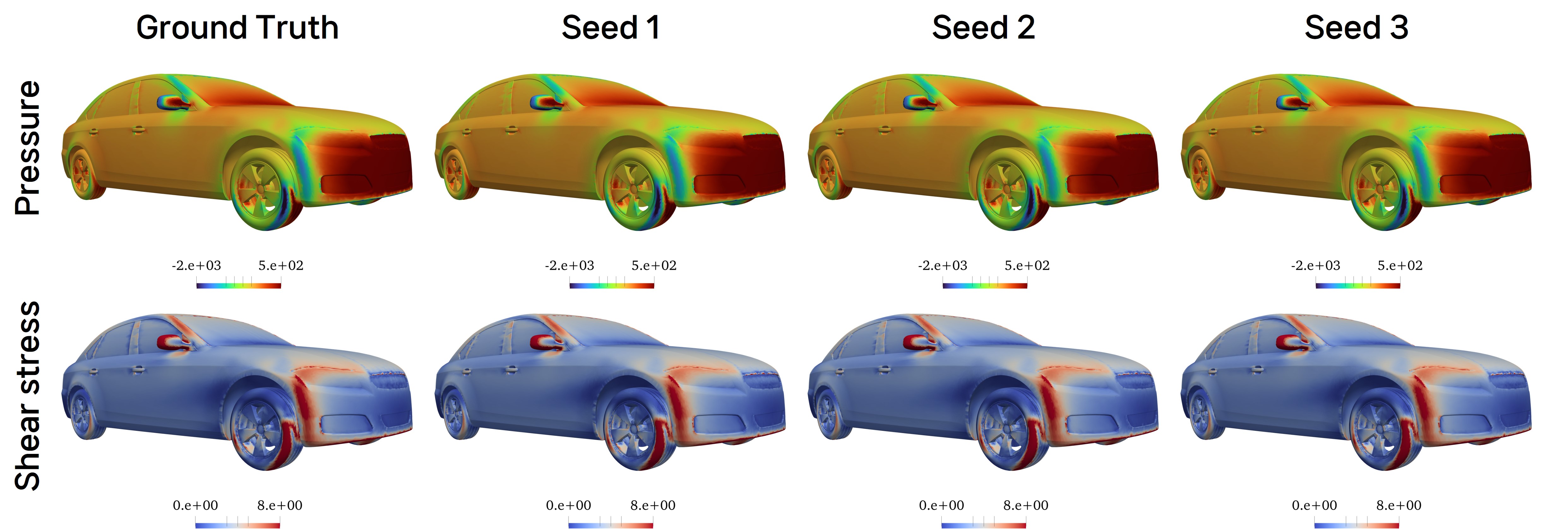}
    \caption{Visualization of denoising results according to initial noise samples with different seeds in a large-scale system (top row: surface pressure, bottom row: wall shear stresses)}
    \label{fig:Fig.5}
\end{figure}
\Cref{fig:Fig.5} demonstrates that our point-wise diffusion model generates visually consistent prediction patterns across different random noise initializations in the road-car external aerodynamics system. The figure shows both surface pressure distribution (top row) and shear stress fields (bottom row) predicted with different random seeds. While the predictions from different seeds are not exactly identical, they exhibit significantly similar physical field distributions and maintain consistent accuracy when compared to the ground truth. This visual consistency indicates that our model substantially reduces the stochastic variability across different noise initializations.

\begin{figure}[H]
    \captionsetup{font=normalsize}
    \centering

    \begin{subfigure}{0.3\textwidth}
        \includegraphics[width=\linewidth]{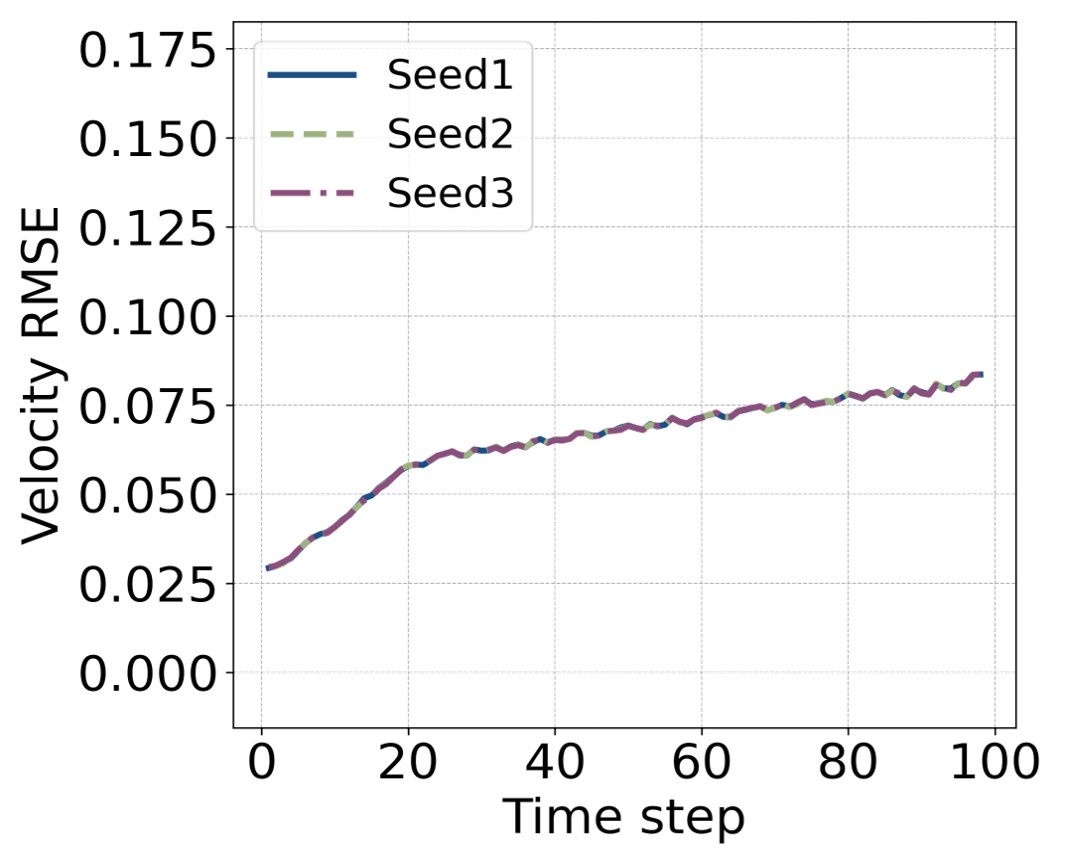}
        \caption{Cylinder fluid flow}
        \label{fig:Fig.4a}
    \end{subfigure}
    \vspace{0.02\textwidth}
    \begin{subfigure}{0.585\textwidth}
        \includegraphics[width=\linewidth]{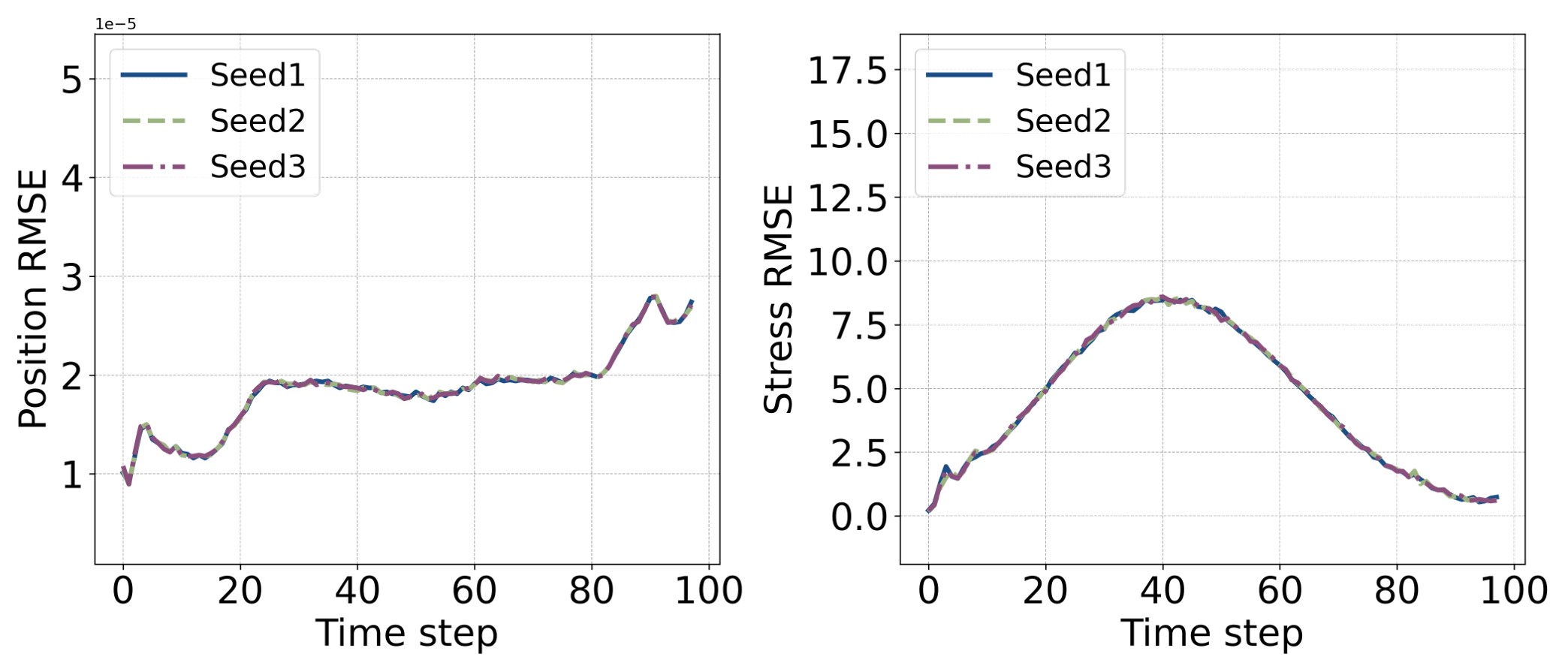}
        \caption{Drop impact}
        \label{fig:Fig.4b}
    \end{subfigure}
    \hspace{0.03\textwidth}
    \begin{subfigure}{0.5\textwidth}
        \includegraphics[width=\linewidth]{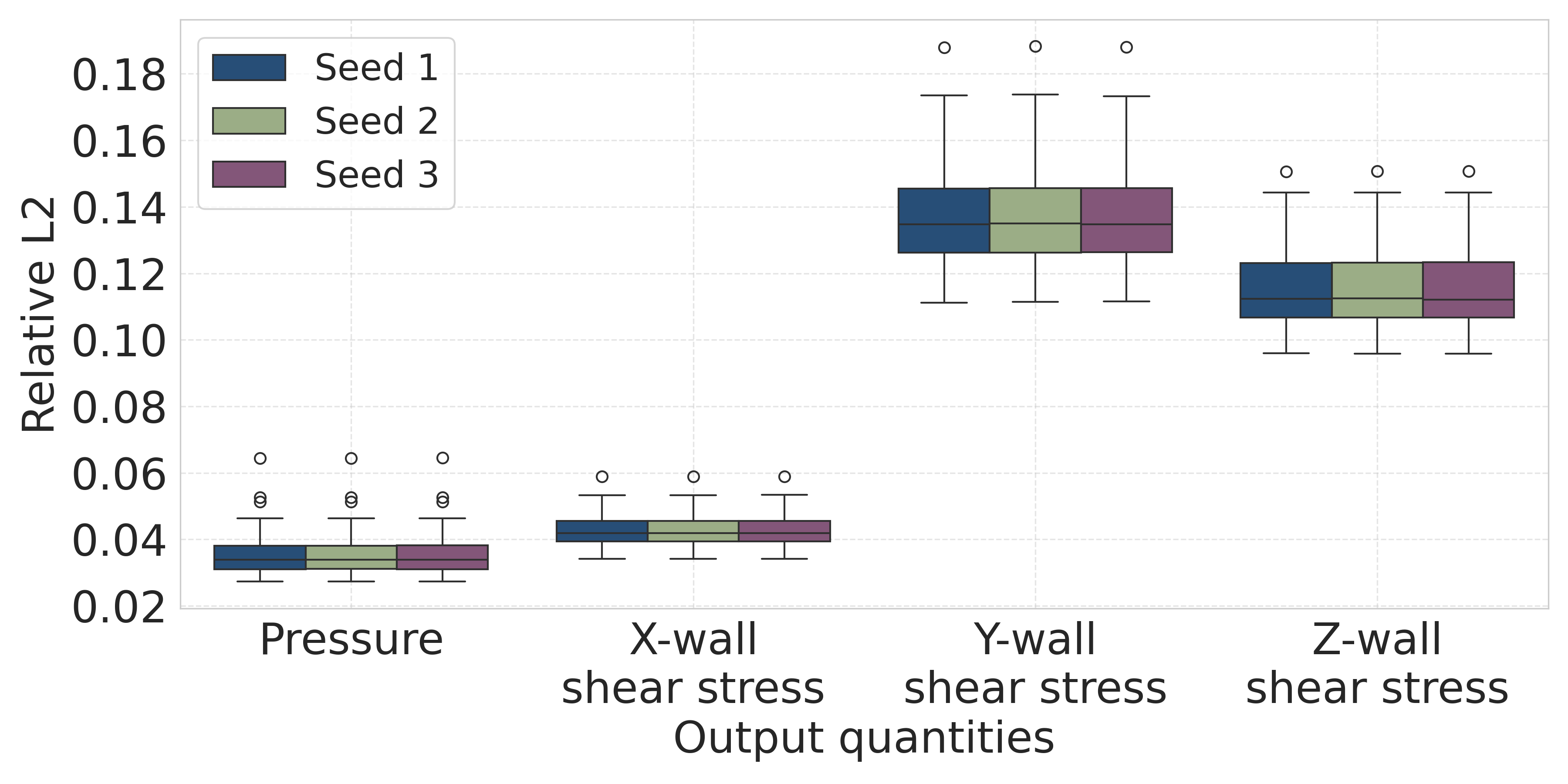}
        \caption{Road-car external aerodynamics}
        \label{fig:Fig.4c}
    \end{subfigure}

    \caption{Consistency analysis of the proposed point-wise diffusion model under different initial noise conditions for three different datasets: (a) cylinder fluid flow, (b) drop impact, and (c) road-car external aerodynamics.}
    \label{fig:Fig.4}
\end{figure}
In addition, to quantitatively validate the consistency of our model, \Cref{fig:Fig.4} presents the error tendencies between ground truth and prediction for the three physical systems on different noise initialization seeds. Remarkably, the error patterns show highly consistent behavior regardless of the initial noise configurations. For cylinder fluid flow, all three seeds exhibit nearly identical velocity RMSE trajectories throughout the physical timestep $t^{\text{phys}}$. Similarly, in the drop impact system, both position and stress RMSE patterns remain consistent across different seeds, demonstrating same error behavior over physical timestep $t^{\text{phys}}$. Furthermore, the road-car aerodynamics system also confirms the consistent results with all measured quantities (surface pressure and XYZ-wall shear stresses).

Therefore, these consistent error patterns across different random noise initializations demonstrate that our model is not affected by seed-dependent stochasticity in all three physical systems, confirming that our diffusion-based prediction framework effectively achieves the key characteristic of deterministic numerical solvers.

\subsection{Comparative analysis between image-based and point-wise approaches}
\label{subsection:Comparative analysis between image-based and point-wise approaches}
\begin{figure}[H]
    \captionsetup{font=normalsize}
    \centering
    \includegraphics[width=1\linewidth]{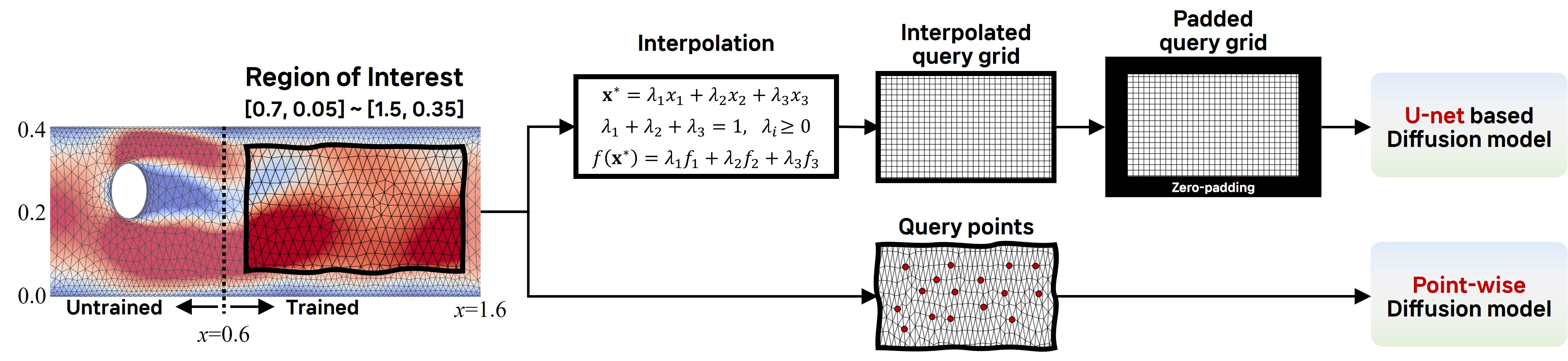}
    \caption{Methodological comparison of two diffusion frameworks: (top) image-based method with grid interpolation and U-Net processing, versus (bottom) point-wise method without data pre-processing.}
    \label{fig:Fig.17}
\end{figure}

Existing diffusion model-based approaches for physics simulation rely on image-based generation methods, processing physical fields through grid structure conversion with additional pre/post-processing steps \citep{zhou2024text2pde, jadhav2023stressd}. However, the conversion process from irregular mesh structures to regular grids inherently causes information loss and requires additional computational resources.

In this regard, we present the necessity of the point-wise approach through performance comparisons between image-based and point-wise methods. We selected cylinder flow system for this comparison because its grid points remain fixed in space, allowing consistent conversion to regular grids required for the image-based method; Lagrangian systems like drop impact create changing spatial patterns that are difficult to map consistently, while a 3D automotive system makes regular grid discretization infeasible due to cubic memory scaling and severe geometric approximation errors when representing complex boundaries on regular grids.

\Cref{fig:Fig.17} clearly illustrates the fundamental differences between the two approaches. In the image-based approach, the physical data obtained from irregular meshes are interpolated into regular grids. Since significant information loss may occur around boundaries (e.g., $x=1.6$ and $y=0.4$) due to interpolation, we selected the region of interest [0.7, 0.05] - [1.5, 0.35] for both models to minimize loss and ensure fair comparison. The interpolated grid is converted to fixed-size images suitable for U-Net processing through zero padding, with padded regions excluded from the training scope. Conversely, our point-wise approach directly utilizes coordinates of original mesh nodes and physical quantities at corresponding locations without any pre-processing. This completely preserves spatial accuracy of original data without interpolation, maintaining precise physical meaning of each node. Furthermore, computational efficiency is improved by eliminating pre- and post-processing steps for complex geometrical shapes.

We applied DDIM in the diffusion process but fundamental differences exist between the two approaches. The image-based method applies uniform noise across each snapshot, while the point-wise method applies noise independently to each individual point throughout trajectories, as described in \Cref{sec: point-wise forward-backward diffusion process for physical system modeling}. 

The image-based model was implemented using a U-Net conditional model \citep{von-platen-etal-2022-diffusers}. The model consists of 3 down/up blocks with cross attention and 1 down/up block, processing 32×16 images with 1 input channel (noisy target) and 1 output channel (predicted noise). For a fair comparison between approaches, three conditions were embedded and injected into the models as in point-wise models: (1) coordinate conditions ($x_i, y_i, t^{\text{phys}}_i$), (2) diffusion timestep $t_{\text{diff}}$, and (3) physical conditions (the initial shape condition $\textit{S}_i$ and the initial velocity field $u_i$). For the image-based approach, conditions are injected through cross-attention mechanisms, while the point-wise approach uses adaLN-zero mechanisms. Furthermore, the number of query points in the point-wise method was matched to the number of regular grid points in the image-based method for the fair comparison.

\begin{table}[H]
\captionsetup{font=normalsize}
\centering
\caption{Performance comparison between image-based and point-wise approach}
\label{table:Performance comparison between image-based and point-wise approach}
\begin{adjustbox}{width=0.8\textwidth, center}
\begin{tabular}{c|c|c|c|c}
\hline
\textbf{Model}& \textbf{Training time}& \textbf{Params}& \textbf{Velocity MAE}& \textbf{Velocity RMSE} \\ \hline
Image-based approach& 25.35h& 17,178,113& 0.095& 0.134\\
Point-wise approach& 1.42h& 1,901,057& 0.061& 0.096\\
\hline
\end{tabular}
\end{adjustbox}
\end{table}
The quantitative evaluation results presented in \Cref{table:Performance comparison between image-based and point-wise approach} demonstrate superior performance of the point-wise approach across all metrics compared to the image-based approach. For velocity prediction, MAE decreased by 35.8\% and RMSE by 28.4\%. In terms of computational efficiency, the point-wise approach reduced training time by 94.4\% compared to the image-based approach. Furthermore, the point-wise diffusion model achieves this performance improvement while demonstrating a lightweight model with substantially fewer parameters, representing a 89.0\% reduction in model size. This substantial parameter reduction demonstrates the efficiency of point-wise approach, eliminating the parameter-heavy cross-attention mechanisms and multi-scale convolutional blocks required for spatial feature extraction in image-based U-Net models, while achieving superior predictive accuracy.
\begin{figure}[H]
    \captionsetup{font=normalsize}
    \centering
    \includegraphics[width=0.6\linewidth]{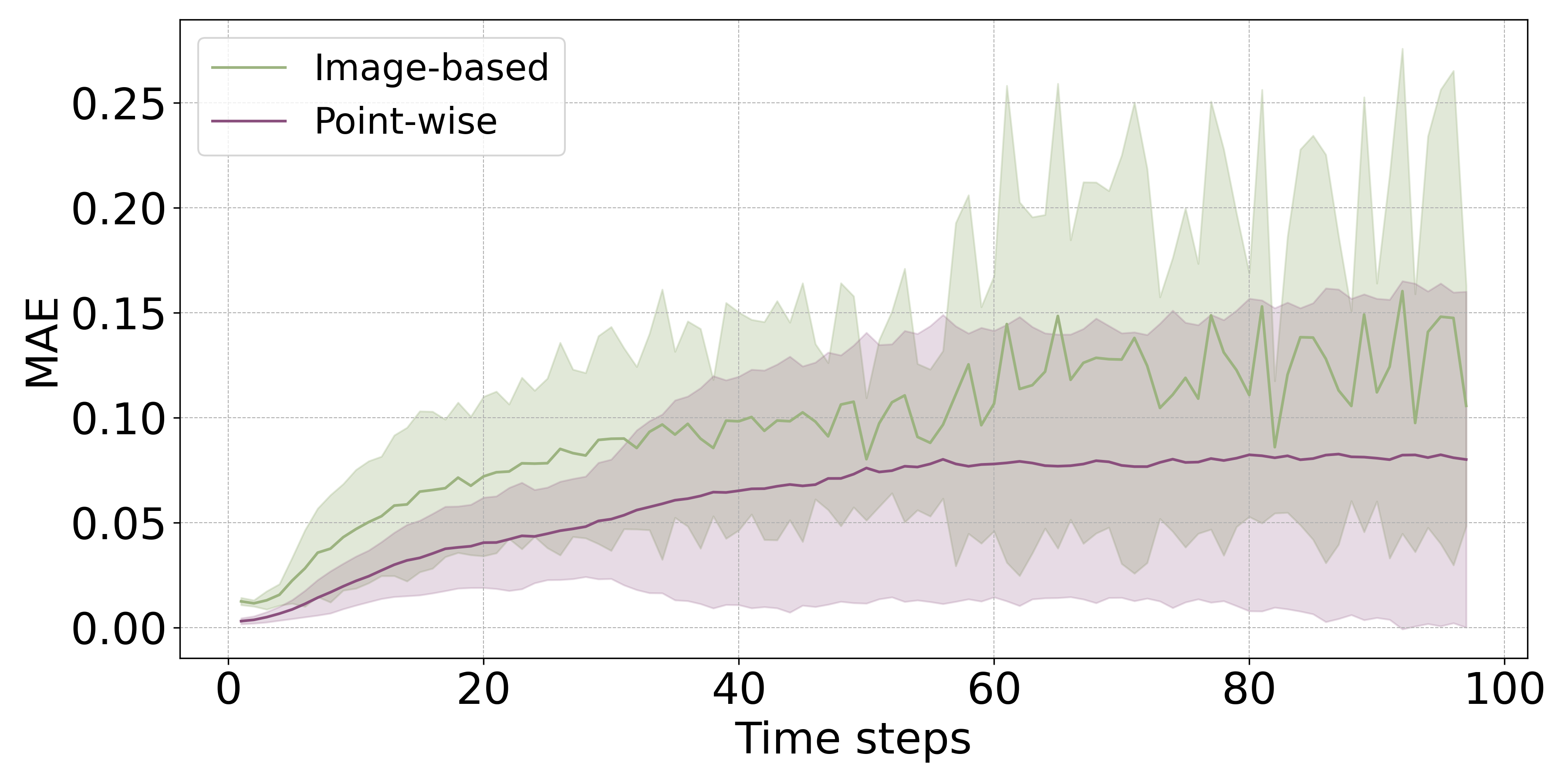}
    \caption{Performance comparison between image-based and point-wise approach across all physical timesteps. At each timestep, the solid lines represent the mean of MAE values computed at all spatial points, while the shaded regions show the standard deviation of these MAE values.}
    \label{fig:Fig.18}
\end{figure}
\begin{figure}[H]
    \captionsetup{font=normalsize}
    \centering
    \begin{subfigure}[t]{0.485\textwidth}
        \centering
        \includegraphics[width=\linewidth]{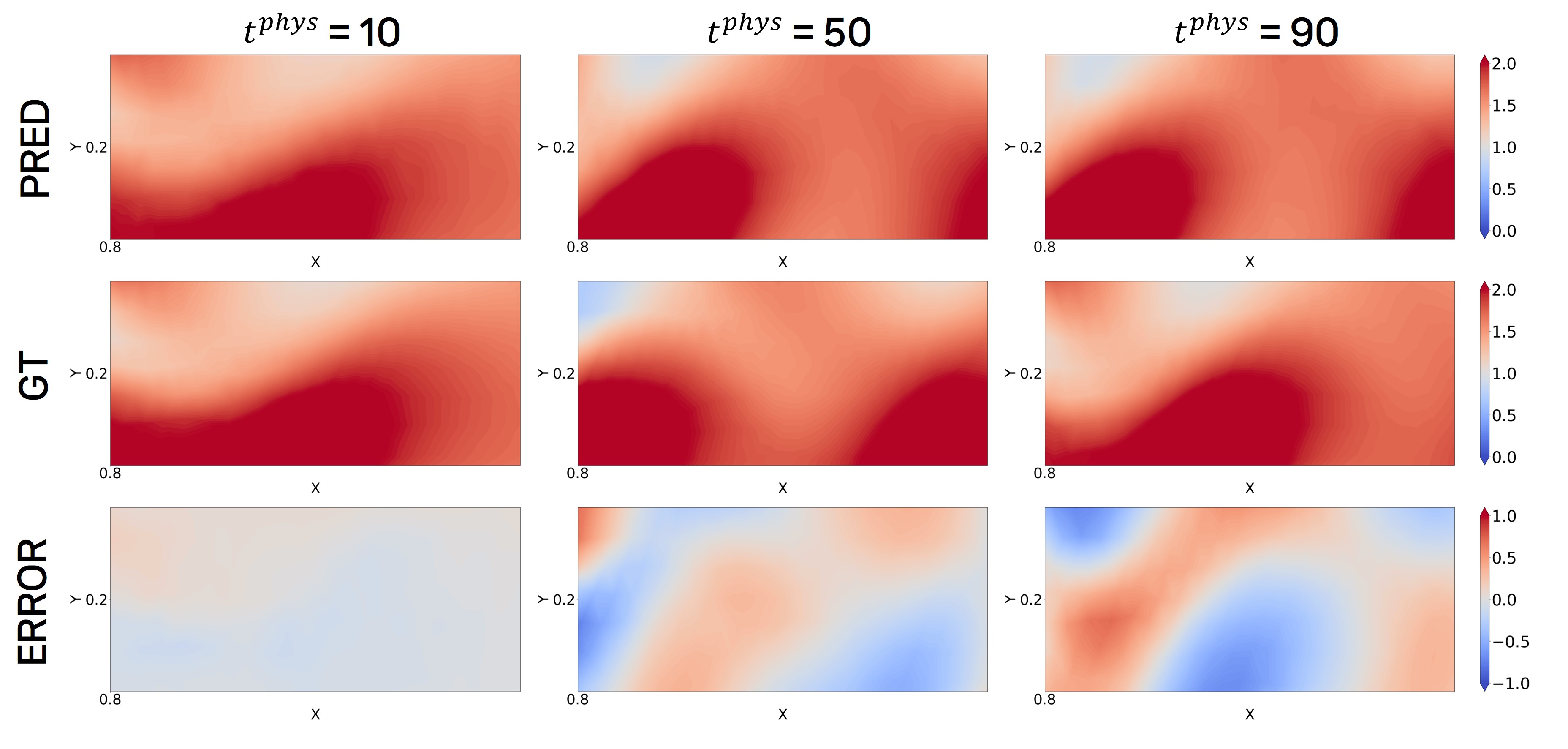}
        \caption{Image-based approach}
        \label{fig:Fig.19a}
    \end{subfigure}
    \hfill
    \begin{subfigure}[t]{0.485\textwidth}
        \centering
        \includegraphics[width=\linewidth]{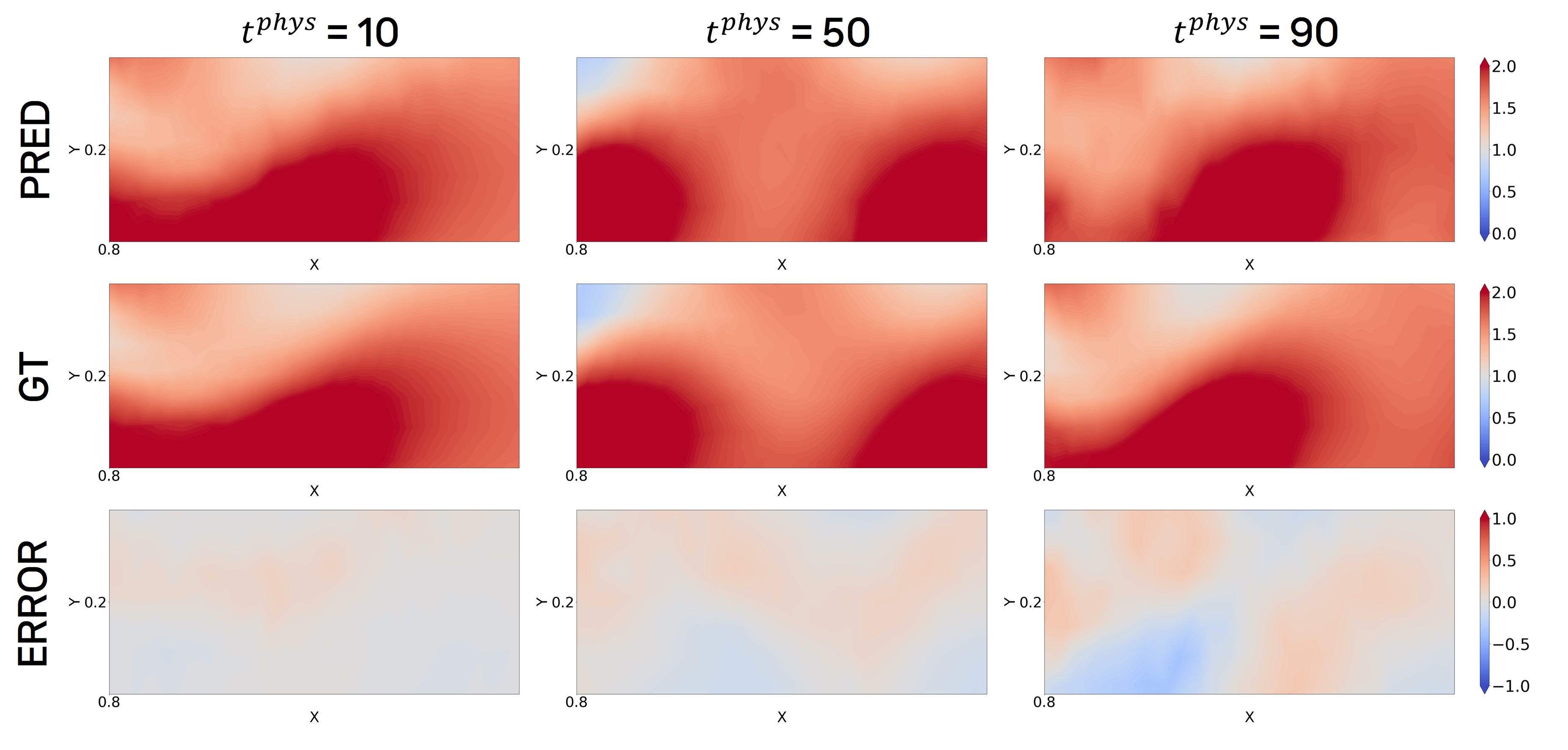}
        \caption{Point-wise approach}
        \label{fig:Fig.19b}
    \end{subfigure}

    \caption{Performance visualization of two approahces across physical timesteps.}
    \label{fig:performance_comparison}
\end{figure}

Examining \Cref{fig:Fig.18}, which compares mean MAE across all spatial points at each physical timestep, the image-based approach exhibits significantly higher error magnitude and larger variability compared to the point-wise approach. The shaded regions representing standard deviation reveal that the image-based method shows inconsistency across different spatial points, with the variability band dramatically widening after 60 physical timestep. In contrast, the point-wise method is evident from its stable error magnitude (below 0.1) throughout the simulation and notably narrow standard deviation bands. This indicates that the image-based method becomes increasingly unreliable as the simulation progresses, showing high sensitivity to different flow configurations and geometric complexities, while the point-wise approach maintains consistently uniform accuracy across all spatial points regardless of temporal variations. Particularly, comparing \Cref{fig:Fig.19a,fig:Fig.19b}, the visual comparison clearly demonstrates the superior performance of the point-wise method in preserving flow physics. The error visualizations show that the image-based approach produces substantial errors in the wake region where vortex shedding occurs as physical time goes on, while the point-wise approach maintains accurate prediction of the unsteady flow patterns and vortical structures. 

\section{Performance investigation: extensive comparison with existing data-flexible surrogate models}
\label{sec:Performance investigation: extensive comparison with existing data-flexible surrogate models}

We evaluate the efficiency, generalizability, and accuracy of our point-wise diffusion model compared to two representative approaches that address the fundamental challenge of handling irregular geometries in physical systems: DeepONet as the coordinate-based neural operator, and MGN as the mesh-based graph neural network framework. Our benchmark analysis spans Eulerian fluid dynamics (cylinder fluid flow), Lagrangian solid mechanics (drop impact simulation), and large-scale aerodynamics (road-car external aerodynamics) applications, demonstrating how our methodology achieves superior results across different physical domains while maintaining computational efficiency. To ensure fair comparison, we performed experiments while keeping the parameter counts of all compared surrogate models within comparable ranges for each system. The performance comparison is presented across these three physical systems in \Cref{subsec:[Eulerian system] Cylinder fluid flow}, \Cref{subsec:[Lagrangian system] Drop impact}, and \Cref{subsec:[Large-scale system] Road-car external aerodynamics}, respectively.

\subsection{Eulerian system: Cylinder fluid flow}
\label{subsec:[Eulerian system] Cylinder fluid flow}

We first present the performance comparison for the cylinder fluid flow problem under an Eulerian formulation, where output quantities are evaluated at fixed spatial locations. As depicted in \Cref{table:Performance comparison of surrogate models for cylinder fluid flow}, our point-wise diffusion model demonstrates superior velocity field prediction compared to conventional surrogate methods---DeepONet and Meshgraphnet (MGN). Quantitatively, our proposed model shows 53\% and 36\% reductions in MAE compared to DeepONet and MGN, respectively. Similarly, RMSE improvements are substantial, with 47\% and 36\% error reductions compared to DeepONet and MGN. These improvements are achieved while maintaining computational efficiency, requiring 50\% less training time than MGN. While DeepONet exhibits the fastest convergence with a training time of only 1.8 hours, its performance is limited by overfitting and difficulties in capturing high-frequency flow features, resulting in consistently higher error metrics. MGN, which employs a message passing scheme to incorporate neighboring node information, achieves better accuracy than DeepONet but incurs a substantial computational burden. Additionally, its autoregressive prediction approach of MGN leads to error accumulation during sequential predictions. In contrast, our point-wise diffusion model strikes a balance between accuracy and efficiency while maintaining reasonable computational requirements, compared to two conventional surrogate models.

\begin{figure}[H]
    \captionsetup{font=normalsize}
    \centering
    \includegraphics[width=0.6\linewidth]{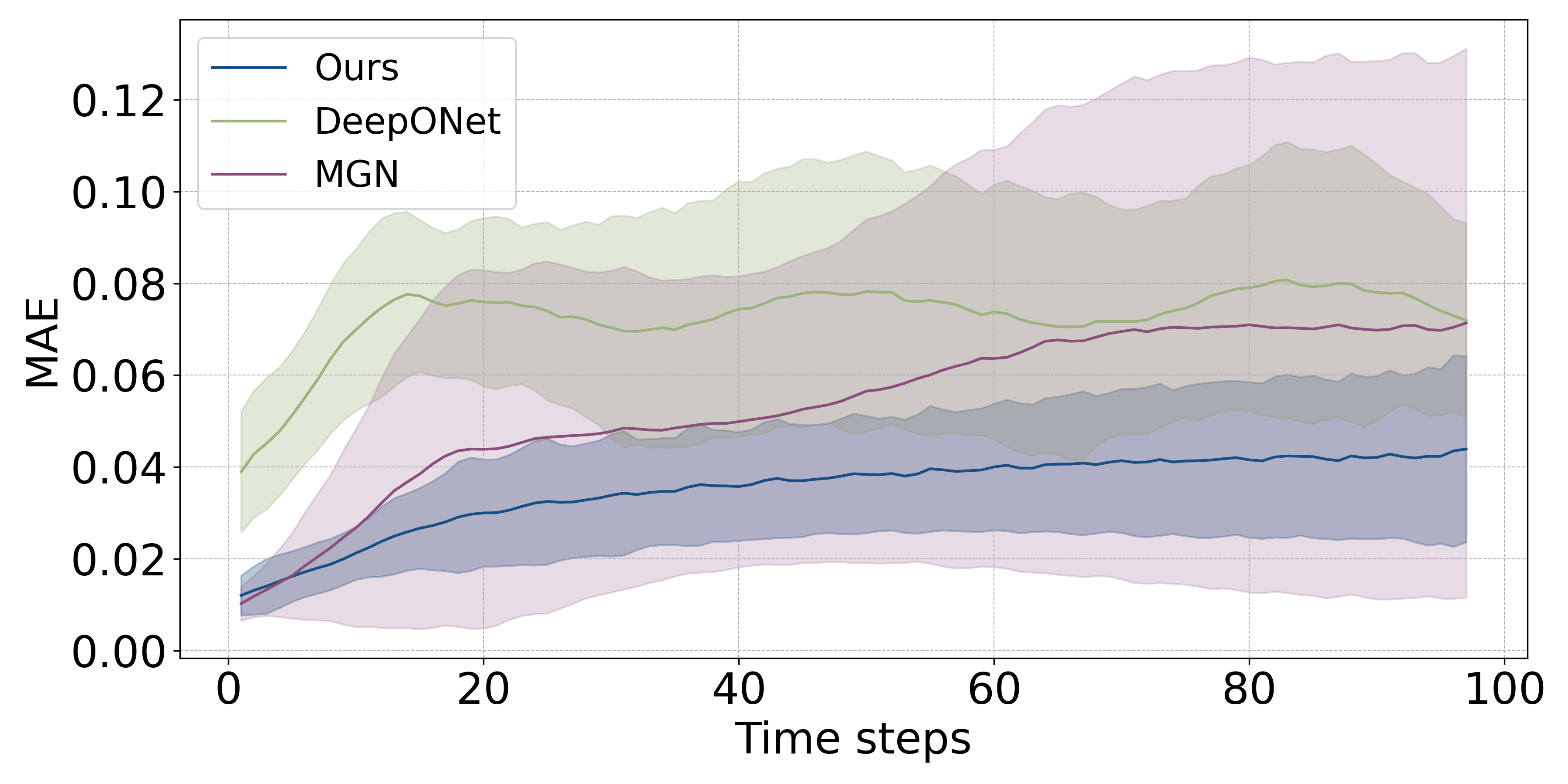}
    \caption{Velocity MAE comparison of surrogate models across physical timesteps for cylinder fluid flow. The solid lines represent mean MAE values across all spatial points at each physical timestep, with the shaded regions representing the standard deviation of these MAE values.}
    \label{fig:Fig.7}
\end{figure}

Furthermore, \Cref{fig:Fig.7} illustrates the average MAE across all nodes over physical time evolution. MGN starts with relatively low errors but exhibits error accumulation over physical timesteps due to its autoregressive scheme. The shaded regions represent standard deviation of MAE values across all spatial points at each physical timestep. The large and widening standard deviation indicates inconsistent prediction quality across different nodes. DeepONet shows high error levels from the beginning. From 15 timestep, it maintains a consistent error level, but the standard deviation gradually increases as physical time progresses. In contrast, our point-wise diffusion model demonstrates consistently low error scales and a narrow standard deviation across all physical timesteps, maintaining robust performance without divergence. 

\begin{figure}[H]
    \captionsetup{font=normalsize}
    \centering
    \includegraphics[width=0.9\linewidth]{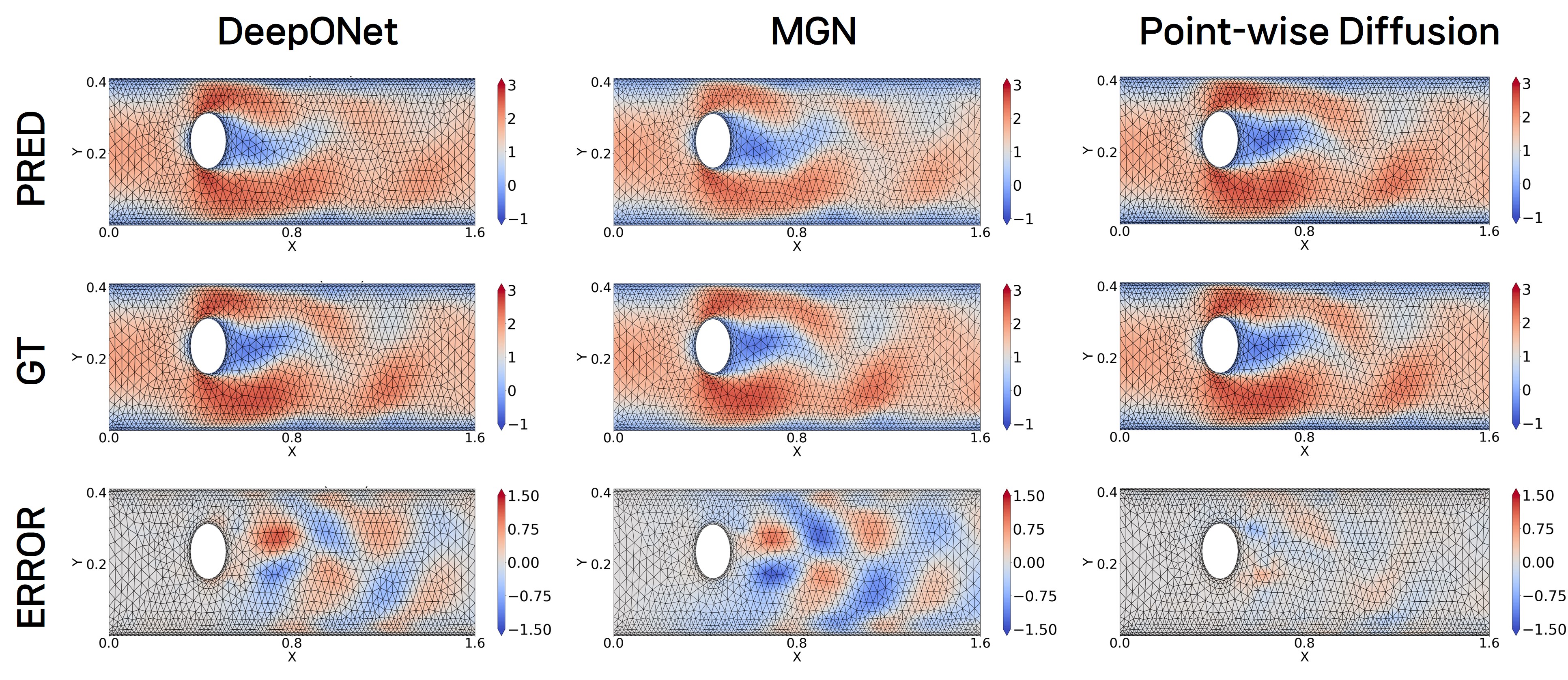}
    \caption{Visual comparison of velocity field predictions from all three models.}
    \label{fig:Fig.8}
\end{figure}

\begin{table}[H]
\captionsetup{font=normalsize}
\centering
\caption{Performance comparison of surrogate models for cylinder fluid flow}
\label{table:Performance comparison of surrogate models for cylinder fluid flow}
\begin{adjustbox}{width=0.85\textwidth, center}
\begin{tabular}{c|c|c|c|c}
\hline
\textbf{Model}& \textbf{Training time}& \textbf{Params}& \textbf{Velocity MAE}& \textbf{Velocity RMSE} \\ \hline
DeepONet & \textbf{1.8h} & 1,798,657 & 0.072 & 0.122 \\
MGN& 20h & 2,332,419 & 0.053 & 0.101 \\ 
Point-wise Diffusion & 8.9h & 1,901,953 & \textbf{0.034}& \textbf{0.065}\\ 
\hline
\end{tabular}
\end{adjustbox}
\end{table}
\vspace{-3mm}
\Cref{fig:Fig.8} provides a visual comparison of velocity field predictions from all three models at $t^{\text{phys}}=90$: predictions (PRED) in the top row, ground truth (GT) in the middle row, and error distribution in the bottom row. The error contour clearly demonstrates that both DeepONet and MGN fail to accurately predict the correct phase of vortex shedding patterns behind the cylinder, as evidenced by the distinct error patterns in the wake region. In contrast, our point-wise diffusion model successfully captures the correct vortex shedding phase, resulting in much lighter blue/red fluctuations in the error contour compared to the pronounced error patterns observed in DeepONet and MGN predictions. A detailed analysis of the model performance in different physical timesteps is provided in \ref{sec:Visualization of surrogate model performance at different timestep}.

\begin{figure}[H]
    \captionsetup{font=normalsize}
    \centering
    \includegraphics[width=\linewidth]{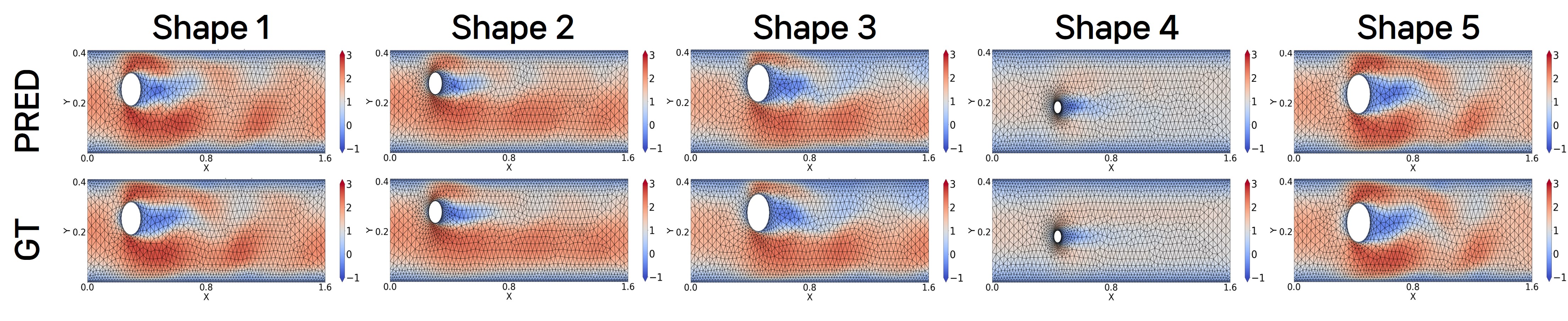}
    \caption{Visualization of point-wise diffusion model predictions across various unseen geometries for cylinder fluid flow system. Each column represents a different geometric configuration (Shape 1-5). Predictions (PRED) are shown in the upper row and ground truth (GT) in the lower row.}
    \label{fig:Fig.8_shape}
\end{figure}

In addition, \Cref{fig:Fig.8_shape} compares the prediction results of our point-wise diffusion model against the numerical solver results for five new geometries that were not included in the training dataset (Shape 1-5) in the Eulerian system. This system evaluates velocity field prediction performance in cylinder flow environments with varying cylinder positions, sizes, and even flow conditions (inlet velocity). Our proposed model accurately predicts the flow characteristics across all these various geometric and flow parameter variations, including challenging cases (Shape 4 and 5) where extreme flow conditions and cylinder geometries lead to different vortex shedding patterns.

\subsection{Lagrangian system: Drop impact}
\label{subsec:[Lagrangian system] Drop impact}

For our second benchmark, we investigate the drop impact simulation, which requires predicting both position and stress of each node in two interacting objects: a falling ball and a multi-layered OLED display panel (see \Cref{[Appendix] Fig.17}). This problem involves a Lagrangian system that tracks physical quantities over time as node positions change, requiring predictions of position and stress for each node at every physical timestep. 

In this system, we present enhanced surrogate models specifically tailored to this problem by addressing the limitations of conventional DeepONet, MGN and by applying specialized adaptations suited to our dataset characteristics, thereby enabling a more rigorous comparison with improved performance and compatibility. For conventional DeepONet, to overcome the limitation that it can only predict a single output function, we implemented the multiple-outputs strategy proposed within the DeepONet framework \citep{lu2022comprehensive} (detailed in \ref{sec:DeepONet Framework}). For MGN, we incorporated physics-constrained loss functions to accurately model the interaction between the ball and the multi-layered OLED display panel, thereby preventing penetration between the two objects \citep{kim2025physics}. This integration of physical constraints constitutes a problem-specific enhancement that simultaneously ensures physical validity and prediction accuracy, which would be difficult to achieve with conventional MGN approaches. Consequently, through these problem-specific adaptations, we conducted a rigorous comparison between models that fully considers the characteristics of our dataset.

As shown in \Cref{table:Performance comparison of surrogate models for drop impact}, DeepONet based on a multiple-outputs strategy demonstrates higher performance compared to physics-constrained MGN for this problem, contrary to previous Eulerian case study. However, our model still demonstrates exceptional accuracy improvements in both position and stress predictions. For position prediction, our approach achieves 73\% and 94\% reductions in MAE compared to DeepONet and physics-constrained MGN, respectively. The improvements are also pronounced in RMSE, with 72\% and 97\% error reductions. For stress prediction, our model shows the best performance gains with 82\% and 87\% MAE reductions compared to DeepONet and physics-constrained MGN, and similarly impressive RMSE improvements of 72\% and 80\%. These substantial accuracy improvements are achieved while maintaining computational efficiency, requiring 68\% less training time than MGN.

\begin{table}[H]
\captionsetup{font=normalsize}
\centering
\caption{Performance comparison of surrogate models for drop impact}
\label{table:Performance comparison of surrogate models for drop impact}
\begin{adjustbox}{width=0.9\textwidth,center}
\begin{tabular}{c|c|c|c c|c c}
\hline
\multirow{2}{*}{\textbf{Model}} & \multirow{2}{*}{\textbf{Training time}} & \multirow{2}{*}{\textbf{Params}} & \multicolumn{2}{c|}{\textbf{Position [mm]}} & \multicolumn{2}{c}{\textbf{Stress [MPa]}} \\
& & & \textbf{MAE} & \textbf{RMSE} & \textbf{MAE} & \textbf{RMSE} \\ \hline
Multi-output DeepONet \citep{lu2022comprehensive} & \textbf{2.7h} & 3,175,427 & 0.064 & 0.065 & 0.411 & 1.570 \\
Physics-constrained MGN \citep{kim2025physics}& 21.6h & 3,852,419 & 0.309 & 0.707 & 0.547 & 2.223 \\ 
Point-wise Diffusion& 8.6h & 2,976,803 & \textbf{0.017} & \textbf{0.018} & \textbf{0.073} & \textbf{0.445} \\ 
\hline
\end{tabular}
\end{adjustbox}
\end{table}

\begin{figure}[H]
    \captionsetup{font=normalsize}
    \centering
    \begin{subfigure}{0.48\textwidth}
        \includegraphics[width=\linewidth]{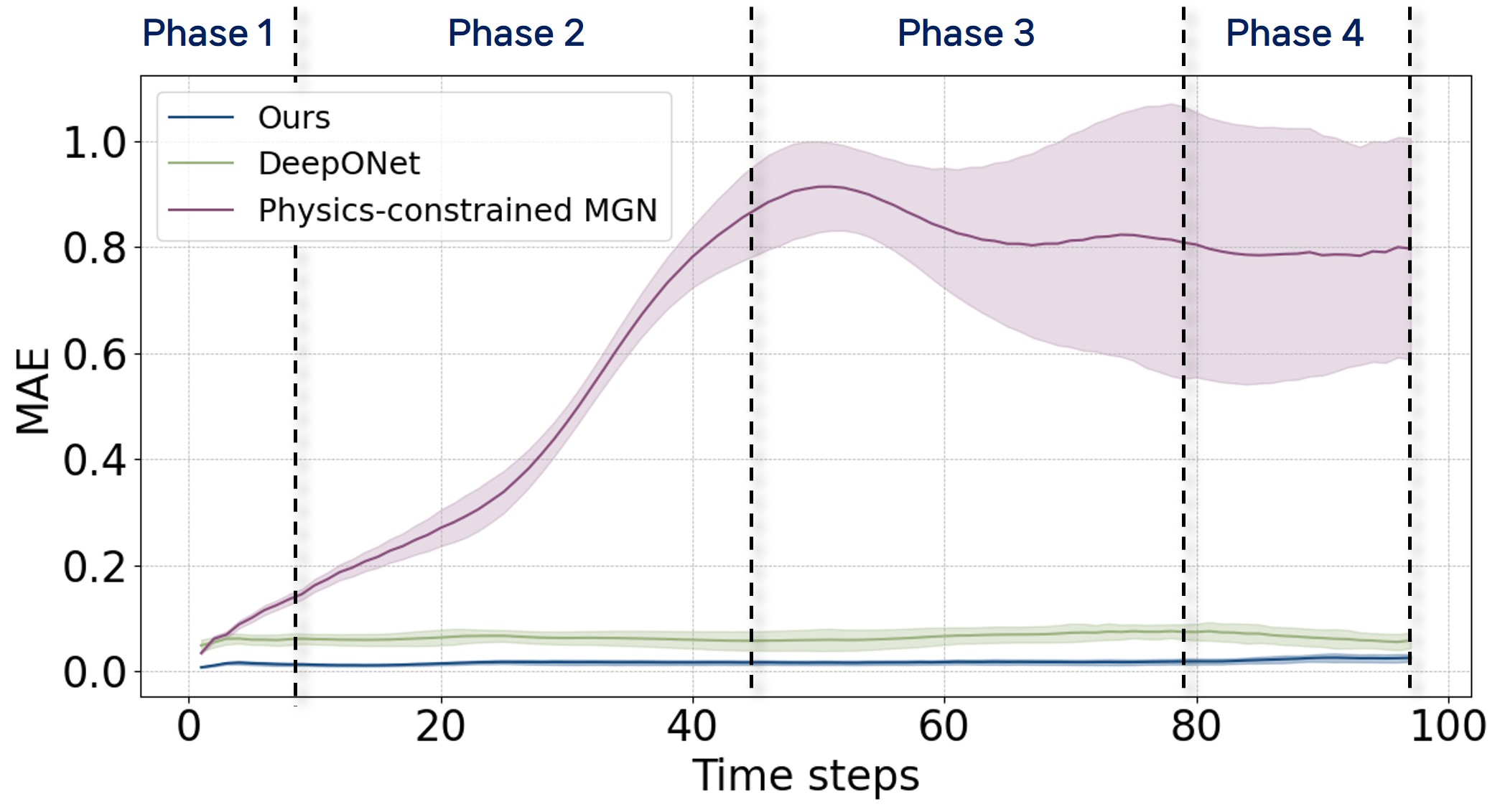}
        \caption{Position error (unit: mm)}
        \label{fig:Fig.9a}
    \end{subfigure}
    \begin{subfigure}{0.48\textwidth}
        \includegraphics[width=\linewidth]{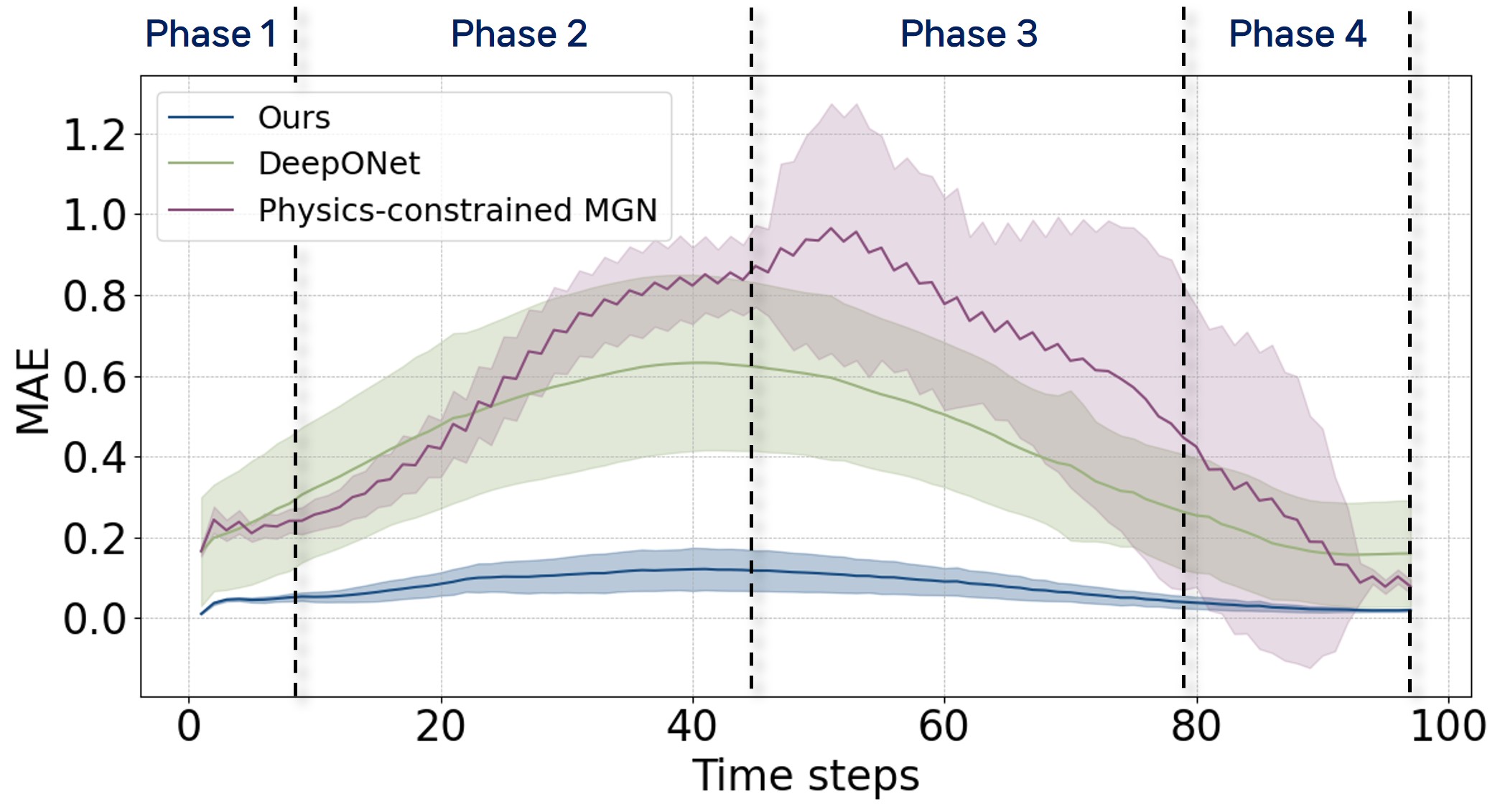}
        \caption{Stress error (unit: MPa)}
        \label{fig:Fig.9b}
    \end{subfigure}

    \caption{Drop impact MAE comparison of surrogate model across physical timesteps}
    \label{fig:Fig.9}
\end{figure}

\Cref{fig:Fig.9} illustrates the MAE for position and stress across different physical timesteps, which is divided into four distinct time regions:

\begin{itemize}
    \item Phase 1: Ball drop, no contact
    \item Phase 2: Ball drop, contact
    \item Phase 3: Ball rebound, contact
    \item Phase 4: Ball rebound, no contact
\end{itemize}

The physics-constrained MGN still suffers significant error accumulation due to its autoregressive scheme, which is particularly evident when analyzing performance in Phase 2 and 3. These phases represent complex physical scenarios involving contact between the ball and panel, making prediction exceptionally challenging. Indeed, severe error accumulation occurs in the physics-constrained MGN within these regions, clearly demonstrating the fundamental limitations of autoregressive prediction approaches. In contrast, both DeepONet and ours exhibit substantially superior performance due to their non-autoregressive scheme. Notably, our point-wise diffusion model maintains consistently excellent predictive performance across the entire physical time range without the abrupt error increase observed in DeepONet during Phase 2. Furthermore, our model demonstrates the most stable results regarding standard deviation of the MAE values across all spatial points at each physical timestep, confirming its robustness throughout all simulation phases. 

\begin{figure}[H]
    \captionsetup{font=normalsize}
    \centering
    \includegraphics[width=0.9\linewidth]{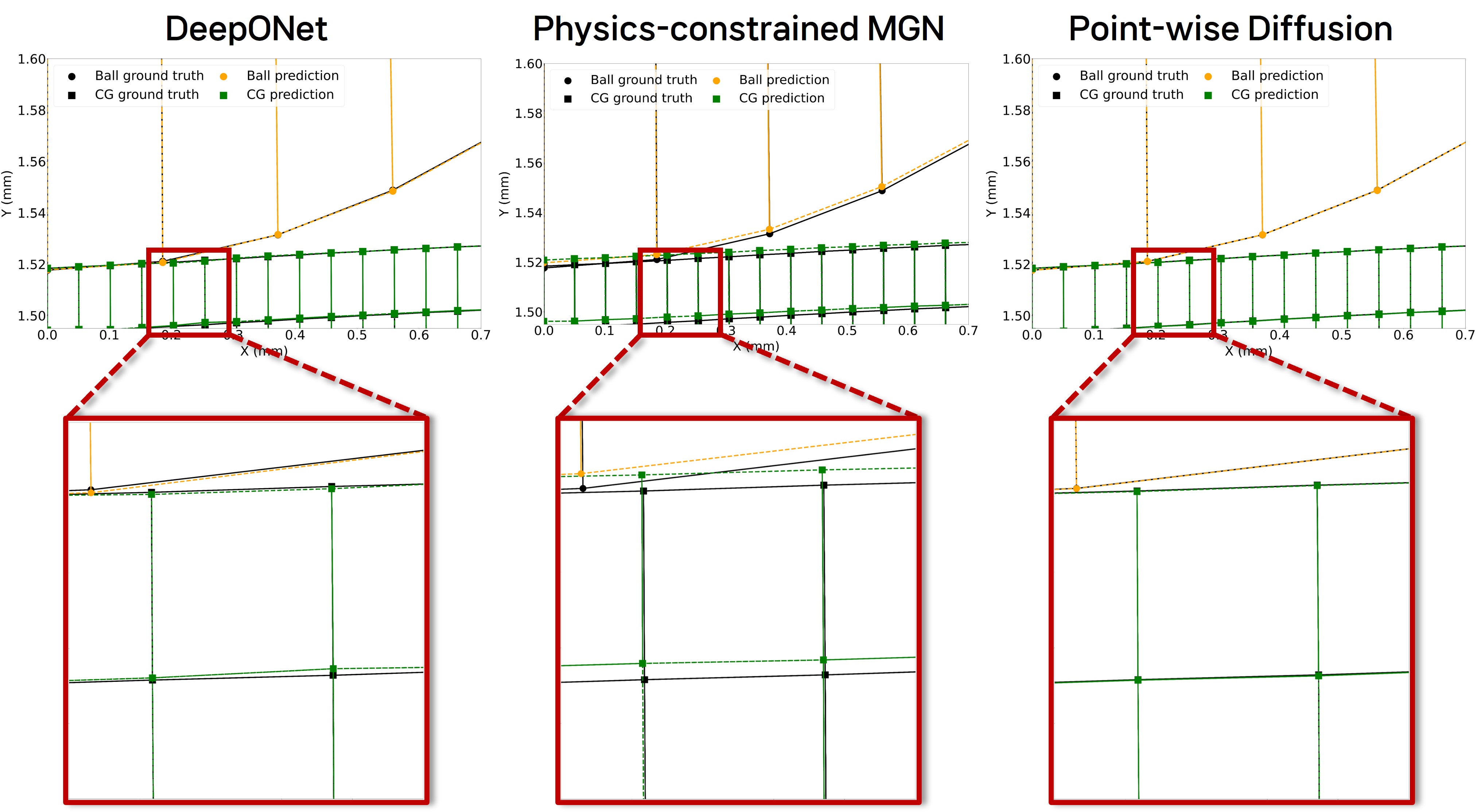}
    \caption{Visual comparison of surrogate models for drop impact: position prediction}
    \label{fig:Fig.10}
\end{figure}

Since it is crucial for this case study to prevent penetration between two objects, we conduct further visual analysis through \Cref{fig:Fig.10} to thoroughly examine how accurately each model can capture penetration phenomena. It shows node position predictions during Phase 3 at $t^{\text{phys}}=70$. Although the physics-constrained MGN prevents penetration through physical constraints, it shows noticeable discrepancies between ground truth and predicted positions. DeepONet demonstrates overall excellent performance but exhibits clear prediction errors, particularly in the panel interior regions. In contrast, our point-wise diffusion model achieves the most accurate predictions by closely matching the ground truth across all node positions, thereby explaining its superior quantitative performance in \Cref{table:Performance comparison of surrogate models for drop impact}.

\begin{figure}[H]
    \captionsetup{font=normalsize}
    \centering
    \includegraphics[width=0.9\linewidth]{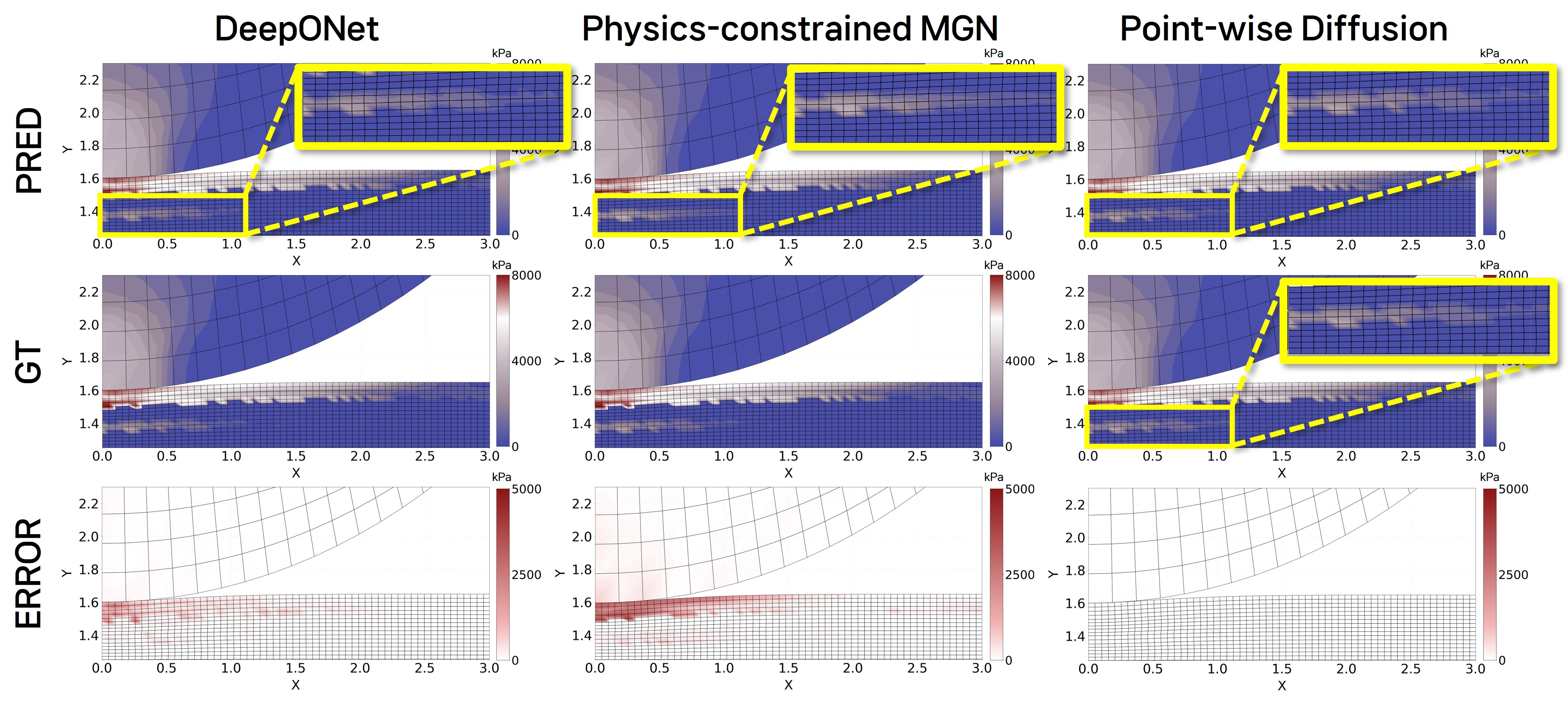}
    \caption{Visual comparison of surrogate models for drop impact: stress prediction}
    \label{fig:Fig.11}
\end{figure}

\begin{figure}[H]
    \captionsetup{font=normalsize}
    \centering
    \includegraphics[width=\linewidth]{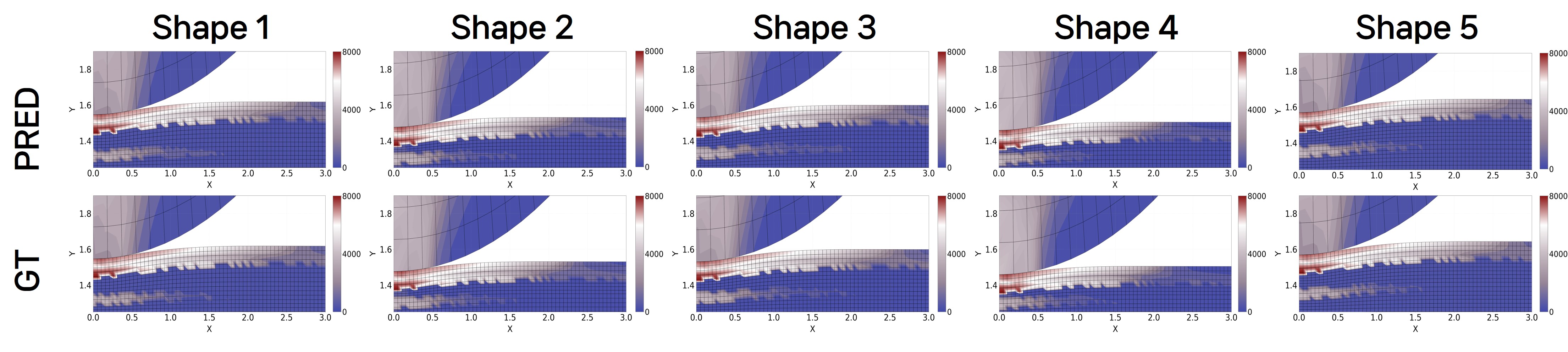}
    \caption{Visualization of point-wise diffusion model predictions across various unseen geometries for drop impact system. Each column represents a different geometric configuration (Shape 1-5). For each system, predictions (PRED) are shown in the upper row and ground truth (GT) in the lower row.}
    \label{fig:Fig.11_shape}
\end{figure}

\Cref{fig:Fig.11} further presents the stress prediction capabilities, revealing even more dramatic differences between the approaches. Our approach not only shows significantly lower error magnitude compared to the other two models, as shown in the error contour, but also captures fine details with better precision (yellow boxes). These results mean that our model accurately reproduces complex stress patterns at boundary interfaces, where physical interactions are most challenging to predict. We validate that our proposed model excels at generalizing to nonlinear dynamic analyses like drop impact simulation, while avoiding the need for complex hyperparameter tuning for object interactions that physics-constrained MGN requires. The model’s prediction performance across different physical timesteps is summarized in \ref{sec:Visualization of surrogate model performance at different timestep}.
Furthermore, \Cref{fig:Fig.11_shape} visualizes prediction performance in a drop impact scenario where a ball drops onto a multi-layer display panel with varying optically clear adhesive. Each shape represents a different thickness combination of two OCA layers (detailed settings are provided in \ref{sec:Data generation in drop impact system}), resulting in varying stress transmission and dispersion patterns within the panel. Our proposed model accurately predicts both the complex stress discontinuities and the dynamic behavior arising from these thickness variations. Particularly in Shapes 2 and 5, the model successfully captures the distinctly different stress distributions and deformations that occur when OCA thickness is at the boundary of training range (min/max).

\subsection{Large-scale system: Road-car external aerodynamics}
\label{subsec:[Large-scale system] Road-car external aerodynamics}

Finally, we focus on road-car external aerodynamics system, which employs hybrid RANS-LES (HRLES) simulations to capture high-fidelity flow characteristics. The system requires the simultaneous prediction of surface pressure and three-dimensional wall shear stress across parametrically varied vehicle geometries. The challenge lies in accurately capturing complex surface aerodynamic quantities around various car configurations while maintaining computational efficiency for large-scale systems. 

Similar to the drop impact system, we apply enhanced surrogate models that are specifically tailored to accommodate this aerodynamics dataset. For DeepONet, we implemented the same multiple-outputs strategy as in the previous case study to handle the prediction of four simultaneous output quantities (detailed in \ref{sec:DeepONet Framework}). For MGN, we adopted the  X-Meshgraphnet (X-MGN) \citep{nabian2024x}, which provides enhanced scalability compared to conventional MGN while effectively handling long-range interactions crucial for large scale dataset. Specifically, X-MGN addresses computational scalability by dividing large graphs into smaller subgraphs, where overlapping boundary regions (halo regions) preserve information exchange between adjacent partitions and gradient aggregation maintains training equivalence to processing the entire graph simultaneously. Additionally, X-MGN captures efficient long-range interactions through multi-scale graph generation that iteratively combines coarse and fine-resolution point clouds. In this study, we employed a 3-level multi-scale graph architecture containing 100k, 200k, and 400k nodes at the respective levels. Each scale is partitioned into 3 subgraphs with halo regions of size 15 to ensure seamless information exchange across partitions, utilizing 6-nearest neighbor connectivity. The model consists of 15 message-passing layers with a hidden dimension of 512. In addition, we conducted training and inference on a single NVIDIA A100 GPU for consistency and fair comparison between surrogate models. Furthermore, relative L1 and L2 error metrics were employed to evaluate prediction accuracy across different vehicle geometries.

\Cref{table:Performance comparison of surrogate models for road-car external aerodynamics} demonstrates that our point-wise diffusion model outperforms other approaches across all output variables. For pressure prediction, our model achieves 35\% and 51\% reductions in relative L2 error compared to DeepONet and X-MGN, respectively. Similar improvements are observed in shear stress predictions, with 30\% and 43\% error reductions in X-wall direction compared to DeepONet and X-MGN, and consistent 29-38\% improvements across Y-wall and Z-wall directions. Relative L1 error analysis reveals even superior performance, with our model consistently achieving 44-68\% error reductions across all predicted quantities compared to both DeepONet and X-MGN. This study achieves these considerable predictive enhancements with enhanced computational efficiency, exhibiting 23\% lower training requirements compared to X-MGN.

\begin{figure}[H]
    \captionsetup{font=normalsize}
    \centering
    \begin{subfigure}{0.24\textwidth}
        \includegraphics[width=\linewidth]{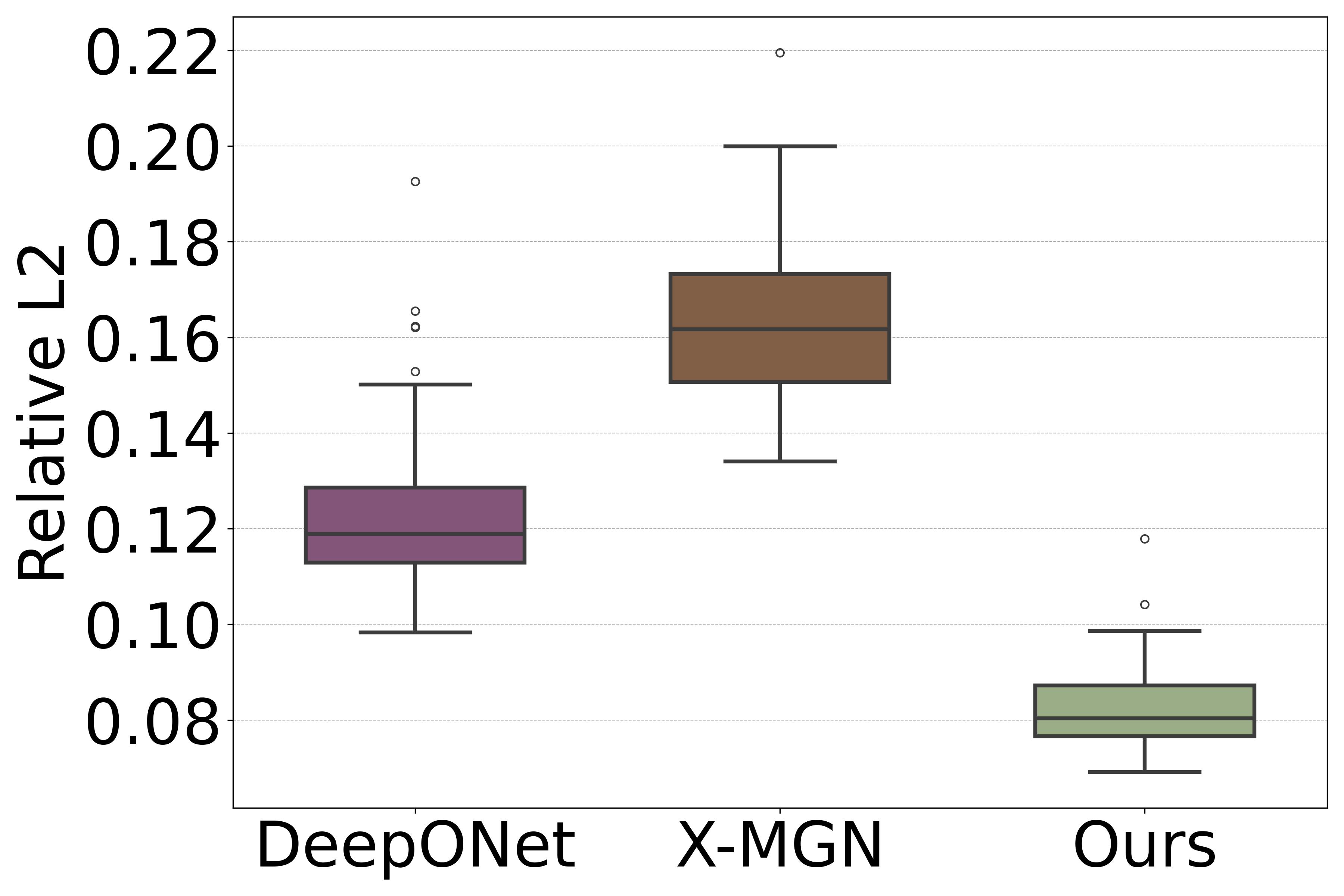}
        \caption{Surface pressure}
        \label{fig:Fig.12a}
    \end{subfigure}
    \begin{subfigure}{0.24\textwidth}
        \includegraphics[width=\linewidth]{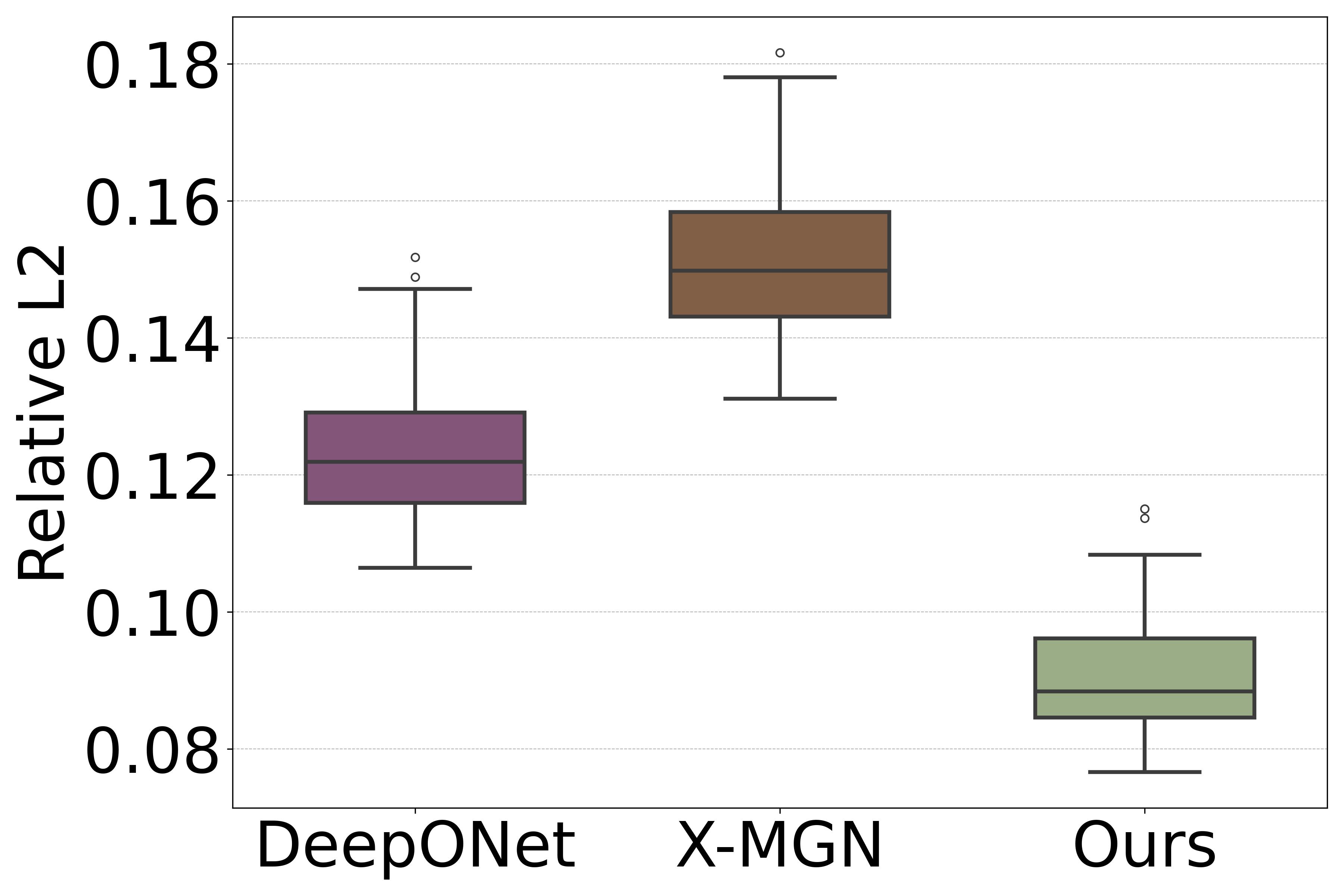}
        \caption{X-wall shear stress}
        \label{fig:Fig.12b}
    \end{subfigure}
    \begin{subfigure}{0.24\textwidth}
        \includegraphics[width=\linewidth]{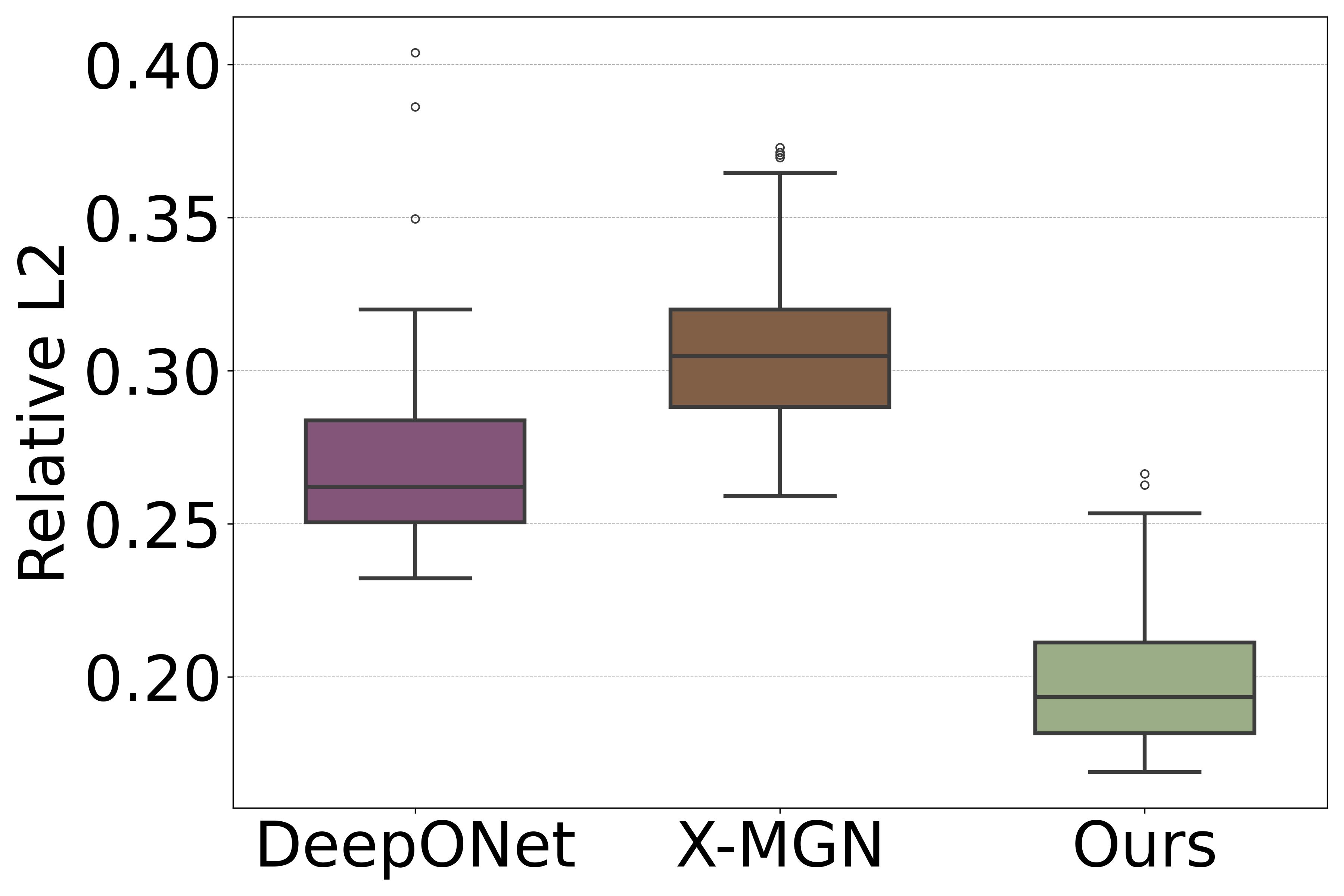}
        \caption{Y-wall shear stress}
        \label{fig:Fig.12c}
    \end{subfigure}
    \begin{subfigure}{0.24\textwidth}
        \includegraphics[width=\linewidth]{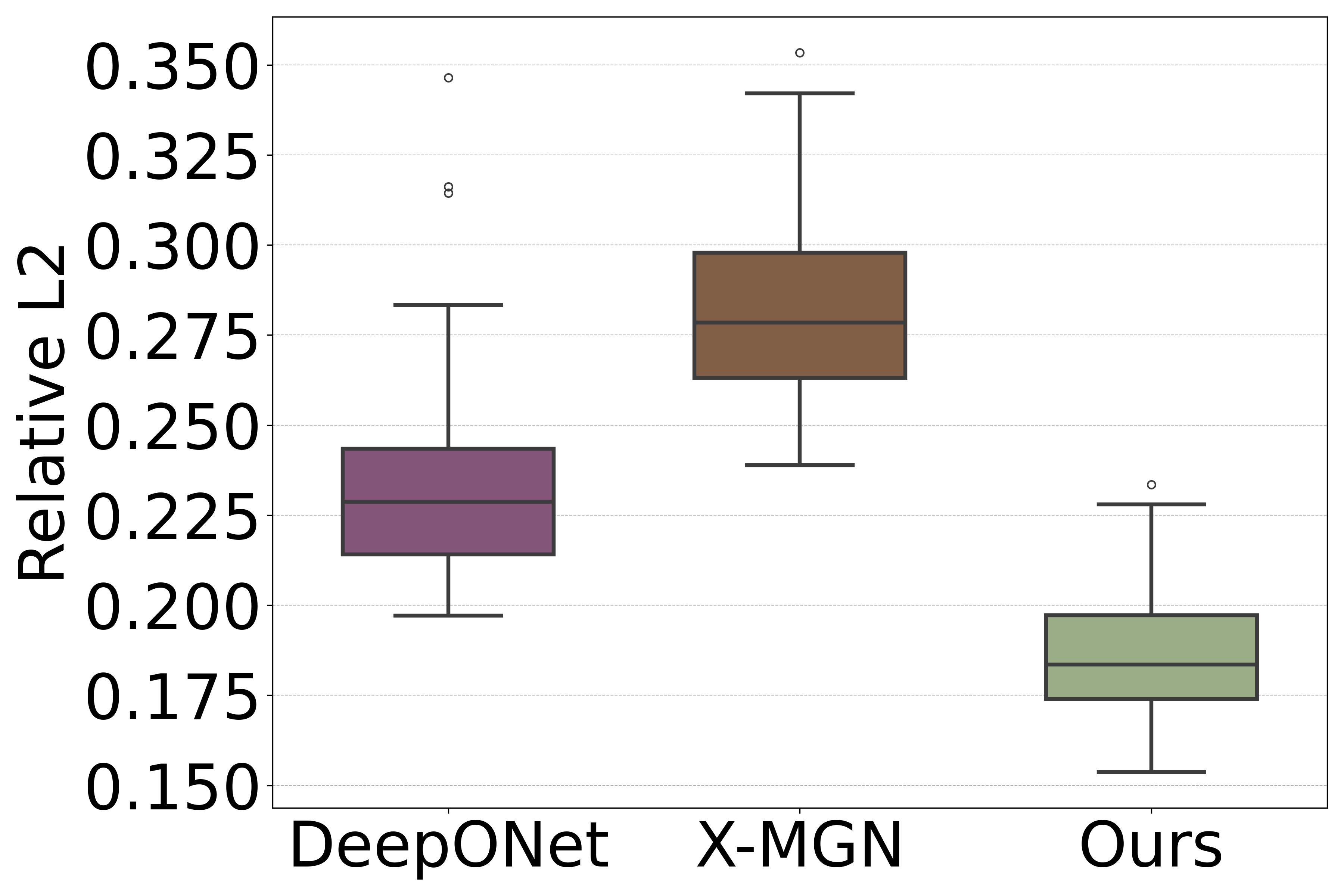}
        \caption{Z-wall shear stress}
        \label{fig:Fig.12d}
    \end{subfigure}
    \caption{Relative L2 comparison of surrogate models for road-car external aerodynamics}
    \label{fig:Fig.12}
\end{figure}

\Cref{fig:Fig.12} provides deeper insights through box plots illustrating performance distributions across models. For surface pressure prediction, DeepONet exhibits intermediate median performance but shows notable outliers, while X-MGN demonstrates the highest median errors with substantial variability. These differences become more evident in shear stress predictions, where DeepONet displays wider interquartile ranges (IQRs), particularly in Y- and Z-wall directions, and X-MGN consistently exhibits the worst median performance with the widest IQRs across all wall shear stress components. In contrast, our approach achieves the lowest median values while maintaining the narrowest IQRs across all quantities with minimal outliers, demonstrating both superior accuracy and consistent prediction reliability.

\begin{table}[H]
\captionsetup{font=normalsize}
\centering
\caption{Performance comparison of surrogate models for road-car external aerodynamics (Rel: Relative)}
\label{table:Performance comparison of surrogate models for road-car external aerodynamics}
\begin{adjustbox}{width=\textwidth,center}
\begin{tabular}{c|c|c|c c|c c|c c|c c}
\hline
\multirow{3}{*}{\textbf{Model}} & \multirow{3}{*}{\textbf{Training time}} & \multirow{3}{*}{\textbf{Params}} & \multicolumn{2}{c|}{\multirow{2}{*}{\textbf{Surface pressure}}} & \multicolumn{6}{c}{\textbf{Shear Stresses}} \\
\cline{6-11}
& & & \multicolumn{2}{c|}{} & \multicolumn{2}{c|}{\textbf{X-Wall}} & \multicolumn{2}{c|}{\textbf{Y-Wall}} & \multicolumn{2}{c}{\textbf{Z-Wall}} \\
\cline{4-11}
& & & \textbf{Rel-L2} & \textbf{Rel-L1} & \textbf{Rel-L2} & \textbf{Rel-L1} & \textbf{Rel-L2} & \textbf{Rel-L1} & \textbf{Rel-L2} & \textbf{Rel-L1} \\
\hline
Multi-output DeepONet \citep{lu2022comprehensive} & \textbf{47.53h}& 3,176,452 & 0.123 & 0.069 & 0.123 & 0.078 & 0.270 & 0.246 & 0.233 & 0.207 \\
X-MGN \citep{nabian2024x}& 84.33h & 3,234,308 & 0.163 & 0.107 & 0.152 & 0.103 & 0.308 & 0.292 & 0.284 & 0.258 \\ 
Point-wise Diffusion& 64.50h & 2,422,804 & \textbf{0.080} & \textbf{0.034} & \textbf{0.086} & \textbf{0.042} & \textbf{0.192} & \textbf{0.137} & \textbf{0.181} & \textbf{0.115} \\ 
\hline
\end{tabular}
\end{adjustbox}
\end{table}

\begin{figure}[H]
    \captionsetup{font=normalsize}
    \centering
    \includegraphics[width=0.9\linewidth]{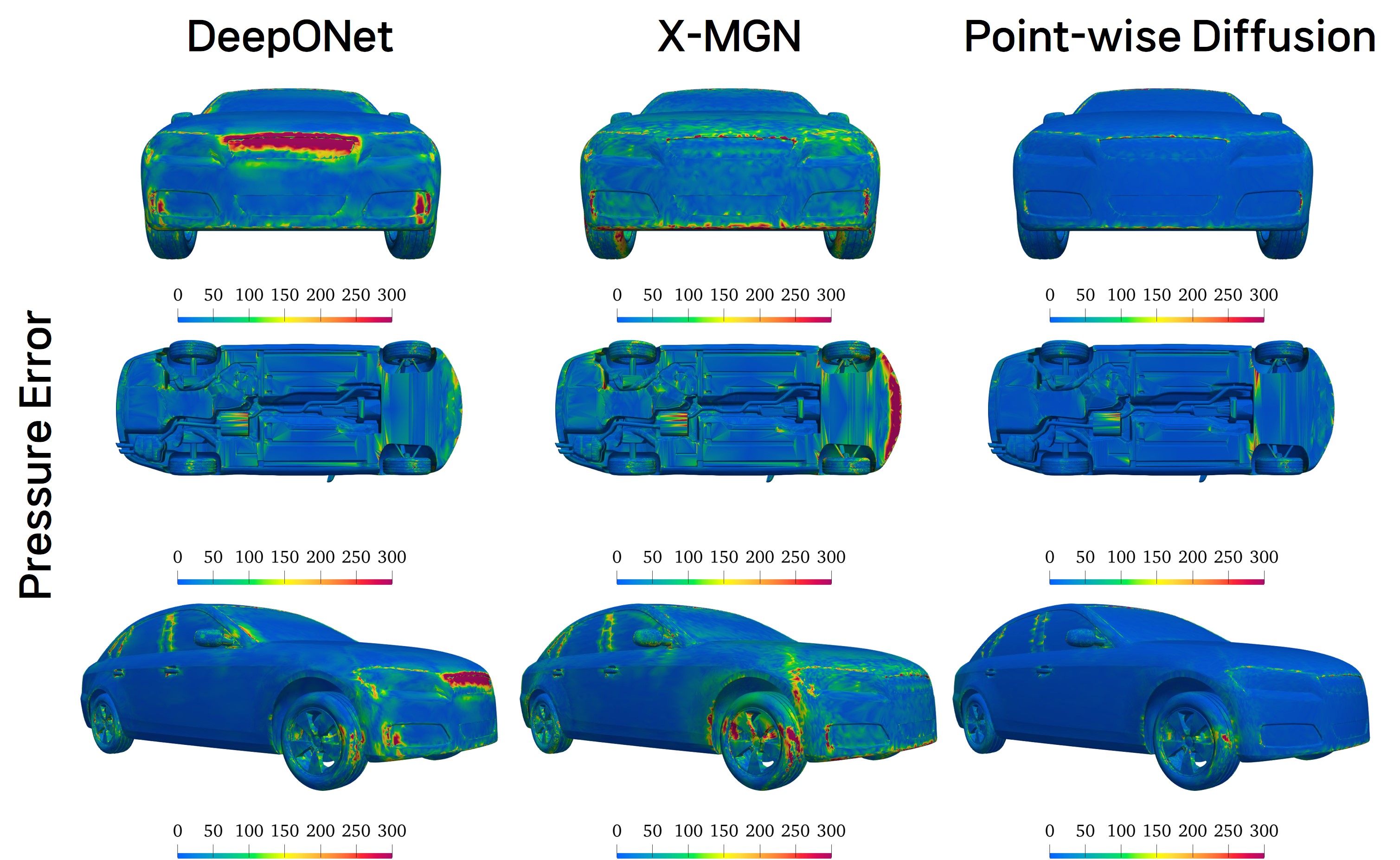}
    \caption{Performance comparison of surrogate models for large-scale automative system: surface pressure prediction}
    \label{fig:Fig.13}
\end{figure}

\begin{figure}[H]
    \captionsetup{font=normalsize}
    \centering
    \includegraphics[width=0.9\linewidth]{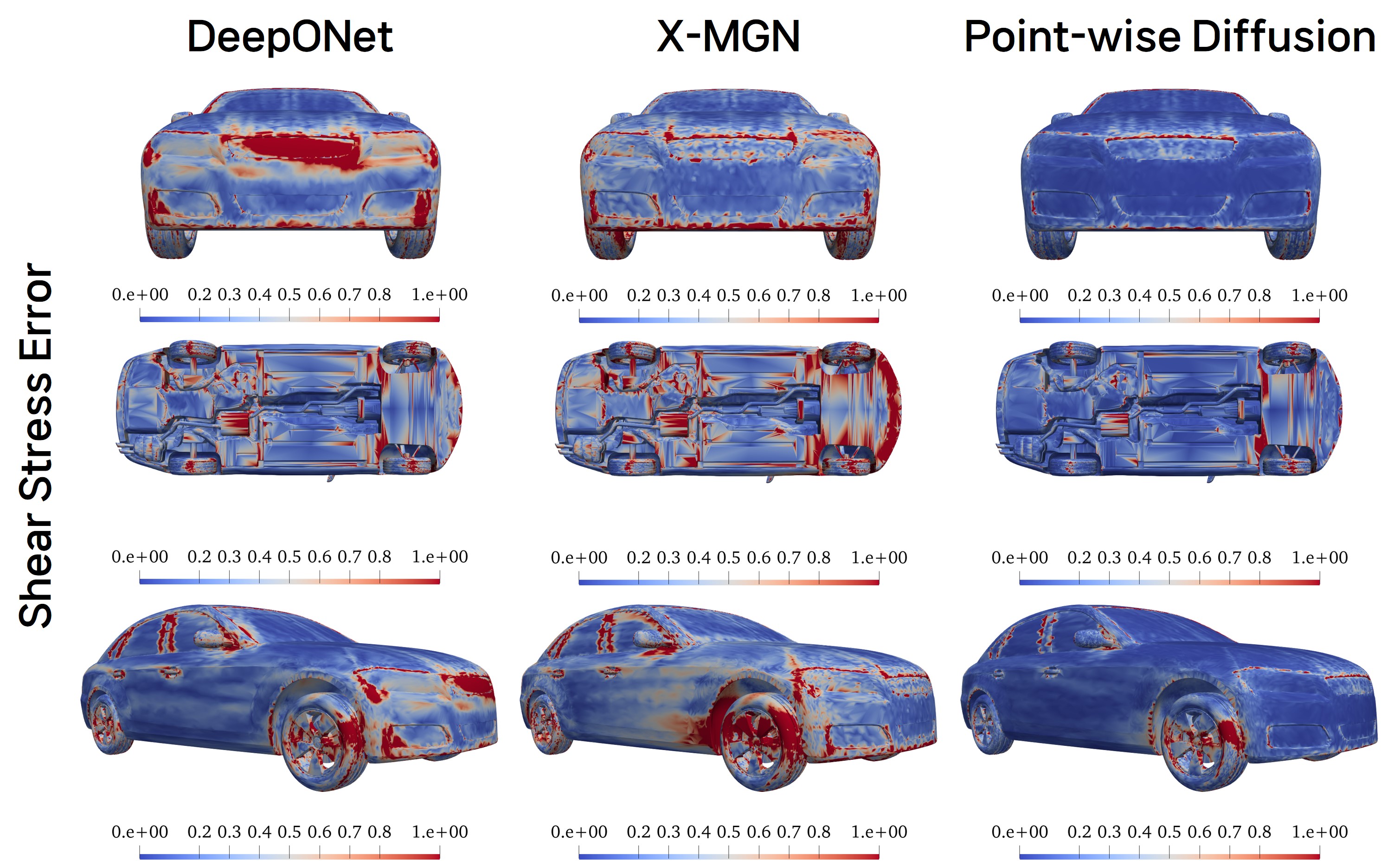}
    \caption{Performance comparison of surrogate models for large-scale automative system: XYZ-wall shear stresses prediction}
    \label{fig:Fig.14}
\end{figure}

\Cref{fig:Fig.13,fig:Fig.14} present spatial error distributions for surface pressure and wall shear stresses predictions across the three surrogate models (detailed prediction results are available in \ref{sec:Visualization of surrogate model performance at different timestep}). Analyzing these error contours from multiple camera angles reveals distinct performance differences between models. In surface pressure prediction (\Cref{fig:Fig.13}), DeepONet exhibits localized error concentrations in front section of vehicle, while X-MGN demonstrates extensive high-magnitude errors across the entire computational domain, with particularly severe inaccuracies in front bumper underbody. Our point-wise diffusion approach maintains consistently low error magnitudes throughout the solution space, with negligible deviations from ground truth. The wall shear stress error analysis (\Cref{fig:Fig.14}) reveals more distinct performance differences. DeepONet exhibits substantial error concentrations (red regions) in the front section of vehicle, the area of underbody, and around the wheel. X-MGN also demonstrates severe error patterns, showing extensive high-magnitude errors that cover large portions of the vehicle surface, including the entire underbody, wheel, side surfaces, and front sections, indicating significant accuracy degradation across the computational domain. In contrast, our methodology demonstrates exceptional spatial accuracy, maintaining predominantly low error levels (blue regions) throughout the vehicle surface, with only minimal localized errors even in geometrically complex areas such as the wheel, side mirrors, and underbody. The consistent low-error performance across all geometric complexities confirms our model's superior capability for accurate and reliable automotive aerodynamic predictions across complex 3D systems.

\begin{figure}[H]
    \captionsetup{font=normalsize}
    \centering
    \includegraphics[width=\linewidth]{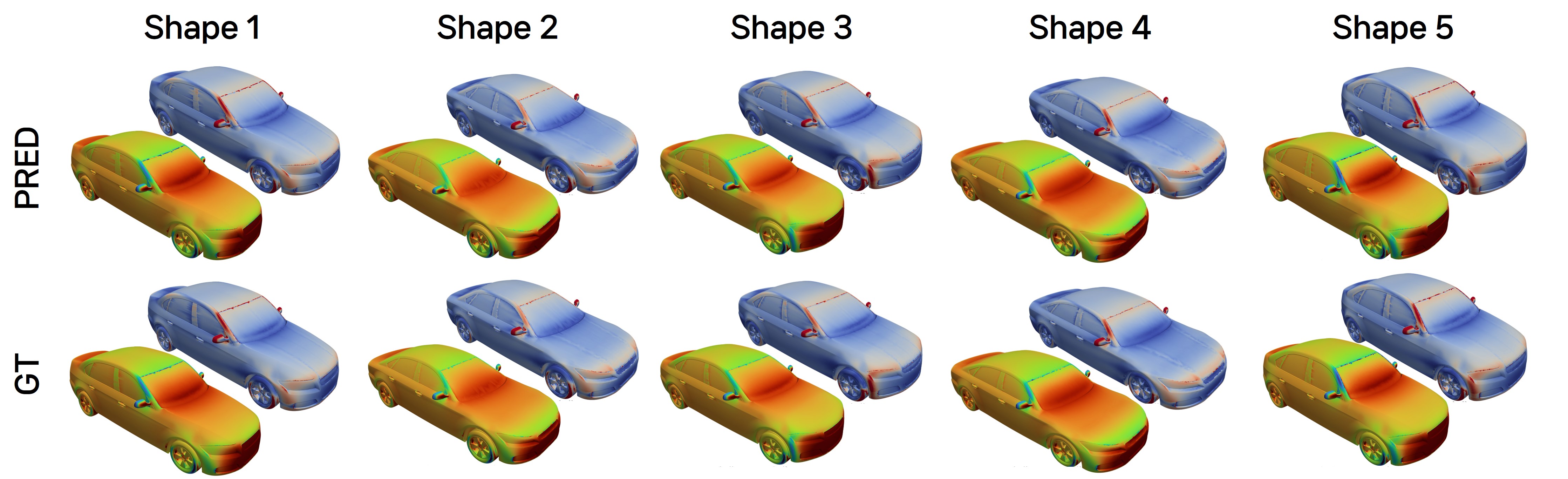}
    \caption{Visualization of point-wise diffusion model predictions across various unseen geometries for the road-car external aerodynamics system. Each column represents a different geometric configuration (Shape 1-5), displaying surface pressure fields on the left vehicle and wall shear stress distributions on the right vehicle. Predictions (PRED) are shown in the upper row and ground truth (GT) in the lower row.}
    \label{fig:Fig.13_14_shape}
\end{figure}

In addition, \Cref{fig:Fig.13_14_shape} demonstrates performance in predicting the aerodynamic characteristics of external flow around vehicles, visualizing surface pressure and wall shear stress distributions across various body shape modifications. Each shape represents a vehicle design with different configurations for 16 design parameters, including front and rear length, front overhang, diffuser angle, and pitch (see \Cref{table:morphing-parameters}). The model accurately predicts key aerodynamic features that emerge from these shape variations, notably reproducing the strong pressure increases observed in the vehicle bumper area and capturing high-shear regions resulting from boundary layer separation around the front corners and side mirrors. Particularly noteworthy are Shapes 4 and 5, which represent extreme geometric configurations with opposing parameter values for vehicle height, hood angle, and rear-end tapering, yet the model maintains robust prediction accuracy across these contrasting design extremes.


\section{Further refinement towards optimization of proposed point-wise diffusion model }
\label{sec:Further refinement towards optimization of proposed point-wise diffusion model}

In this section, we explore further refinement strategies to optimize the performance and computational efficiency of our point-wise diffusion model. Specifically, we examine two critical aspects: (i) comparative analysis between direct and residual prediction strategies across spatio-temporal physical systems, and (ii) model efficiency across varying point sampling ratios for computational scalability.

\subsection{Direct versus residual prediction schemes in spatio-temporal physical systems: a comparison}
\label{subsec:Direct versus residual prediction schemes in spatio-temporal physical systems: a comparison}

We compare direct state prediction and residual prediction schemes in spatio-temporal physical systems to evaluate their relative performance. In residual prediction, rather than directly predicting absolute states $q_t$, the model learns to estimate incremental changes $\Delta q_t = q_t - q_0$ from the initial state $q_0$. This approach is more physically reasonable than direct prediction, as it aligns with conventional PDE solvers that typically compute incremental changes from the current state rather than predicting absolute field values directly.

\begin{table}[H]
\captionsetup{font=normalsize}
\centering
\caption{Performance comparison between direct and residual prediction with point-wise diffusion model}
\label{table:residual_vs_direct}
\begin{adjustbox}{width=0.8\textwidth,center}
\begin{tabular}{cc cc cc cc}
\hline
\multirow{2}{*}{\textbf{System}} & \multirow{2}{*}{\textbf{Prediction Type}} & \multicolumn{2}{c}{\textbf{Velocity}} & \multicolumn{2}{c}{\textbf{Position}} & \multicolumn{2}{c}{\textbf{Stress}} \\
& & \textbf{MAE} & \textbf{RMSE} & \textbf{MAE} & \textbf{RMSE} & \textbf{MAE} & \textbf{RMSE} \\ \hline
\multirow{2}{*}{Cylinder Fluid Flow} & Direct& 0.036 & 0.068 & \multicolumn{2}{c}{N/A} & \multicolumn{2}{c}{N/A} \\ \cline{2-8}
& Residual & \textbf{0.034} & \textbf{0.065} & \multicolumn{2}{c}{N/A} & \multicolumn{2}{c}{N/A} \\ \hline
\multirow{2}{*}{Drop Impact} & Direct& \multicolumn{2}{c}{N/A} & 0.501 & 0.376 & \textbf{0.062} & \textbf{0.416}\\ \cline{2-8}
& Residual& \multicolumn{2}{c}{N/A} & \textbf{0.017} & \textbf{0.018} & 0.073 & 0.445\\ \hline
\end{tabular}
\end{adjustbox}
\end{table}

To evaluate the effectiveness of residual prediction, we conducted a comparative analysis against direct prediction with two distinct time-dependent physical systems: cylinder fluid flow and drop impact dynamics (Table~\ref{table:residual_vs_direct}). For velocity prediction in the cylinder flow system, residual prediction demonstrated modest improvements, reducing MAE and RMSE by 5.6\% and 4.4\%, respectively.  Quantitative analysis of the output distributions reveals that the original velocity field spans the range [-0.8421, 2.8224] while the corresponding velocity residuals span [-1.9481, 1.1739]. The range magnitude decreased only slightly from 3.66 (2.8224 + 0.8421) to 3.11 (1.9481 + 1.1739), representing merely a 15\% reduction. This limited range reduction accounts for the modest performance gains observed in this system.

In contrast, the drop impact system demonstrated dramatic improvements with residual prediction for position estimation. Residual learning achieved substantial error reductions of 96.6\% in MAE and 95.2\% in RMSE compared to direct prediction. Analysis of target distributions reveals a striking difference: while the original $y$-axis positions span from 1.0998 to 4.2351, the corresponding displacement residuals range only from -0.0725 to 0.0354 approximately 3.4\% of the original range. This substantial reduction in target variability appears to facilitate more effective learning due to its similarity with conventional PDE solvers' operations, as the model focuses on predicting incremental changes rather than absolute spatial coordinates. For stress prediction in the drop impact system, the residual and direct formulations are mathematically equivalent since initial stress values are zero ($\Delta \sigma_t = \sigma_t$). However, empirical results show slight performance differences between the two approaches. These subtle differences likely stem from the multi-output prediction setup, where stress is predicted simultaneously with position in the direct approach versus displacement in the residual approach.

These findings demonstrate that residual prediction effectiveness is system-dependent, with substantial benefits observed for drop impact systems but limited advantages for cylinder fluid flow system. This underscores the need for careful consideration of target variable properties when designing prediction frameworks for physical systems.

\subsection{Efficiency analysis across different sampling ratios for computational scalability}
\label{subsec:Efficiency analysis across different sampling ratios for computational scalability}

\begin{figure}[H]
    \captionsetup{font=normalsize}
    \centering
    \includegraphics[width=\linewidth]{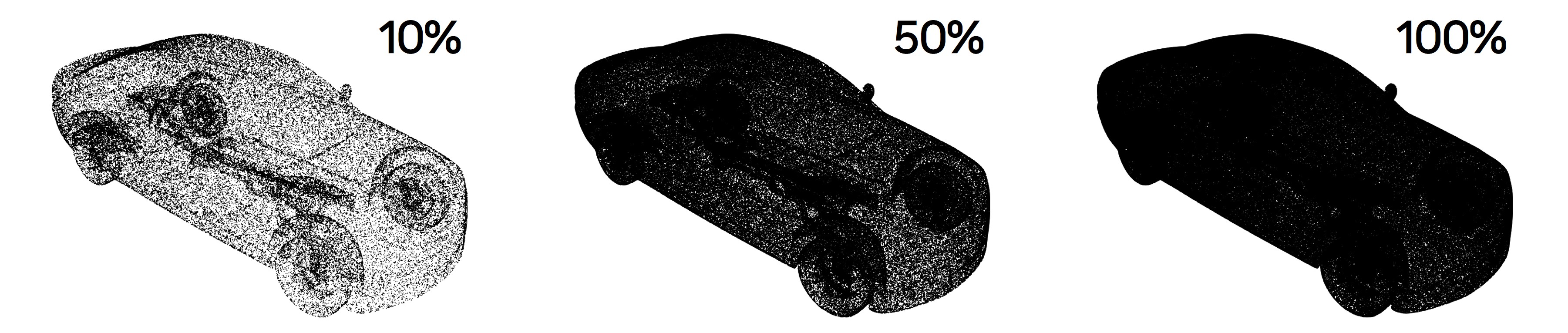}
    \caption{Surface point distribution at different sampling ratios (10\%, 50\%, 100\%) for road-car external aerodynamics.}
    \label{fig:Fig.15}
\end{figure}

Our point-wise diffusion model operates on individual spatio-temporal point, making computational cost directly sensitive to the number of points processed. For the road-car external aerodynamics problem specifically, each vehicle comprises approximately 900,000 nodes, leading to prohibitively high computational costs when processing 100\% of nodes across all 387 training samples in our dataset. To investigate the impact of point sampling ratios on model performance, we performed a comprehensive analysis across various sampling ratios.

In this experiment, we selected only DeepONet and point-wise diffusion model for comparison. Meshgraphnet was excluded because it faces topological constraints when constructing new edges from subsampled nodes. This limitation highlights an inherent drawback of mesh-based graph networks and indirectly demonstrates the importance of flexibility offered by point-wise approaches.

\Cref{fig:Fig.15} illustrates the visual representation of different sampling ratios (10\%, 50\%, and 100\%) for the road-car model, showing how the point distribution becomes progressively denser with higher sampling rates. We generated datasets with varying fidelity levels (10\%, 30\%, 50\%, 70\%, and 100\% of nodes) through uniform sampling and evaluated how effectively models trained on these reduced datasets could predict high-fidelity (100\% nodes) results. In all cases, inference was performed on the complete set of nodes to assess model scalability, with results summarized in Table~\ref{table:sampling_rates}.

The experimental results demonstrate that our point-wise diffusion model outperforms DeepONet even with significantly lower sampling ratios. At just 30\% sampling, our model already achieves superior performance (approximately 19\% lower average error across all physical quantities) compared to DeepONet trained on 100\% of the node sampling, while requiring only 38\% of the training time. This demonstrates an excellent balance between reduced computational cost and maintained high prediction accuracy. When increasing to 50\% sampling, our model further improves performance (approximately 30\% lower average error across all physical quantities) compared to DeepONet's full node sampling results, while still requiring only 65\% of the training time—offering an optimal balance between enhanced accuracy and computational efficiency. These findings highlight the inherent advantage of our point-based framework, which enables efficient learning from limited data while maintaining scalability to full-resolution inference, potentially saving substantial computational resources when applied to large-scale industrial problems.

\begin{table}[H]
\captionsetup{font=normalsize}
\centering
\caption{Performance evaluation with different point sampling ratios}
\label{table:sampling_rates}
\begin{adjustbox}{width=\textwidth,center}
\begin{tabular}{c|c|c|cc|cc|cc|cc}
\hline
\multirow{2}{*}{\textbf{Model}} & \multirow{2}{*}{\begin{tabular}[c]{@{}c@{}}\textbf{Training time}\\ {\textbf{[h]}}\end{tabular}} & \multirow{2}{*}{\begin{tabular}[c]{@{}c@{}}\textbf{Point Sampling}\\ {\textbf{[\%]}}\end{tabular}} & \multicolumn{2}{c|}{\textbf{Surface pressure}} & \multicolumn{2}{c|}{\textbf{X-Wall Shear Stress}} & \multicolumn{2}{c|}{\textbf{Y-Wall Shear Stress}} & \multicolumn{2}{c}{\textbf{Z-Wall Shear Stress}} \\
&  &  & \textbf{Rel-L2} & \textbf{Rel-L1} & \textbf{Rel-L2} & \textbf{Rel-L1} & \textbf{Rel-L2} & \textbf{Rel-L1} & \textbf{Rel-L2} & \textbf{Rel-L1} \\ \hline
\multirow{5}{*}{\begin{tabular}[c]{@{}c@{}} Multi-output DeepONet \citep{lu2022comprehensive} \end{tabular}} & 11.43 & 10 & 0.143 & 0.080 & 0.145 & 0.091 & 0.326 & 0.289 & 0.270 & 0.239 \\
& 25.55 & 30 & 0.129 & 0.072 & 0.130 & 0.081 & 0.288 & 0.260 & 0.246 & 0.218 \\
& 34.32 & 50 & 0.124 & 0.070 & 0.127 & 0.079 & 0.276 & 0.252 & 0.239 & 0.211 \\
& 42.12 & 70 & 0.123 & 0.069 & 0.125 & 0.078 & 0.274 & 0.248 & 0.236 & 0.208 \\
& 47.53 & 100 & 0.123 & 0.069 & 0.123 & 0.078 & 0.270 & 0.246 & 0.233 & 0.207 \\ \hline
\multirow{5}{*}{Point-wise Diffusion} & 6.15 & 10 & 0.152 & 0.062 & 0.159 & 0.076 & 0.412 & 0.273 & 0.308 & 0.210 \\
& 17.97 & 30 & 0.097 & 0.042 & 0.109 & 0.052 & 0.258 & 0.176 & 0.211 & 0.142 \\
& 30.83 & 50 & 0.084 & 0.036 & 0.094 & 0.045 & 0.213 & 0.148 & 0.190 & 0.123 \\
& 43.45 & 70 & 0.081 & 0.036 & 0.089 & 0.044 & 0.200 & 0.143 & 0.181 & 0.119 \\
& 64.50 & 100 & 0.080 & 0.034 & 0.086 & 0.042 & 0.192 & 0.137 & 0.181 & 0.115 \\ \hline
\end{tabular}
\end{adjustbox}
\end{table} 

\section{Conclusion}
\label{sec:conclusion}

This study introduces a novel point-wise diffusion model that processes spatio-temporal points independently to efficiently predict spatio-temporal and large-scale physical systems with complex geometric variations. Our methodological contribution lies in the development of a point-wise diffusion framework that applies forward and backward diffusion processes at individual spatio-temporal points, coupled with a point-wise DiT architecture for the denoising process. This approach fundamentally differs from conventional image-based diffusion models that operate on structured data representations, as it enables training on arbitrary spatial data types without any preprocessing constraints.

Our comprehensive experimental validation demonstrates improvements across multiple performance metrics and physical domains. The proposed methodology achieves 100-200$\times$ computational speedup through DDIM sampling while maintaining prediction accuracy, establishing its viability for real-time inference applications. Comparative analysis reveals that our point-wise approach outperforms conventional image-based diffusion methods, yielding 35.8\% reduction in mean absolute error with 94.4\% less training time and 89.0\% fewer parameters. Performance evaluations across three distinct physical systems—Eulerian fluid dynamics, Lagrangian solid mechanics, and large-scale aerodynamics—consistently demonstrate superior accuracy, with error reductions ranging from 53\% to 94\% compared to established surrogate models including DeepONet and MGN. Furthermore, the framework exhibits remarkable data efficiency in large-scale automotive aerodynamic systems, maintaining superior performance with only 50\% subsampled training data, and demonstrates robust generalization capabilities to previously unseen geometric configurations. The superior performance of our proposed approach stems from two fundamental design principles that address key limitations of existing approaches: 1) point-wise processing methodology eliminates the need for data preprocessing steps such as grid conversion or mesh connectivity requirements, thereby preserving geometric fidelity and enabling direct handling of complex, irregular geometries; 2) non-autoregressive prediction strategy circumvents temporal error accumulation inherent in sequential methods, facilitating stable long-term predictions for spatio-temporal systems.

However, there are several limitations based on our experimental findings. Although our model demonstrates strong generalization within parametric design spaces encountered during training, its performance for geometric and temporal extrapolation beyond training bounds remains constrained. Additionally, the current implementation focuses primarily on parametric design variations with predefined geometric parameters (e.g., cylinder diameter and position, OCA thickness variations, vehicle morphing parameters), which, while effective for many engineering applications, cannot ensure high accuracy in non-parametric design configurations. 

To address these limitations, following research directions require investigation. Developing enhanced geometric extrapolation capabilities could involve incorporating physics-informed constraints or geometric reasoning mechanisms that enable reliable predictions beyond training boundaries. Extending temporal modeling performance might benefit from integrating long-term stability constraints or hybrid approaches that combine learned dynamics with conservation laws. Furthermore, advancing toward non-parametric geometric handling would enable the framework to accommodate arbitrary design modifications through geometry-aware encoding schemes or adaptive sampling strategies. Such developments would establish a more universal surrogate modeling framework capable of delivering high performance across diverse design scenarios.

\section*{CRediT authorship contribution statement}

\textbf{Jiyong Kim:} Conceptualization, Data curation, Formal analysis, Investigation, Methodology, Visualization, Writing - Original draft, Supervision, Writing – Review \& Editing, Software. \textbf{Sunwoong Yang:} (Co-corresponding authors) Methodology, Supervision, Writing – Review \& Editing. \textbf{Namwoo Kang:} (Co-corresponding authors) Funding acquisition, Project administration, Resources, Supervision, Writing – Review \& Editing.

\section*{Declaration of competing interest}
The authors declare that they have no known competing financial interests or personal relationships that could have appeared to influence the work reported in this paper.

\section*{Data availability}
Data will be made available on request.

\section*{Acknowledgements}
This work was supported by the Ministry of Science and ICT of Korea grant (No. 2022-0-00969, No. 2022-0-00986, and No. RS-2024-00355857), the Ministry of Trade, Industry \& Energy grant (RS-2024-00410810, RS-2025-02317327), and the National Research Council of Science \& Technology (NST) grant by the Korea government (MSIT) (No. GTL24031-000).

\bibliographystyle{elsarticle-num}
\bibliography{cas-refs}

\clearpage
\appendix
\renewcommand{\thetable}{\arabic{table}}  
\renewcommand{\thefigure}{\arabic{figure}}  

\section{Data generation in drop impact system}
\label{sec:Data generation in drop impact system}
\begin{table}[H]
\captionsetup{font=normalsize}
\centering
\caption{Material properties of components in the drop impact system dataset}
\label{table:material_properties}
\begin{adjustbox}{width=0.8\textwidth,center}
\begin{tabular}{cccc}
\hline
Layer & \begin{tabular}[c]{@{}c@{}}Elastic\\Modulus \lbrack GPa\rbrack\end{tabular} & \begin{tabular}[c]{@{}c@{}}Poisson's\\Ratio \end{tabular} & \begin{tabular}[c]{@{}c@{}}Thickness\\\lbrack$\mu$m\rbrack\end{tabular} \\
\hline
Ball & 200 & 0.3 & 5,000 (radius) \\
Cover glass (CG) & 77 & 0.21 & 100 \\
Optically clear adhesive 1 ($\text{OCA}_1$) & 0.01 & 0.45 & 50$\sim$150 \\
Polarizer (POL) & 4 & 0.33 & 50 \\
Optically clear adhesive 2 ($\text{OCA}_2$) & 0.01 & 0.45 & 50$\sim$150 \\
organic light emitting diodes (OLED) & 5.15 & 0.3 & 30 \\
Aluminum plate (PLATE) & 68.9 & 0.33 & 1,200 \\
\hline
\end{tabular}
\end{adjustbox}
\end{table}
\begin{figure}[H]
    \captionsetup{font=normalsize}
    \centering
    \includegraphics[width=0.5\linewidth]{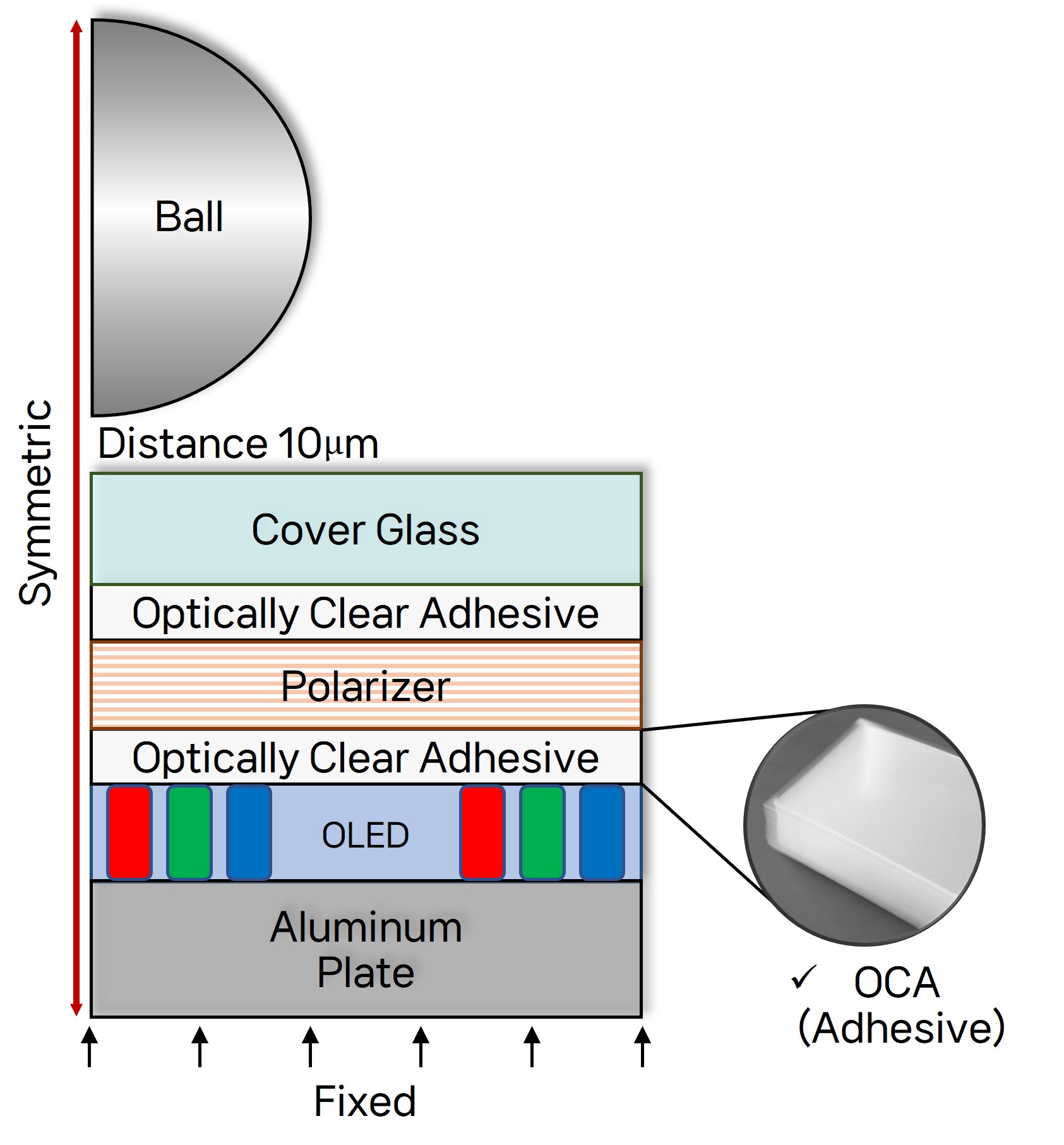}
        \caption{Data configuration of drop impact simulation.}
\label{[Appendix] Fig.17}
\end{figure}

\section{DeepONet Framework}
\label{sec:DeepONet Framework}

\textbf{Single output} In the Eulerian cylinder fluid flow system, we predict only the $x$-velocity field using the standard DeepONet architecture with branch-net processing physical conditions $(u_t, n_t, S_t)$ and trunk-net handling coordinate conditions $(x_t, y_t, t_t^{\text{phys}}$ or $x_t, y_t, z_t)$.
\begin{figure}[H]
    \captionsetup{font=normalsize}
    \centering
    \includegraphics[width=0.6\linewidth]{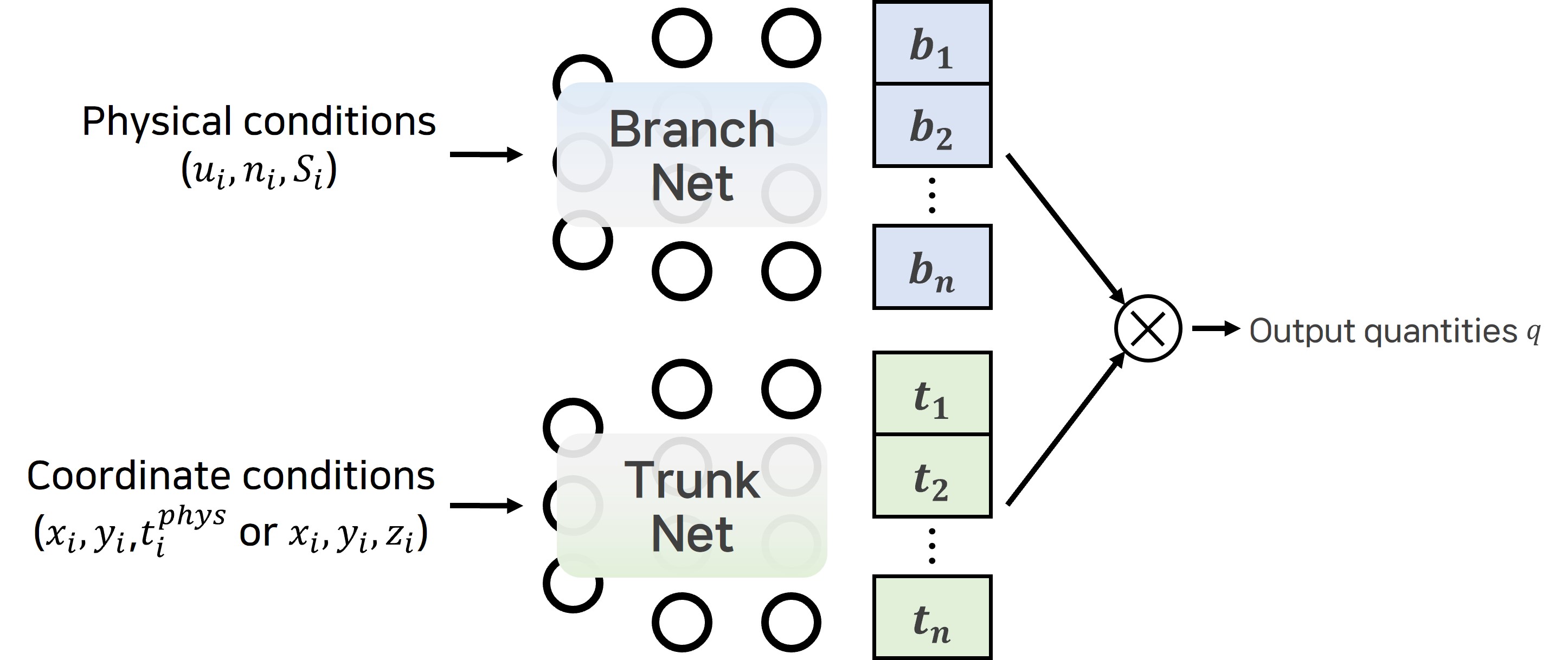}
    \caption{DeepONet architecture for single output prediction.}
    \label{[Appendix] Fig.18}
\end{figure}

\textbf{Multiple outputs} For systems requiring simultaneous prediction of multiple physical quantities such as drop impact simulation (position and stress) and road-car external aerodynamics (surface pressure and wall shear stresses), we implemented the multiple-outputs strategy proposed by \citep{lu2022comprehensive} in the DeepONet framework. The architecture partitions the branch and trunk network outputs into $k$ segments, where each segment performs separate dot products to predict individual output quantities, and $k$ denotes the total number of output quantities.
\begin{figure}[H]
    \captionsetup{font=normalsize}
    \centering
    \includegraphics[width=0.8\linewidth]{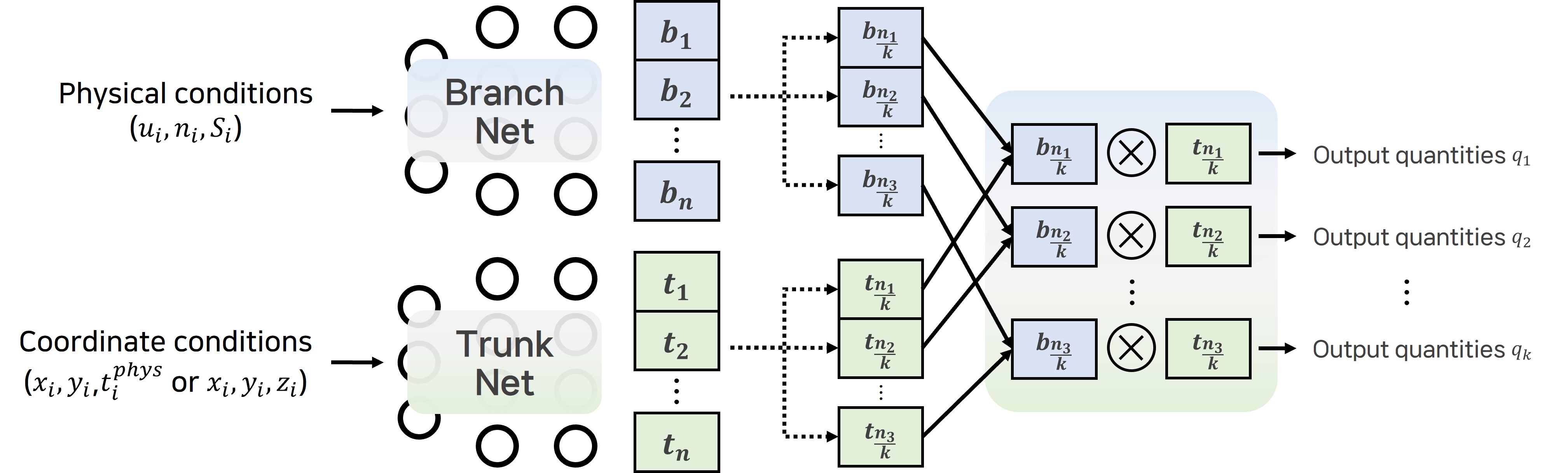}
    \caption{DeepONet architecture for multiple outputs prediction.}
    \label{[Appendix] Fig.19}
\end{figure}

\section{Detailed visualization of surrogate model performance at three physical systems}
\label{sec:Visualization of surrogate model performance at different timestep}

\begin{figure}[H]
    \captionsetup{font=normalsize}
    \centering
    \includegraphics[width=\linewidth]{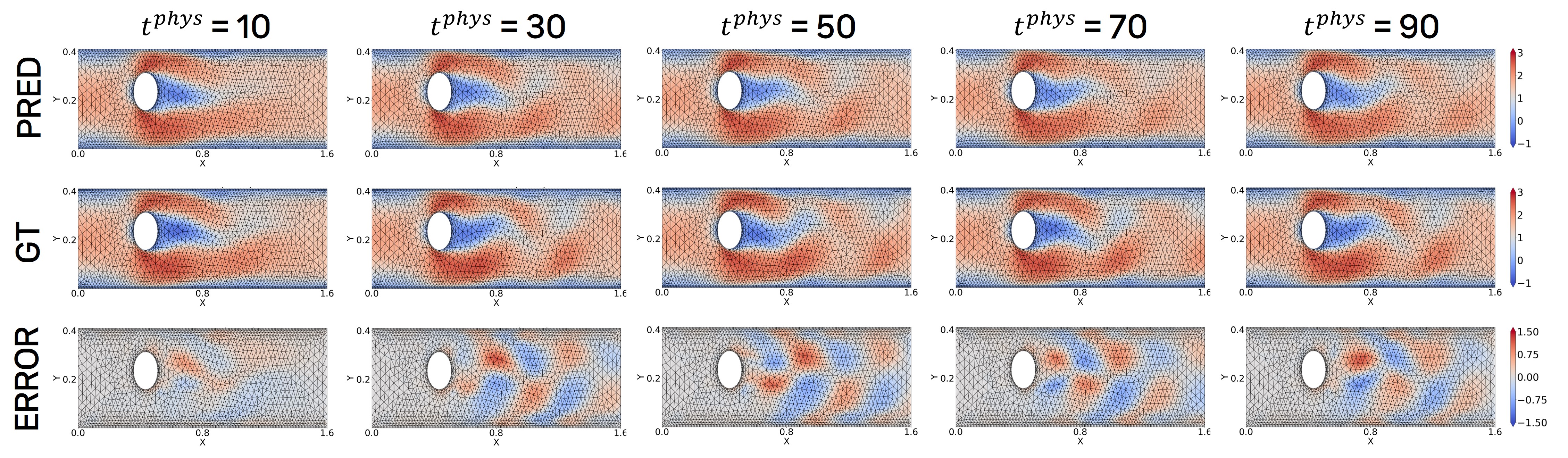}
    \caption{Visual performance comparison of DeepONet for cylinder fluid flow across different physical timesteps ($t^{\text{phys}} = 10, 30, 50, 70, 90$). Top row: predictions (PRED), middle row: ground truth (GT), bottom row: error distribution (ERROR).}
    \label{APPENDIX. [Eulerian] DeepONet}
\end{figure}
\begin{figure}[H]
    \captionsetup{font=normalsize}
    \centering
    \includegraphics[width=\linewidth]{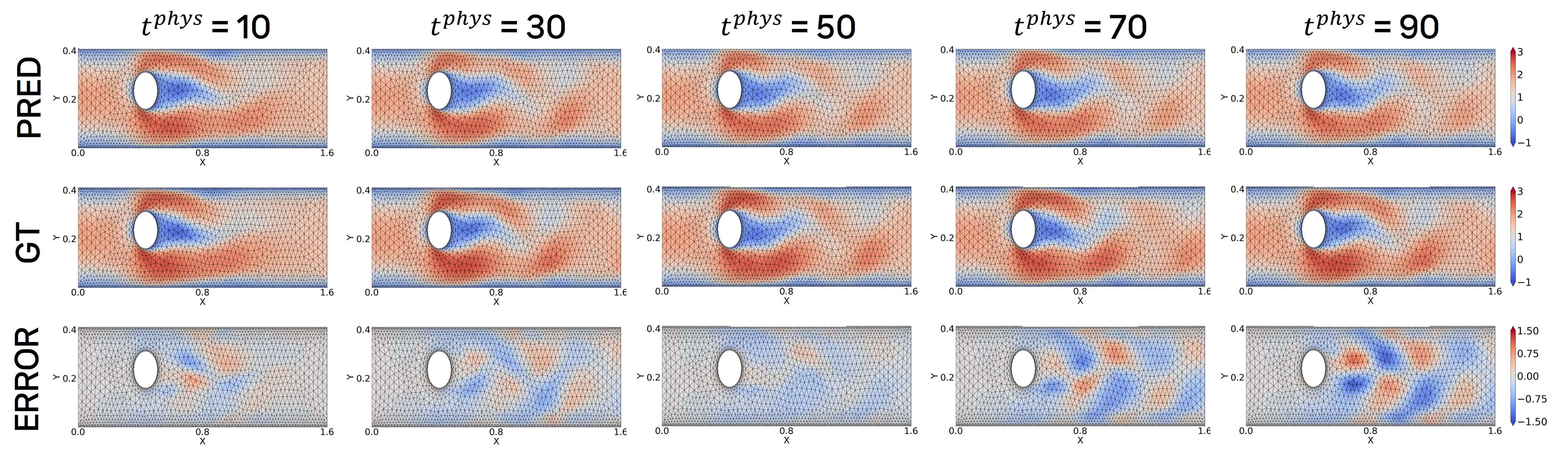}
    \caption{Visual performance comparison of MGN for cylinder fluid flow across different physical timesteps ($t^{\text{phys}} = 10, 30, 50, 70, 90$). Top row: predictions (PRED), middle row: ground truth (GT), bottom row: error distribution (ERROR).}
    \label{APPENDIX. [Eulerian] MGN}
\end{figure}
\begin{figure}[H]
    \captionsetup{font=normalsize}
    \centering
    \includegraphics[width=\linewidth]{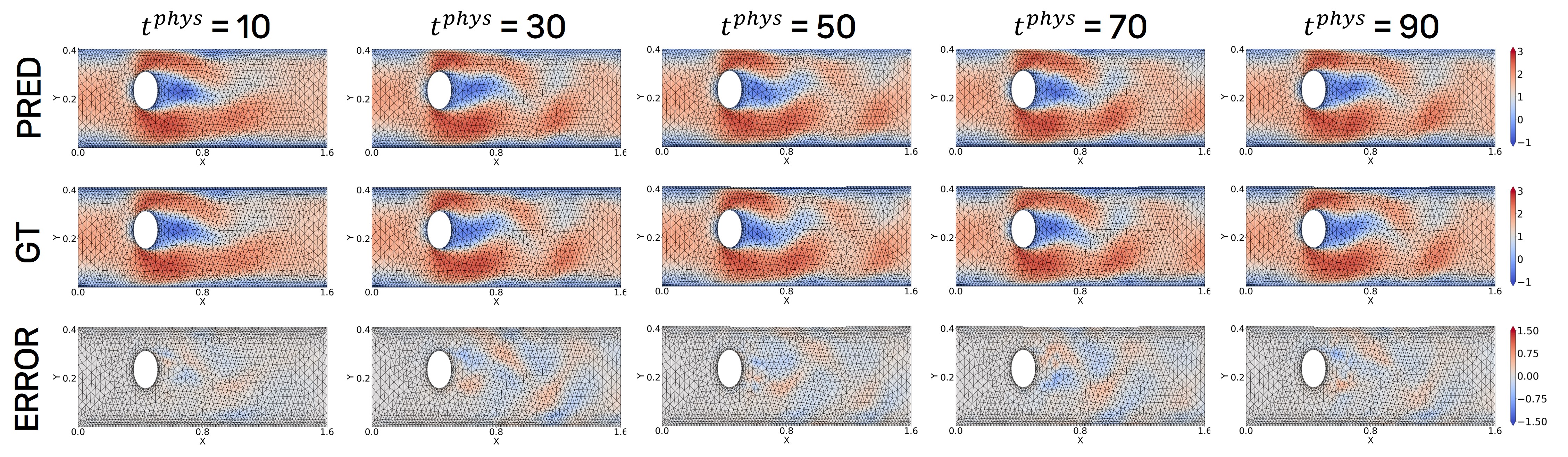}
    \caption{Visual performance comparison of point-wise diffusion model for cylinder fluid flow across different physical timesteps ($t^{\text{phys}} = 10, 30, 50, 70, 90$). Top row: predictions (PRED), middle row: ground truth (GT), bottom row: error distribution (ERROR).}
    \label{APPENDIX. [Eulerian] Point-wise Diffusion}
\end{figure}
\begin{figure}[H]
    \captionsetup{font=normalsize}
    \centering
    \includegraphics[width=\linewidth]{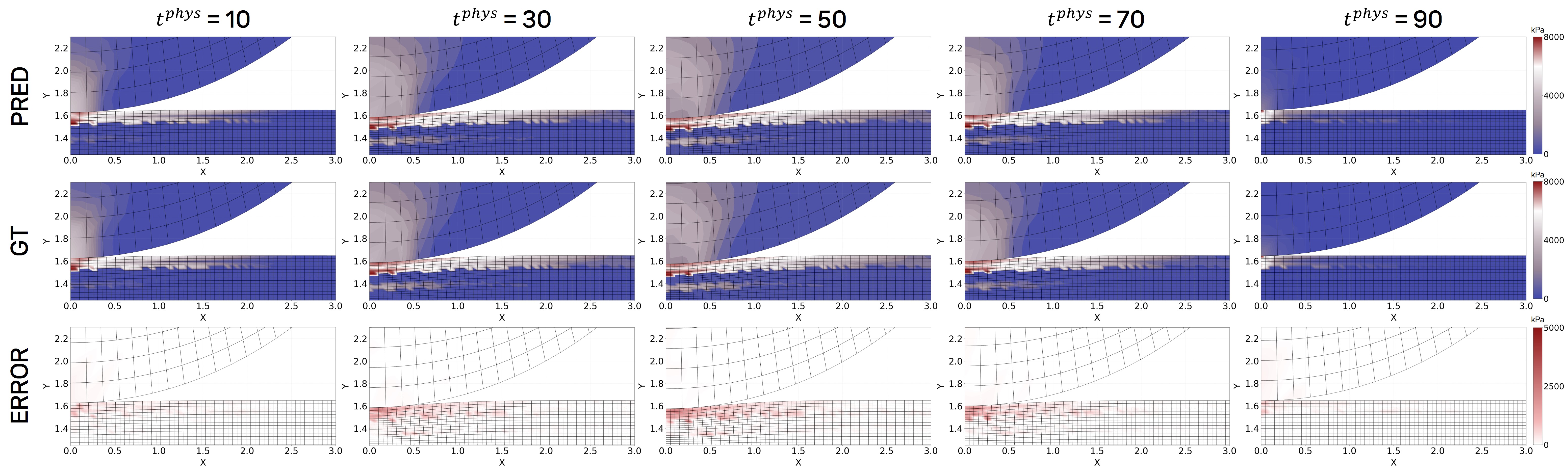}
    \caption{Visual performance comparison of DeepONet for drop impact simulation across different physical timesteps ($t^{\text{phys}} = 10, 30, 50, 70, 90$). Top row: predictions (PRED), middle row: ground truth (GT), bottom row: error distribution (ERROR).}
    \label{APPENDIX. [Lagrangian] DeepONet}
\end{figure}
\begin{figure}[H]
    \captionsetup{font=normalsize}
    \centering
    \includegraphics[width=\linewidth]{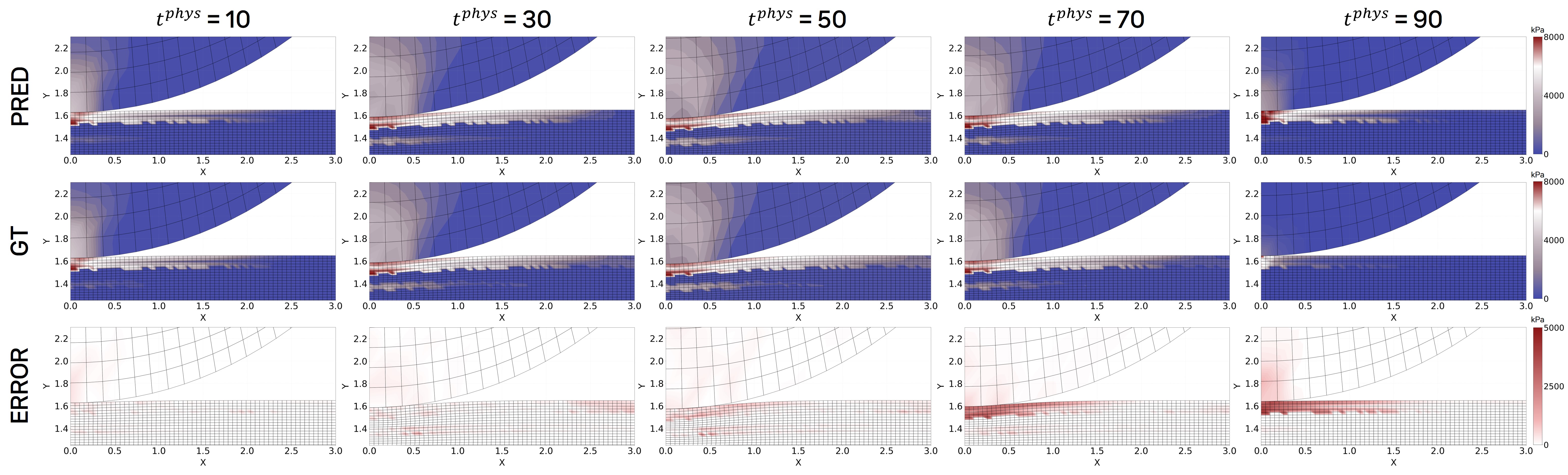}
    \caption{Visual performance comparison of MGN for drop impact simulation across different physical timesteps ($t^{\text{phys}} = 10, 30, 50, 70, 90$). Top row: predictions (PRED), middle row: ground truth (GT), bottom row: error distribution (ERROR).}
    \label{APPENDIX. [Lagrangian] MGN}
\end{figure}
\begin{figure}[H]
    \captionsetup{font=normalsize}
    \centering
    \includegraphics[width=\linewidth]{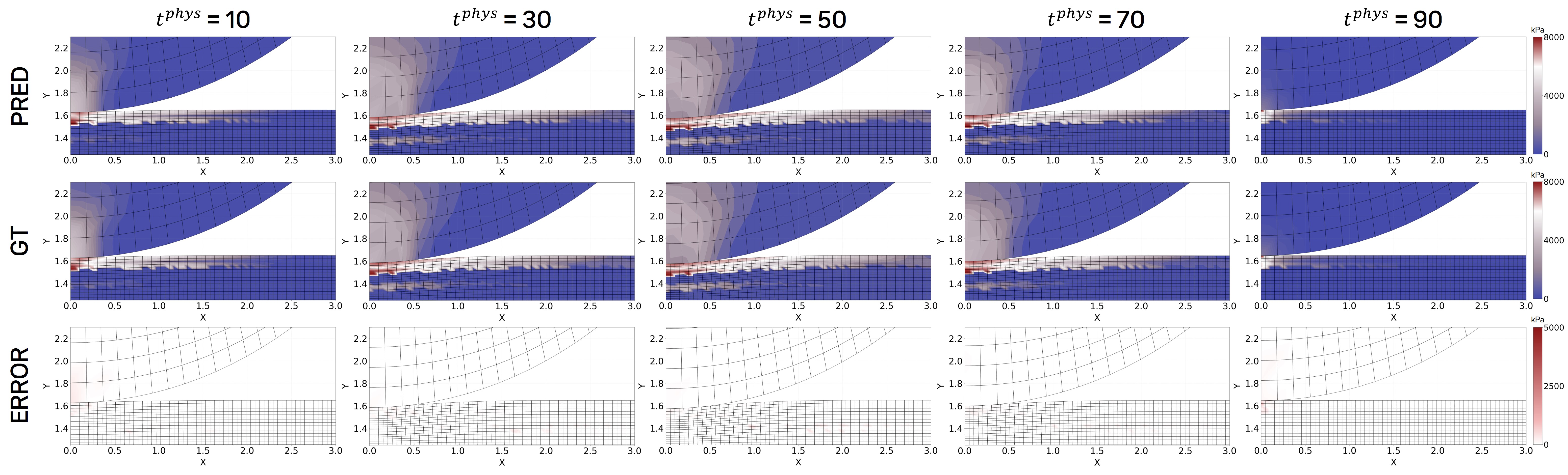}
    \caption{Visual performance comparison of point-wise diffusion model for drop impact simulation across different physical timesteps ($t^{\text{phys}} = 10, 30, 50, 70, 90$). Top row: predictions (PRED), middle row: ground truth (GT), bottom row: error distribution (ERROR).}
    \label{APPENDIX. [Lagrangian] Point-wise Diffusion}
\end{figure}
\begin{figure}[H]
    \captionsetup{font=normalsize}
    \centering
    \includegraphics[width=\linewidth]{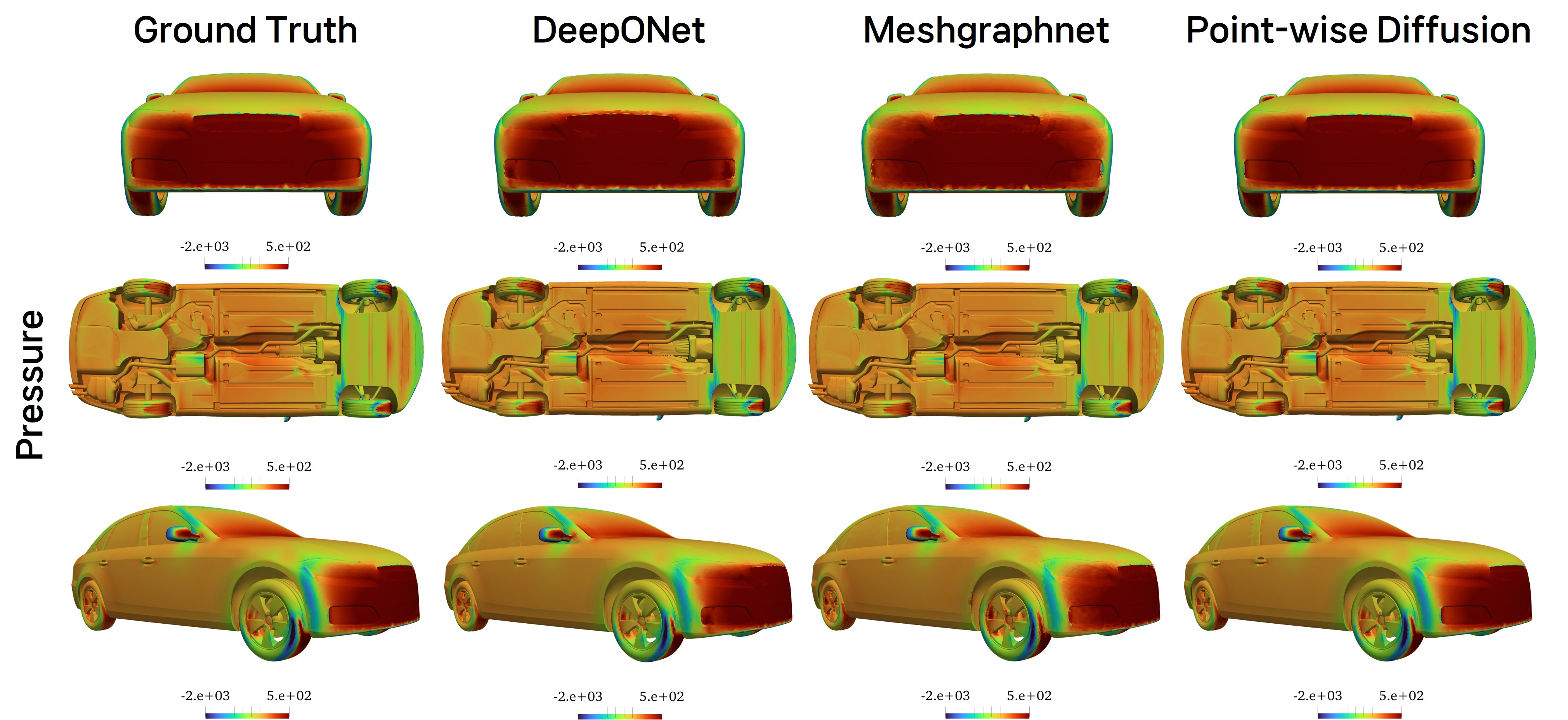}
    \caption{Comparative visualization of surface pressure prediction for road-car external aerodynamics across different models. Each column shows results from Ground Truth, DeepONet, Meshgraphnet, and Point-wise Diffusion models respectively.}
    \label{APPENDIX. [Pressure] Road-car external aerodynamics}
\end{figure}
\begin{figure}[H]
    \captionsetup{font=normalsize}
    \centering
    \includegraphics[width=\linewidth]{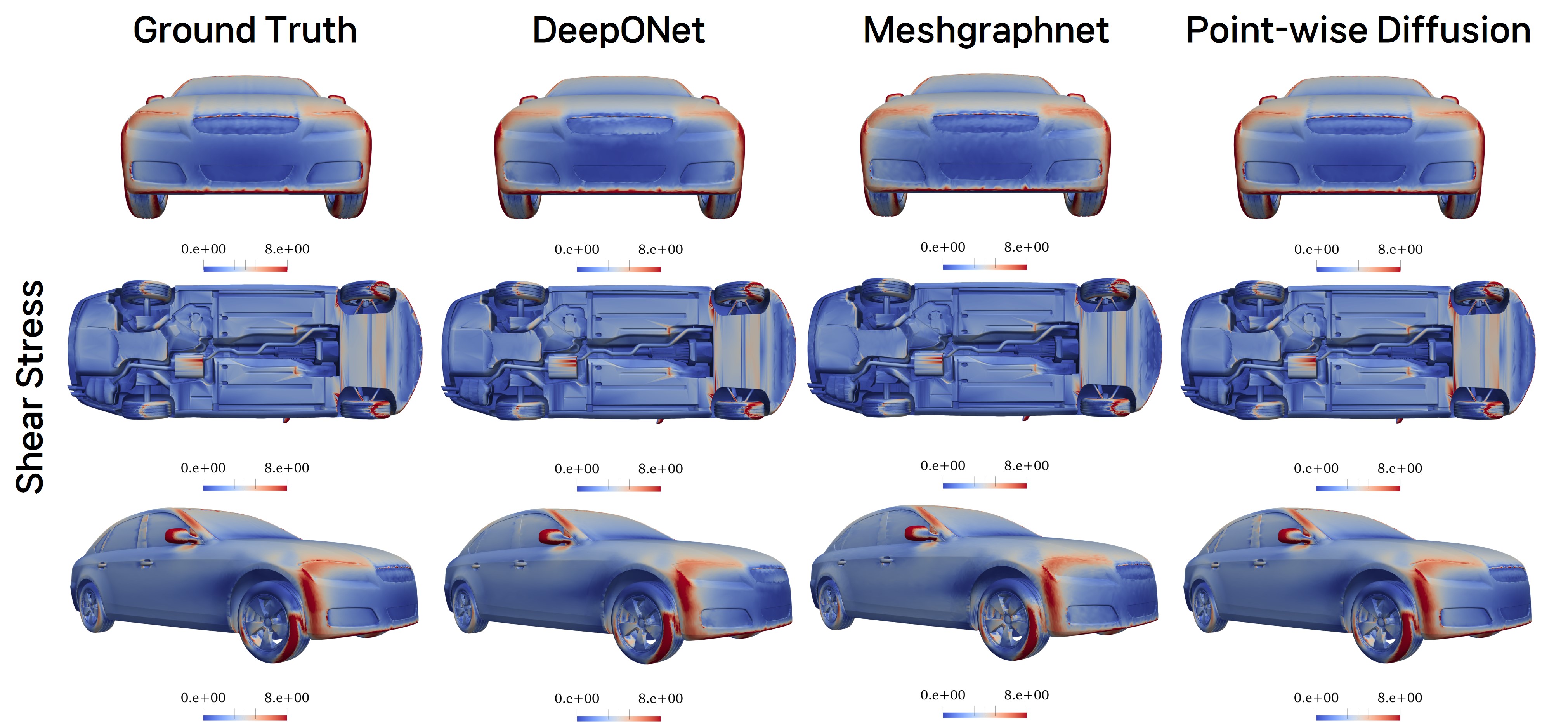}
    \caption{Comparative visualization of wall shear stress prediction for road-car external aerodynamics across different models. Each column shows results from Ground Truth, DeepONet, Meshgraphnet, and Point-wise Diffusion models respectively.}
    \label{APPENDIX. [Shear Stress] Road-car external aerodynamics}
\end{figure}

\end{document}